\documentclass[botnum, fleqn, final]{unmeethesis}
\usepackage{subfigure}
\usepackage{wrapfig}
\usepackage{amsmath,amsfonts,amssymb,amstext}
\usepackage[dvipdfm]{graphicx}
\usepackage[colorlinks,linkcolor=blue]{hyperref}
 \usepackage{float}
\newcommand{\bra}[1]{\ensuremath{\langle #1 \vert}}
\newcommand{\ket}[1]{\ensuremath{\vert #1 \rangle}}
\newcommand{\braket}[2]{\ensuremath{\langle #1 \vert #2 \rangle}}
\newcommand{\mbf}[1]{\mathbf{#1}}

\newcommand{\be}{\begin{equation}}
\newcommand{\ee}{\end{equation}}
\newcommand{\ben}{\begin{eqnarray}}
\newcommand{\een}{\end{eqnarray}}
\newcommand{\bes}{\begin{subequations}}
\newcommand{\ees}{\end{subequations}}
\newcommand{\bF}{\begin{figure}}
\newcommand{\eF}{\end{figure}}

\newcommand{\avg}[1]{\langle #1 \rangle}
\def\ket#1{ | #1 \rangle}
\def\bra#1{{\langle #1 |  }}
\def\tr{ {\rm{Tr }}\,}
\newcommand{\proj}[1]{\mbox{$|#1\rangle \!\langle #1 |$}}
\newcommand{\proja}[1]{\mbox{$|#1\rangle_A\langle #1 |$}}
\newcommand{\projb}[1]{\mbox{$|#1\rangle_B\langle #1 |$}}

\newcommand{\arxiv}[2][arxiv:]{\href{http://arxiv.org/abs/#1#2}{#1#2}}

\begin{document}


\frontmatter



\title{Quantum Correlations, Chaos and Information}

\author{Vaibhav Madhok}

\degreesubject{Ph.D., Physics}
    
\degree{Doctor of Philosophy \\ Physics}

\documenttype{Dissertation}

\previousdegrees{B.S., Computer Science, University of Pune, 2002 \\
                 M.S., Physics, Colorado State University, 2007}

\date{July, \thisyear}

\maketitle

\makecopyright

\begin{dedication}
  To my mother and grandparents for their love, patience and sacrifices.
      \\[3ex]
   ``O Life, you put many traps in my way\\
Dare to try, is what you clearly say\\
All that is, thy command must obey\\
You lead me away and call me astray.''\\
         -- the Rubaiyat of Omar Khayyam
\end{dedication}
\nopagebreak[4]
\begin{acknowledgments}
   \vspace{1.1in}
   I would like to express my gratitude for my advisor, Professor Ivan H. Deutsch, for the education, training and financial support I received during my stay at the UNM. Ivan encouraged my creative endeavors by giving me the freedom to pursue my own research interests and directions while helping me have focus at the same time. I also thank him for his patience by letting me figure out things at my own pace and his understanding for allowing me to have an eccentric work schedule. All these years, it has been a pleasure to work with him in an environment of mutual trust, respect and honesty driven by a common desire to do creative and meaningful research. Above all, I hope to inherit his wonderful tastes in judging what constitutes good science.  
  
  I also thank my colleagues, Animesh Datta, Collin Trail and Carlos Riofrio for their participation in the research projects I undertook. Thanks to Bryan Eastin, Josh Combes, Krittika Goyal, Anil Shaji, Rob Cook, Jonas Anderson, Ben Baragiola, Leigh Norris, Matthias Lang, Zhang Jiang, Alex Tacla, Xuefang Zhang, Seth Mekel, Cris Ceasre and other CQUIC members serving as critique of my work and having numerous enlightening discussions.
  
  At UNM, I benefited greatly from taking courses and having interactions with Prof. Kevin Cahill, Prof. Daniel Finley, Prof. Sudhakar Prasad, Prof. Nitant Kenkre, Prof. Susan Atlas, Prof. Andrew Landahl and Prof. Cris Moore.
      \newpage
Outside UNM, I thank Dr. Salman Habib and Dr. Tanmoy Bhattacharya for being my mentors during my stint as a summer fellow at LANL. Prof. Arul Lakshminarayan has been a great help, always willing to discuss and provide his inputs on a wide range of topics spanning quantum chaos. I also thank Prof. Richard Eykholt at the Colorado State University for initiating me into research in classical chaos. Many thanks to Prof. Shohini Ghose for serving on my committee. It was her projects and research that got me interested in the field of quantum chaos. 

I thank my friends Navin, Pramod, Amitabh, Arnab, Japji, Rajeev, Dibyendu, Tom, Aaron, Anushka, Ashwin, Manju, Srikanth, Niranjan, Prabhakar, Hari, Ravi, Pankaj, Proma, Soumik, Manoj, Shilpa, Iniyan, Pearlson, Vijay and Wen Yun for their companionship all these years.
I would also like to acknowledge the support and advice of my friends Ankit, Vipul, Inderaj, Pranesh, Pankaj, Atul, Mustali, Anita and Tej. Thanks also to Roy and Irene, Daniel Mirell and Desiree.

 Dr. Abhijat Vichare, Dr. Shiraz Minwalla and Dr. Mustansir Barma deserve a special mention for reminding me, the thing we all know but fail to realize, - ``to follow the heart".

A journey of this magnitude involving uncertainty and setbacks requires nourishment for the soul. For this, I thank the Nur Akshi Jerrahi Sufi order, The Art of Living team in Albuquerque, especially Robert and Diego, and the Albuqueruqe Sikh gurudwara by always welcoming me as one of them.   

Back home in India, I express gratitude to family and friends. I bow and thank to Giridharijee, Prem, Jagdeesh, Manraj, Edmond and Daulatjee who worked with love and affection in making my life comfortable. 
There is so much I learnt from my parents, not by listening to what they say as I seldom did that, but by their very presence. I thank my father, Dr. Ramesh Kumar Madhok, for being an example of generosity, strength of character and courage. I thank my mother, Mrs. Shashi Madhok, for teaching me the value of sincerity, creativity and for being my first teacher. At the same time, most of what I have done is by not conforming to their or anyone else's world view and opinions. 
This is one message I would share with the reader. Follow your heart and you shall be fine.

Lastly, I thank my mates from the CSU, UNM and Colorado cricket teams for the wins, the losses, the battles fought together and above all the camaraderie. 
\end{acknowledgments}

\maketitleabstract 
\begin{abstract}

  Quantum chaos is the study of quantum systems whose classical description is chaotic. How does chaos manifest itself in the quantum world? 
In this spirit, we study the dynamical generation of entanglement as a signature of chaos in a system of periodically kicked coupled-tops, where chaos and entanglement arise from the same physical mechanism.   The long-time  entanglement as a function of the position of an initially localized wave packet very closely correlates with the classical phase space surface of section -- it is nearly uniform in the chaotic sea, and reproduces the detailed structure of the regular islands.  The uniform value in the chaotic sea is explained by the random state conjecture. As classically chaotic dynamics take localized distributions in phase space to random distributions, quantized versions take localized coherent states to pseudo-random states in Hilbert space. Such random states are highly entangled, with an average value near that of the maximally entangled state. For a map with global chaos, we derive that value based on new analytic results for the entropy of random states.  For a mixed phase space, we use the Percival conjecture to identify a ``chaotic subspace" of the Hilbert space.  The typical entanglement, averaged over the unitarily invariant Haar measure in this subspace, agrees with the long-time averaged entanglement for initial states in the chaotic sea. In all cases the dynamically generated entanglement is that of a random complex vector, even though the system is time-reversal invariant, and the Floquet operator is a member of the circular orthogonal ensemble.

 Continuing on our journey to find the footprints of chaos in the quantum world, we explore quantum signatures of classical chaos by studying the rate of information gain in quantum tomography.  The measurement record is obtained as a sequence of expectation values of a Hermitian operator evolving under repeated application of the Floquet operator of the quantum kicked top on a large ensemble of identical systems. We find an increase in the rate of information gain and hence higher fidelities in the process when the Floquet maps employed increase in chaoticity. We make predictions for the information gain using random matrix theory in the fully chaotic regime and show a remarkable agreement between the two. Finally we discuss how this approach can be used in general as a benchmark for information gain in an experimental implementation based on nonlinear dynamics of atomic spins measured weakly by the Faraday rotation of a laser probe.
        
The last part of this thesis is devoted to the study of the nature of quantum correlations themselves.
Quantum correlations are at the heart of the weirdness of quantum mechanics and at the same time serve as a resource for the potential benefits quantum information processing might provide. For example, Einstein described quantum entanglement as ``spooky action at a distance"~\cite{epr}. However, even entanglement does not fully capture the complete quantum character of a system. Quantum discord aims to fill this gap and captures essentially all the quantum correlations in a quantum state~\cite{oz02}. 
There is a considerable interest in the research community about quantum discord, since there is evidence
showing this very quantity as responsible for the exponential speed up of a certain class of quantum algorithms over classical ones~\cite{dsc08}.

Now, an important question arises: Is discord just a mathematical construct or does it have a definable physical role in information processing?
  This thesis provides a link between quantum discord and an actual physical task involving communication between two parties~\cite{md10}.
We present an operational interpretation of quantum discord based on the quantum state merging protocol. Quantum discord is the markup in the cost of quantum communication in the process of quantum state merging, if one discards relevant prior information. We further derive a quantitative relation between the yield of the fully quantum Slepian-Wolf protocol in the presence of noise and the quantum discord of the state involved. This protocol is the most general known in the family of protocols in quantum information theory, a unification of essentially all bipartite, unidirectional and memoryless quantum communication protocols. The significance of quantum discord in noisy versions of teleportation, super-dense coding, entanglement distillation and quantum state merging are discussed. We also demonstrate similar roles for quantum discord in quantum computation and correlation erasure. Our work shows that quantum discord captures and quantifies the advantage of quantum coherence in quantum communication.
\clearpage 

\end{abstract}

\tableofcontents

\listoffigures


\mainmatter

\chapter{Introduction}
  
 Information is physical. As first emphasized by Landauer \cite{l61}, this indicates that the information processing capabilities of a device are not separate from the physics that governs its operation. On the one hand, over the last few decades, the field of quantum information science has shown that the devices governed by the laws of quantum mechanics can have information processing capabilities superior to their classical counterparts. It has also shed light on what properties of  quantum systems should be harnessed for building future technology and engineering applications. Here, the role of  information, quantified suitably as entanglement and other measures of quantum correlations between sub-systems, provides vital clues to the superior information processing capabilities of devices based on quantum mechanics. 
 
 On the other hand, and at a more fundamental level, understanding how physical systems process and exchange information is crucial to gaining insights into the workings of our universe. For example, the connections between entropy, information and thermodynamics form the cornerstone of statistical mechanics. An example more relevant to us is Feynman's path-integral formulation of non-relativistic quantum mechanics \cite{f48}.  In this approach, physical phenomena are described by \emph {events}. An \emph{event} can occur through various alternatives/paths, each of which is characterized by a complex probability amplitude, that has both real and imaginary components in general. If these paths are in principle indistinguishable, i.e. there is no information whatsoever in the universe that can help us distinguish between them, then in that case the corresponding probability amplitudes add up causing  interference. For example, consider the 2-slit experiment. As long as there is no information available as to which slit the photon takes, we see interference fringes on the screen. The photon is considered to be in a superposition of two wave packets, centered around the classical path out of each slit. The question as to why does the universe care about which ``path" the photon takes is both philosophical and intriguing. It certainly tells us that the universe cares about certain kinds of information or the lack of it in formulating its laws.

We saw above that a photon is considered to be in a superposition of two wave packets, centered around the classical path out of each slit. In general, the system is described by a vector in the Hilbert space. This vector in general is a superposition of $N$ basis vectors (this comes from the linearity of Schr$\ddot{\rm{o}}$dinger's equation), where $N$ is the dimension of the Hilbert space. For an infinite dimensional Hilbert space, the sum in the superposition is replaced by an integral. The macroscopic world around us does not permit the majority of these state vectors. For example, a tennis ball is never seen in a superposition of two values of position simultaneously. This leads us to another fundamental question : How does the classical physics governed by Newton's laws emerge out of the underlying quantum mechanics? The flow of information in quantum physics is at the heart of understanding the emergence of the
classical world and the disappearance of quantum coherence via decoherence. Decoherence, which is the leaking of information to the environment, suppresses quantum interference effects and enables the quantum to classical transition. Thus we see that information plays a fundamental role, perhaps as fundamental as that of energy and momentum, in the workings of our universe.
         
The connections between information and dynamical properties of quantum systems is a recurring theme of this thesis.
The first half of this thesis looks at quantum chaos from a quantum information science perspective.

\section{Classical and quantum chaos}
 
Broadly speaking, classical chaos is the long term aperiodic behavior in a deterministic dynamical system that exhibits
sensitive dependence on initial conditions \cite{s94}. Aperiodic long term behavior means that there are trajectories which do not settle down to fixed points, periodic orbits, or quasiperiodic orbits in the limit $t \rightarrow \infty$, where $t$ is the time of evolution of the trajectory. Sensitive dependence on initial conditions means that nearby trajectories separate exponentially fast; the rate of separation is given by the Lyapunov exponent, $\lambda$, which 
characterizes the dynamics of the system. A system with $N$ degrees of freedom having $N$ constants of motion is said to be integrable and its dynamics is regular. When one has fewer than $N$ constants of motion, then the individual trajectories can explore the phase space in a complex manner and the system can exhibit chaos. 
   
It is not difficult to see that the above ``definition" of chaos fails in the quantum domain. A quantum state is not a point in the phase space but is described by a state vector. The time evolution of the state vector due to the  Schr$\ddot{\rm{o}}$dinger's equation is unitary. This means that the overlap of two state vectors undergoing evolution is \textit{constant} with time.
Furthermore, while classical chaos can lead to infinitely fine structures in the phase space, in quantum mechanics, Planck's constant, $\hbar$, sets the scale for such structures. According to Heisenberg's uncertainty principle, our resolution of the phase space is determined by Planck's constant.
 This is often stated as the key reason for the absence of chaos in the quantum domain. This however is not the complete story. An alternate description of classical mechanics, involving the evolution of classical probability densities, preserves the distance between two probability densities as a function of time.  Hence, the distance between two probability densities does not show exponential sensitivity even for classical mechanics.
        
 All this leads to two interesting questions: 
\begin{enumerate}
\item How does classically chaotic dynamics inform us about certain properties of quantum systems, e.g. the energy spectrum, nature of eigenstates, correlation functions, and more recently, entanglement. Alternatively, what features of quantum systems arise due to the fact that their classical description is chaotic?
\item  Since all systems are fundamentally quantum, how does the classical chaos, with trajectories sensitive to initial conditions, arise out of the underlying quantum equations of motion?   
 \end{enumerate}
 These two questions are not unrelated. However, the first question deals mainly with finding the signatures of chaos by studying the properties of the quantum Hamiltonian, while the second concerns with the dynamical behaviour of quantum states and the emergence of classically chaotic behaviour.    
 
 One of the  first steps in addressing question (1) was taken by Gutzwiller ~\cite{g70} by developing the trace formula.  The ``trace formula" relates the density of states of the quantum system in the semi-classical regime to the properties of the classical periodic orbits.  A central result of quantum chaos is the connection with the theory of random matrices \cite{Haake}.  In the limit of large Hilbert space dimensions (small $\hbar$), for parameters such that the classical description of the dynamics shows global chaos, the eigenstates and eigenvalues of the quantum dynamics have the statistical properties of an ensemble of random matrices.  The appropriate ensemble depends on the properties of the quantum system under time-reversal \cite{Haake}. 
 In this direction an important quantum signature of chaos was obtained by Bohigas and collaborators \cite{Bohigas}, describing the spectral statistics of quantum Hamiltonians whose classical counterparts exhibit complete chaos using random matrix theory. Such signatures of quantum chaos have mainly focussed on the time-independent Schr$\ddot{\rm{o}}$dinger's equation and features like energy spectra and eigenstates. The connections between quantum chaos, time reversal symmetry and Random Matrix Theory will be important to us, and hence we devote the next chapter giving an overview of these aspects. 
 
 In studying the dynamical aspects of quantum chaos, it is worth mentioning the conflict between the classical and quantum theories in describing the future dynamics of system. Ehrenfest's correspondence principle gives us an estimate of how well the quantum expectation values follow Hamilton's canonical equations of motion. 
The ``quantum break time" is the characteristic time at which the two diverge.  This correspondence follows when the width of a well localized quantum state is extremely narrow as compared to the characteristic variations in the macroscopic potential. For systems having regular dynamics, significant deviations between the quantum expectation values and classical dynamics occur on very late time scales \cite{Liboff, Messiah}. These deviations scale roughly with the size of the system in action relative to $\hbar$. This, however, is not true for chaotic systems. In the case of chaotic systems, the break time can be logarithmic in the size of the action relative to $\hbar$, and thus even for macroscopic systems, this time will be surprisingly short. Zurek took this argument to its absurd conclusion and showed how the trajectory of Hyperion, a moon of Saturn, will diverge from classical behavior after only 20 years!  He concluded that decoherence resolves the paradox \cite{oz02, z98}.   
     
 Habib and collaborators picked up this problem almost a decade ago and studied it from the point of view of continuous measurement and quantum trajectories \cite{bhj03}.  In particular, they asked the following question: Under what conditions does a quantum trajectory that tracks a measurement of a given observable follow the classical trajectory?  For the case of a single degree of freedom they showed conditions under which this was true.  
 In particular they showed that if the backaction was strong enough that the wave packet remain sufficiently localized that interference across different classical trajectories is negligible, but weak enough that the backaction noise is small compared to the quantum uncertainty, then the system follows essentially classical trajectories.  These conditions could be satisfied if the system is sufficiently macroscopic and the measurement backaction is sufficiently strong. Ghose and collaborators extended this to study the quantum to classical transition in a bipartite system consisting of coupling between spin and motional degrees of freedom \cite{Ghose03}. They concluded that even when the measured motional degree of freedom can be treated classically, the entanglement between the two subsystems causes strong measurement backaction on the quantum spin subsystem and one does not recover classical trajectories in this regime. It is important for both subsystems to be sufficiently macroscopic for the recovery of classical trajectories. Habib et al. extended this to study the Lyapunov exponents of quantum trajectories even when the system was not macroscopic \cite{bhj00}.
  
  Though quantum systems show no exponential separation under the evolution of a known unitary evolution, they do show a sensitivity to the parameters in the Hamiltonian \cite{per84}. Peres \cite{per84} showed that the evolution of a quantum state is altered when a small perturbation is added to the Hamiltonian. As time progresses, the overlap of the perturbed and unperturbed states gives an indication of the stability of quantum motion. It was shown that if a quantum system has a classically chaotic analog, this overlap has a very small value. On the other hand, if the classical analog is regular, the overlap remains appreciable.
  In another perspective, as seen in the work of Schack and Caves, quantum systems exhibit chaos when they are perturbed by the environment. They become hypersensitive to perturbations \cite{sc96}, as seen in the information theoretic studies of the cost to maintain low entropy in the face of loss of information to the environment. This particular feature of quantum chaotic systems has several interesting consequences. For example, Shepelyansky has done extensive work on the issue of many-body quantum chaos in the quantum computer hardware and its effect on the accuracy of quantum computation \cite{Shepelyansky}. Recently classical simulations of quantum dynamics have been connected to integrability and chaos \cite{pz07}.

It is imperative to mention the role played by quantum information theory in the above journey.
Quantum information science has added a whole new perspective to the study of quantum mechanics. This has resulted in a better understanding of quantum phenomena like entanglement and decoherence, and given us the tools to view certain quantum properties of physical systems as a resource. This has also enabled us to address the key questions in quantum chaos from a new perspective. As mentioned above, this has led to an information theoretic characterization of quantum chaos \cite{sc96} and enabled the exploration of the behavior of chaotic quantum systems in the presence of environment induced decoherence \cite{Zurek/Paz} along with its connection to the quantum to classical transition. The study of quantum chaos from a quantum information perspective is also closely related to the theory and application of random quantum circuits \cite{e03}. 
  
A major part of this thesis is devoted to the study of signatures of quantum chaos as seen in the light of quantum information theory.  We explore the connections between entanglement, information generation in quantum tomography, and chaos. Our studies throw light on the connections between ergodicity, entropy production and chaos. Moreover, employing continuous measurement quantum tomography, we develop an entirely novel paradigm to study quantum chaos.

Here I introduce and motivate the work undertaken in this thesis in the above quest. I also mention the chapters where the reader can find the details of the projects undertaken.

\section{Dynamically generated entanglement as a signature of quantum chaos}

The connections between complexity, nonlinear dynamics, ergodicity, and entropy production, have long been at the heart of the foundations of statistical physics. A central goal of ``quantum chaos" has been to extend this foundation to the quantum world. In the last decade, the tensor product structure of quantum mechanics, essential for understanding systems with multiple degrees of freedom, has come to the fore.  In that context, one is naturally led to consider how the dynamical generation of entanglement between quantum subsystems is connected with the chaotic dynamics of coupled classical degrees of freedom.  Such studies address fundamental issues of complexity in quantum systems and are potentially applicable in quantum information processing, where entanglement is considered to be an essential resource.  

The connection between chaos in the classical description of Hamiltonian dynamics and entanglement in the quantum description has been the subject of extensive study over the last decade.  The original motivation of Zurek and Paz  was to address the quantum-to-classical transition \cite{Zurek/Paz}.  By conjecturing that chaotic systems decohere exponentially fast through their entanglement with the environment, they hoped to resolve a paradox in which a macroscopic system would exhibit the effects of quantum coherence on a time scale logarithmic in $\hbar$.  

  
  To address these issues, we consider a model system of kicked coupled-tops,  described in detail in Chapter 3.  This system is motivated by its connection to possible experimental realizations, our ability to easily visualize the classical phases space, and to analyze the Floquet map.  We use this system as a forum to explore questions (1) and (2) above -- how is chaos in the classical dynamics of the joint system correlated with the long-time averaged dynamically generated entanglement? 
\section{Quantum signatures of chaos in quantum tomography}
At a fundamental level chaos represents ``unpredictability", so this seems at odds with the goal of gaining information in order to estimate an unknown quantum state. However, on the flip side, this unpredictability represents the potential ``information to be gained" in an estimation process. If everything is predicted and known, we learn nothing new.  The missing information in deterministic chaos is the \textit{initial condition}. A time history of a trajectory at discrete times is an archive of information about the initial conditions given perfect knowledge about the dynamics. Moreover, if the dynamics is chaotic the rate at which we learn information increases due the rapid Lyapunov divergence of distinguishable trajectories and we expect unbiased information because of the ergodic mixing of phase space.  That is, if the information is generated by chaotic dynamics, the trajectory is random, and all initial conditions are equally likely until we invert the data and discover the initial state.  
  
The standard way to perform quantum tomography is to make projective measurements of an ``informationally complete" set of observables and repeat them many times. The statistics obtained are used to estimate the expectation values of the observables and hence the unknown initial state.
The projective measurements pose a hurdle in exploring the connections between information gain in tomography and chaos due to large measurement back-action on the system. However, we overcome this by employing the protocol
for tomography via weak continuous measurement developed by Silberfarb et al. \cite{sjd05}.
In this protocol, the ensemble is collectively controlled and probed in a time dependent manner to obtain an ``informationally complete" continuous measurement record. We consider the case of a very weak measurement such that the back-action is negligible.
 Our protocol for quantum tomography via continuous measurement of a driven system \cite{sjd05} gives us a window into the complexity of quantum dynamics and its relationship to chaos. This work is intimately related to the protocols that have recently been implemented in the laboratory \cite{jessen}.
 
 As in the case of deterministic chaos, the missing information in tomography is the \textit{initial condition}. We try to accurately model all of the quantum dynamics occurring in the system, and then use the measurement time history to give us information about the initial quantum state.  The dynamics is ``informationally complete" if the time history contains information about an arbitrary initial condition. Our goal is to characterize and quantify the performance of tomography, when the dynamics driving the system are chaotic in the classical limit. We use this to draw comparisons 
 between the role played by regular and chaotic dynamics in the information gain in the tomography procedure.
 In chapter 4, we report the novel and intriguing quantum signatures of classical chaos we have under this paradigm. 
      
 The second half of the thesis is concerned with the study of quantum correlations and exploring their role
in quantum information processing protocols.

 \section{Nonclassical correlations and their role in \\
 quantum communication}
 Quantum information science is primarily aimed at harnessing the quantum structure of nature for information processing and computing tasks~\cite{nielsen00a}. This quest has met with considerable success over the last 20 years, but there has been substantial progress in the other direction as well. Information theory has provided a novel framework for unraveling the intricacies of quantum mechanics. Quantum correlations, as well as classical ones are now viewed as resources, whose interconvertibility is governed by quantum information theory~\cite{dhw08}. Foremost amongst these is evidently entanglement, which provides enhanced performance in several important tasks like communication, computation, metrology and others~\cite{pv07}. However, even entanglement does not fully capture the complete quantum character of a system. \\
Quantum discord aims to fill this gap and captures essentially all the quantum correlations in a quantum state~\cite{oz02}. 
There is a considerable interest in understanding the meaning of quantum discord, since there is evidence
showing this very quantity as responsible for the exponential speed up of a certain class of quantum algorithms over classical ones~\cite{dsc08}.
Now, an important question arises: Is discord just a mathematical construct or does it have a definite physical role in information processing?
  This thesis provides a link between quantum discord and an actual physical task involving communication between two parties~\cite{md10}.
In chapter 5, we present an operational interpretation of quantum discord based on the quantum state merging protocol. Quantum discord is the markup in the cost of quantum communication in the process of quantum state merging, if one discards relevant prior information. \\
In chapter 6, we further derive a quantitative relation between the yield of the fully quantum Slepian-Wolf protocol \cite{adhw09} in the presence of noise and the quantum discord of the state involved. This protocol is the most general known in the family of protocols in quantum information theory, a unification of essentially all bipartite, unidirectional and memoryless quantum communication protocols. The significance of quantum discord in noisy versions of teleportation, super-dense coding, entanglement distillation and quantum state merging are discussed. We also demonstrate similar roles for quantum discord in quantum computation and correlation erasure. Our work shows that quantum discord captures and quantifies the advantage of quantum coherence in quantum communication. 

\section{Related Work and List of Publications}

Several new results are presented in this thesis.
The work presented here has been published in refereed journals and conferences. 

\begin{enumerate}

\item Entanglement and generation of random states in the quantum chaotic dynamics of kicked coupled tops, Collin M. Trail, Vaibhav Madhok, and Ivan H. Deutsch, \textit{Phys. Rev. E}, \textbf{76}, 046211, (2008). (Chapter 3)

\item Quantum signatures of chaos in quantum tomography, Vaibhav Madhok, Carlos Riofrio, and Ivan H. Deutsch in preparation. (Chapter 4)

\item Interpreting quantum discord through quantum state merging,  Vaibhav Madhok and Animesh Datta, \textit{Phys. Rev. A} \textbf{83}, 032323, (2011). (Chapter 5)

\item  Role of quantum discord in quantum communication, Vaibhav Madhok and Animesh Datta, arxiv:1107.0994 (2011),  6th Conference on Theory of Quantum Computation, Communication and Cryptography ( Springer's Lecture Notes in Computer Science, 2011). (Chapter 6)

\item  Quantum discord as a resource in quantum protocols,
    Vaibhav Madhok and Animesh Datta, submitted to  \textit{Phys. Rev. Lett.} (Chapter 6)

\item Quantum discord : The advantage of coherence in quantum communication, Vaibhav Madhok and Animesh Datta \textit{invited} article to appear in the International Journal of Modern Physics B, (Chapter 2 and Chapter 6).

\end{enumerate}

In addition, I have worked on quantum metrology during my graduate career at the UNM. 
The goal here is to study the fundamental bounds imposed upon precision measurements by quantum mechanics and
to investigate the nature of quantum states and resources that might help us attain those bounds. In particular, we showed that mode entangled, two mode, Gaussian states can be used to estimate a phase $\phi$ with a measurement uncertainty  $\delta \phi$ that scales as $1/N$, where $N$ is the average photon number in the state. We also considered the effect of photon loss on the measurement uncertainty. The results of this investigation will be reported elsewhere.

\chapter{Background}

The purpose of this chapter is to build a foundation to aid understanding of the original work presented in the remaining part of this thesis. This chapter gives a pedagogical treatment of some technical aspects of this research, which will be helpful for understanding the work done and the results in the subsequent chapters. 

  Broadly, studies in quantum chaos are divided into kinematics and dynamics.  The cornerstone of the former, denoted by Berry as ``quantum chaology", is random matrix theory \cite{Mehta}, connecting the statistics of the spectrum and eigenvectors of Hamiltonians and Floquet maps of chaotic systems with those of random matrix ensembles.  Classic works on the subject including level statistics \cite{Berry77}, properties of Wigner functions \cite{Berry77a}, and quantum scars in ergodic phase spaces \cite{HellerScar} have tended to focus on the properties of wave mechanics, e.g. the dynamics of single particle billiards \cite{McDonald} (also seen in the properties of classical waves, e.g. microwave cavities \cite{Stockmann}).

The connections between quantum chaos, time reversal symmetry and the random matrix theory will be important to us, and hence we start by reviewing these aspects of our study.
 
 \section{Quantum chaos}
 
 \subsection{Time reversal} 
    Roughly speaking, if one considers the time evolution of a physical system to be a movie being played, then the absence/presence of the time reversal symmetry is our ability/inability to tell which way, forward or backwards, the movie is running. For classical systems, time reversal symmetry implies that for any given solution $\textbf{x}(t)$, $\textbf{p}(t)$ of Hamilton's equations an independent solution $\textbf{x}'(t')$, $\textbf{p}'(t')$ can be obtained with $t' = -t$ and some algebraic operation relating $\textbf{x}'$ and $\textbf{p}'$ to the position $\textbf{x}$ and momenta $\textbf{p}$ \cite{Haake}. It is important to mention the distinction between irreversibility and the absence of time reversal invariance symmetry. Reversibility is related to whether or not we have lost information, and whether the past condition can be recovered from the present. One can have reversibility without having time reversal invariance symmetry. For example, the motion of a charged particle in a magnetic field is not time-reversal invariant but it is reversible since there is no dissipation or loss of information. 
    
      In quantum mechanics, the quantum system obeying Schr$\ddot{\rm{o}}$dinger's equation
\begin{equation}
i \hbar \dot{\psi}(\textbf{x}, t) = H \psi(\textbf{x},t)
\end{equation}
 is time reversal invariant, if for any given solution $\psi(\textbf{x}, t)$, there is another solution, $\psi'(\textbf{x}', t')$, with $t' = -t$ and $\psi'$ uniquely related to $\psi$ \cite{Haake}. For example, consider a spinless particle with the Hamiltonian described solely by a scalar potential
\begin{equation}
H(\textbf{x},\textbf{p}) = \frac{\textbf{p}^{2}}{2m} + V(\textbf{x}), V(\textbf{x}) = V^{*}(\textbf{x}).
\end{equation}
If $\psi(\textbf{x},t)$ is a solution satisfying the Schr$\ddot{\rm{o}}$dinger's equation with time $t$ then, for $t'= - t $, $\psi^{*}(\textbf{x},t')$ is a solution. Hence $\psi'(\textbf{x},t) = K \psi(\textbf{x},-t)$ solves the Schr$\ddot{\rm{o}}$dinger's equation where $K$ is the complex conjugation operator defined with respect to the position basis. In general, a complex conjugation operator $K'$ can be defined with respect to any representation.
 
 Wigner showed that all time reversal operators $T$ must be antiunitary \cite{Haake}
\begin{equation}
\langle T \psi | T \phi \rangle = \langle \phi | \psi \rangle.
\end{equation}
Therefore, any time reversal operator $T$ can be given the standard form
\begin{equation}
\label{sf}
T = UK
\end{equation}
where $U$ is a suitable unitary operator and $K$ is the complex conjugation with respect to a standard representation.
Another requirement for $T$ is that any quantum state should be reproduced, to within a phase factor, 
on application of $T$ twice, i.e
\begin{equation}
\label{norm1}
T^2 = \alpha, |\alpha| = 1.
\end{equation}
Using Eq. \ref{sf} and Eq. \ref{norm1} one gets a useful condition on $T$,
\begin{equation}
T^2 = \pm 1.
\end{equation}
When we have a time reversal symmetry, and the antiunitary operator $T$ is such that 
\begin{equation}
\label{T_invariance}
[H, T] = 0,  T^{2} = 1,  
\end{equation}
then the Hamiltonian can always be given a real matrix representation and such a representation can be found without diagonalizing $H$.
 In such a scenario, a T-invariant basis, $\{\psi_{1}, \psi_{2}, ...\psi_{n}\}$ can be constructed such that $T \psi_{1} = \psi_{1}$. With respect so this basis, the Hamiltonian $H = THT$ is a real matrix \cite{Haake}. The canonical transformations for such Hamiltonians are the orthogonal matrices $O$, $O O^{T} = 1$.

 In the field of quantum chaos, time reversal becomes important in the following way.
 In the limit of large Hilbert space dimensions (small $\hbar$), for parameters such that the classical description of the dynamics shows global chaos, the eigenstates and eigenvalues of the quantum dynamics have the statistical properties of an ensemble of random matrices.  The appropriate ensemble depends on the properties of the quantum system under time-reversal \cite{Haake}. The ensemble of random matrices used to describe the Hamiltonians unrestricted by the time reversal symmetry is the Gaussian Unitary Ensemble (GUE). Similarly, the ensemble of random matrices used to describe the Hamiltonians having a time reversal symmetry are given by the Gaussian Orthogonal Ensemble (GOE).
Since the random matrix theory is crucial to our understanding of the quantum chaos, we take look at it in the next section.
 
\subsection{Random Matrix Theory}       

In 1984, Bohigas, Giannoni, and Schmit (BGS) conjectured that random matrix theory (RMT), originally introduced by Wigner and Dyson to explain the spectra of the complex many-body  physics of nuclei, applied also to few body (even one body with hard wall boundaries) systems when the dynamics is completely chaotic.  Using RMT to describe the properties of eigenvalues/eigenstates in completely chaotic system is the foundation of much of modern ``quantum chaology".
There is overwhelming evidence for the existence of universality in local fluctuations in the quantum energy spectra for systems that display global chaos in their classical phase spaces. All such Hamiltonian matrices of sufficiently large dimension have spectral fluctuations which can be described by studying such fluctuations for an appropriate ensemble of random matrices \cite{Haake}. Any member of such an ensemble can serve as a model of the Hamiltonian.           
  
In the field of quantum chaos, we typically look at twok kinds of random matrices -- random Hermitian matrices and random unitary matrices. Random Hermitians are of interest when one wants to understand energy levels of a complex system. The ensemble of random matrices used to describe Hamiltonians unrestricted by the time reversal symmetry is the Gaussian Unitary Ensemble (GUE). Similarly, the ensemble of random matrices used to describe those Hamiltonians having a time reversal symmetry are given by the Gaussian Orthogonal Ensemble (GOE). The other class of random matrices typically studied are the random unitary matrices. They are employed for periodically driven systems, as models of the unitary ``Floquet" operators, $F$, describing the change of the quantum state during one cycle of the driving. Powers of the ``Floquet" operator, $F^{n}$, gives us a stroboscopic description of the dynamics.
The ensemble of random unitaries are also known as the ``circular ensembles", originally introduced by Dysan. 
  As was the case for random Hermitian matrices, time reversal symmetry arguments play a similar role in the choice of the appropriate ensemble of random unitaries employed to model the ``Floquet" operator
to study the properties of the chaotic system. 
Depending on whether the system has time reversal symmetry or not, the appropriate ensemble of random unitaries is called the Circular Orthogonal Ensemble (COE) or the Circular Unitary Ensemble (CUE) respectively.    
The eigenvectors of the COE and CUE have the same properties as that for the respective GOE and GUE, but the eigenvalues are distributed differently. 
The circular unitary ensemble (CUE) is just the ensemble of random unitary matrices picked from U(n) according to the Haar measure. CUE eigenvalues lie on the unit circle in the complex plane, hence the name.
We look at an example of the construction of a Gaussian ensemble of Hermitian matrices below.

\subsubsection{An example of Gaussian Ensemble of Hermitian Matrices}

To illustrate the theory discussed above, we provide an example of construction of the Gaussian ensemble by constructing real symmetric $2 \times 2$ matrices with the orthogonal group as their group of canonical transformations as discussed in \cite{Haake}. Consider the matrix, 
$$ \left[
  \begin{array}{ c c }
     H_{11} & H_{12} \\
     H_{21} & H_{22}
  \end{array} \right]
$$
    
We want to determine the probability density, $P(H)$, for the three independent matrix elements $H_{11}$, $H_{22}$, 
and $H_{12}$ normalised as
\begin{equation}
\int_{-\infty} ^{+\infty} P(H) dH_{11}dH_{22}dH_{12} = 1
\end{equation}
To determine $P(H)$ uniquely, we demand

(1) $P(H)$ must be invariant under any canonical transformation, i.e., 
\begin{equation}
P(H) = P(H') ,\text{ }  H' = O^{T} H O , \text{ } O^{T} = O^{-1}.
\end{equation}

(2) The three independent elements must be uncorrelated. This implies that $P(H)$ be of the form
\begin{equation}
P(H) = P_{11}(H_{11})P_{22}(H_{22})P_{12}(H_{12}).
\end{equation}
Considering infinitesimal orthogonal change of basis, 
$$ O = \left[
  \begin{array}{ c c }
     1 & \theta \\
     -\theta & 1
  \end{array} \right]
$$
for which $H' = O^{T} H O$ gives
  \begin{subequations}
\begin{gather}
H'_{11} = H_{11} - 2 \theta H_{12}\\
H'_{22} = H_{22} + 2 \theta H_{12}\\
H'_{12} = H_{12} + \theta ( H_{12} - H_{22} ).
\end{gather}
\end{subequations}   
Using the above equations, for infinitesimal $\theta$, $P_{11}(H'_{11}) = P_{11}(H_{11}) - 2 \theta H_{12} \frac{dP(H_{11})}{ dH_{11}}$, with similar expressions for $P_{22}(H'_{22}) $ and $P_{12}(H'_{12})$.
  Since $P(H)$ is invariant under the orthogonal transformation we get, 
\begin{equation}
P(H) = P(H) \{ 1 - \theta [  2 H_{12} \frac{d \ln P_{11}}{d H_{11}} - 2 H_{12} \frac{d \ln P_{22}}{d H_{22}} - (H_{11} - H_{22})   \frac{d \ln P_{12}}{d H_{12}}]\}.  
\end{equation}
  Since the angle $\theta$ is arbitrary, its coefficient in the above equation must vanish,
\begin{equation}
\frac{1}{H_{12}} \frac{d \ln P_{12}}{d H_{12}} - \frac{2}{H_{11} - H_{22}} [\frac{d \ln P_{11}}{d H_{11}} - \frac{d \ln P_{22}}{d H_{22}}] = 0.
\end{equation}
This gives three differential equations for three independent functions $P_{ij}(H_{ij})$. The solutions are Gaussian and have the product
\begin{equation}
P(H) = C \exp [ - A (H_{11}^2 + H_{22}^2 + 2 H_{12}^2) - B(H_{11} + H_{22})].
\end{equation}
The constant $B$ can be eliminated by choosing an appropriate zero of energy. Then $P(H)$ can be written as 
\begin{equation}
P(H) = C \exp ( - A Tr \{H^2\} ).
\end{equation}
   
In a similar way, we can describe the construction of Gaussian Hermitian matrices where the probability $P(H)$ is invariant under the unitary transformation. To summarize \cite{Mehta, Haake}, different ensembles of random matrices follow from demanding (1) invariance of $P(H)$ under the different possible groups of canonical transformations and (2) complete statistical non-correlations between all matrix elements. 
We now discuss briefly the circular ensembles

\subsection{Circular Ensembles}					

 The Gaussian ensembles are useful in studying the properties of the energy levels of the system. To model the dynamical behaviour of chaotic systems, study of the properties of random unitaries is useful. As mentioned above, the ensemble of random unitaries are also known as the ``circular ensembles". 	The ensemble of random matrices used to describe the unitary evolution unrestricted by the time reversal symmetry is called the Circular Unitary Ensemble (CUE). Similarly, the ensemble of random matrices used to describe the unitary evolution having a time reversal symmetry are known as the Circular Orthogonal Ensemble (COE). 
    
 To discuss the properties of the eigenvalues of random unitary matrices, it will be useful to look at an example of a $2 \times 2$ random unitary matrix from the orthogonal ensemble \cite{nb09}. A general symmetric unitary matrix can be expressed as:
 
 $$ \left[
  \begin{array}{ c c }
     \sqrt{R} e^{i \gamma} & \sqrt{T} e^{i \eta}  \\
     \sqrt{T} e^{i \eta} & -\sqrt{R} e^{i (2 \eta-\gamma)} 
  \end{array} \right]
$$
 	and $T + R = 1$.	 The phases $\gamma$ and $\eta$  are assumed to be independent and uniformly distributed between $0$
 	and $2\pi$. The eigenvalues of the matrix lie on the unit circle in the complex plane and are given by:
 	\begin{equation}
\lambda_{1,2} = \pm e^{i (\eta \pm \phi)} = e^{i \alpha_{1,2}},  
\end{equation}	
 where $\sin \phi = \sqrt{R} \sin (\gamma - \eta)$.			
We see that the difference between the phases, $\alpha = \alpha_{1} - \alpha_{2} = \pi - 2\phi$, depends only on the the difference $\gamma - \eta$. The probability distribution function of $\gamma - \eta$ is a constant given by $P(\gamma - \eta) = \frac{1}{2\pi}$. Thus, the distribution function of the phase difference is given by:
\begin{equation}
P(\alpha) = \frac{1}{2 \pi} \left |  \frac{\partial (\gamma - \eta)}{ \partial{\alpha}} \right | = \frac{\sin(\alpha/2)}{4 \pi \sqrt{R - \cos^{2} (\alpha/2)}}.
\end{equation}					

 We see that as the phase differences become smaller, the probability distribution function is suppressed. This is commonly known as the \textit{repulsion} of eigenvalues for random unitary matrices. Similar repulsion occurs for matrices in higher dimensions. Level repulsion also occurs for random unitary matrices invariant under the unitary transformation \cite{nb09}. Properties of eigenvalues like the level repulsion, as predicted by random matrix theory, agree remarkably well with the Hamitonians and unitaries describing a quantized chaotic map. 
					   
   \section{Shannon and Fisher Information}
   
In information theory, entropy is a measure of the uncertainty associated with a random variable. In this context, the term usually refers to the Shannon entropy, which quantifies the expected value of the information contained in a message, usually in units such as bits. In this context, a ``message" means a specific realization of the random variable.
In classical information theory, the Shannon entropy, $H$ is associated to a probability distribution in the following way:
   \begin{equation}
 H= - \sum_{i} p(i) \log p(i),
 \label{Eq:HKCT}
 \end{equation}
where $\{p(1), p(2), ..., p(i)...\}$ are the probability values taken by some random variable.
The Shannon entropy is a measure of the amount of information and is related to the physical resources required to solve certain information processing tasks. For example, Shannon's noiseless channel coding theorem \cite{s48} says the following - \textit{ Suppose ${X_{i}}$ is an i.i.d information source with entropy rate $H(X)$ (bits per symbol). Suppose $R \geq H(X)$. Then there exists a reliable compression scheme of rate $R$ for the source. Conversely, if $R \leq H(X)$ then any compression scheme will not be reliable.} 
 Thus,  Shannon's entropy represents an absolute limit on the best possible lossless compression of any communication.
 This \textit{operational} interpretation of Shannon entropy in terms of data compression is the cornerstone of classical information theory. 
 
 The Fisher information is a way of measuring the amount of information that an observable random variable $X$ carries about an unknown parameter $θ$ upon which the probability of $X$ depends. Thus, the Fisher information tells us how well we can estimate a parameter given a sequence of data.

We start with a few definitions taken from \cite{ct}. Let $\{f(x; \theta)\}$, $\theta \in \Theta $, denote an indexed family of probability densities, $f(x;\theta) \geq 0$, $\int f(x; \theta) dx  =  1$ for all $\theta $. Here $\theta$ is the parameter set.
\\   
\textit{\textbf{Definition:}} An \textit{estimator} for $\theta$ for sample size $n$ is a function $T$, whose input is the sequence of $n$ data values and the output is an estimate, $\theta_{est}$, of the parameter $\theta$. More formally, $T : \digamma^{n} \rightarrow \Theta. $
The estimator gives us the approximate value of the parameter based upon the data $X$. It is desirable to have an idea of how good the estimator is in estimating the value of the parameter.
Here, as we shall see, the Fisher information plays a crucial role here.
     
 The Fisher information is defined as 
     \begin{equation}
J_{\theta} = E_{\theta}[\frac{\partial}{\partial \theta} \ln f(x; \theta)]^{2}.
\end{equation}   
The subscript $\theta$ means that the expectation is with respect to the density $f(  .   ; \theta)$.
The Fisher information gives a lower bound on the error in estimating $\theta$ from the data. We state this result, which shows the significance of the Fisher information, without proof \cite{ct}.  \\ 
 \textbf{Theorem :} The mean squared error of any unbiased estimator $T(X)$ of the parameter $\theta$ is lower bounded by the reciprocal of the Fisher information, i.e.,     
\begin{equation}
$var$ (T) \geq \frac{1}{J(\theta)}.
\end{equation}   
This is known as the Cramer-Rao bound \cite{c46}. The bound states that the variance of any unbiased estimator is at least as high as the inverse of the Fisher information. An unbiased estimator which achieves this lower bound is said to be \textit{efficient}.
 Hence, the Fisher information gives us the idea of the goodness of the estimator. 

\section{Entanglement}      

 Entanglement is a uniquely quantum phenomena. It is the property of composite quantum mechanical systems that describes the non-separability of the constituents. Suppose, for example, Alice has an electron in spin up state, $| \uparrow\rangle$, while Bob has a
an electron in the spin down state $|\downarrow\rangle$. Then we say that the joint state of Alice's and Bob's electron is,
\begin{equation}
\label{EPR}
| \uparrow\rangle \otimes  |\downarrow\rangle = |\uparrow \downarrow \rangle.
\end{equation}
These are called \textit{tensor product} states.
  The Hilbert space that encompasses tensor product states is, however, not restricted by these
states. For example the state, $\frac{|\uparrow \downarrow \rangle - |\downarrow \uparrow \rangle}{\sqrt{2}}$, known as the EPR pair (after Einstein, Podolsky and Rosen) is a perfectly valid physical state. Entanglement tries to capture this inability to express a quantum state as a tensor product.
Thus, entanglement can be regarded as a joint property of the system. Entanglement cannot be increased under local operations and classical communication and therefore has a non-local character to it.

Quantum entanglement has shed light on some of the most fundamental aspects of quantum mechanics. For example, entangled states like the ``EPR" pair can be used to show that ``locality" (the assumption that measurements made at a location A cannot influence the measurement outcomes at another location B that is``space-like" separated from A) and ``realism" (physical properties have definite values which exist independent of measurement) are incompatible with each other. The assumptions of ``locality" and ``realism" are together known as local realism. John Bell \cite{b66}, in 1964,  demonstrated that if there is a realistic interpretation of quantum theory, in the sense defined above, it has to be nonlocal. Nature is \textit{not} locally realistic!  

Not only does entanglement teach us important lessons in the foundations of quantum mechanics, it enables certain tasks 
which either cannot be performed classically or would otherwise require far more resources. Entanglement has a crucial role in quantum teleportation \cite{b93}, quantum dense coding \cite{bw92}, cryptography with the Bell theorem \cite{e91}, and quantum computation using pure quantum states \cite{jl03}.

 Now the question arises: how do we quantify the entanglement between two quantum systems? For this we look at an important result known as the Schmidt decomposition in the next section.


\subsection{Quantifying Entanglement}
 We consider entanglement of pure states of the bipartite system.  Entanglement is then uniquely determined by the coefficients in the Schmidt decomposition of the joint state of the system,
\begin{equation}
\ket{\Psi}_{IJ}=\sum_{i} \sqrt{\lambda_i} \ket{u_i}_I \ket{v_i}_J ,
\end{equation}
where $\lambda_i$ are the eigenvalues of the reduced density matrix of either subsystem, and the Schmidt basis vectors $\{\ket{u_i}_I, \ket{v_i}_J\}$ are their respective eigenvectors. The entanglement $E$ is the Shannon entropy of the squares of the Schmidt coefficients,
\begin{equation}
E = -\sum_i \lambda_i  \log( \lambda_i).
\end{equation}
 
Unfortunately, quantification of entanglement for mixed states is not as simple. 
 Extending the definition of separability from the pure case, we say that a mixed state is separable if it can be written as
\begin{equation}
\rho =\sum_{i} p_{i} \rho_{i}^{A} \otimes \rho_{i}^{B}
\end{equation}
  where $\rho^{A}$ and $\rho^{B}$ are themselves states on the subsystems A and B respectively. Therefore, a state is separable if it is a probability distribution over uncorrelated states, or product states. $\rho^{A}$ and $\rho^{B}$ can be taken to be pure states without any loss in generality. A state is then said to be entangled if it is not separable. As mentioned above, in general, finding out whether or not a mixed state is entangled is considered difficult. In general, the problem of determining whether a bipartite system in a mixed quantum state is entangled or not is NP hard \cite{g03}.
For a detailed review of mixed state and multi-partite entanglement we refer the reader to \cite{hhhh09}.
   
The theory of entanglement has been very useful in helping us understand certain features of quantum mechanics. 
Entanglement also serves as a uniquely quantum resource which can be harnessed to accomplish various quantum information processing tasks. This is the focus of our next section.
              
\section{The quantum advantage behind quantum information processing}      
      
      The crowning achievement of quantum information science has been the discovery of algorithms and communication protocols based on the laws of quantum mechanics that accomplish tasks much more efficiently than the known classical methods. For example the field of quantum computation has supplied us with quantum algorithms for integer factorization \cite{ej96, shor} and for searching an unsorted database \cite{grov} that offer significant improvements over the best known classical algorithms.     
    
    Characterizing the resources behind the enhancements and speedups provided by quantum mechanics over best known classical procedures is one of the most fundamental questions in quantum information science. Quantum entanglement~\cite{hhhh07} is generally seen to be the key resource that gives quantum information processors their power. There are, however, quantum processes which provide an exponential advantage in the presence of little or no entanglement~\cite{kl98, dqc1}. In the realm of mixed-state quantum computation, for example, quantum discord~\cite{oz02, hv01} has been proposed as a resource~\cite{dsc08} and there has been progress in this direction in the last few years~\cite{adv}. It has also been shown to be a resource in quantum state discrimination~\cite{rra11, lfwf11} and quantum locking~\cite{lock}. 
     
Quantum mechanics also provides an advantage in communication.
For example, sending secret messages between two parties in the presence of adversaries in a protocol known as quantum cryptography \cite{e91, grtz02} provides security well beyond any known classical protocol.
The advantage of quantum cryptography is that it enables certain cryptographic tasks that are impossible using only classical communication. In particular, quantum mechanics guarantees that measuring quantum data disturbs that data; this can be used to detect an adversary's interference with a message.   

Quantum cryptography is not the only setting where we utilize the features of quantum mechanics to get an advantage over best known classical communication protocols. Research in quantum information science has led to a discovery of 
new protocols and tasks which are classically impossible. The most relevant example is that of quantum teleportation \cite{b93}. Quantum teleportation, or entanglement-assisted teleportation, is the process by which an unknown quantum state can be transmitted from one location to another, without the state being transmitted through the intervening space. 

We first introduce the notation typically used in quantum communication protocols: $[q \rightarrow q]$ represents one qubit of communication between two parties and $[qq]$ represents one shared ebit (one maximally entangled two qubit state) between two parties. Similarly, $[c \rightarrow c]$ represents one classical bit of communication between the parties.
Expressing teleportation as a resource inequality \cite{dhw08} we get,
\be
\label{teleport}
 [qq]  + 2 [c \rightarrow c] \succeq  [q \rightarrow q].
\ee
This inequality says that one can employ a shared ebit and 2 bits of classical communication to communicate a single \textit{unknown} quantum bit. 
To communicate an unknown quantum bit by a classical procedure will take exponentially large amount of resources as compared to teleportation.

Another protocol which shows the advantage of quantum communications is the superdense coding \cite{bw92}.
Expressing it as a resource inequality \cite{dhw08},
\be
\label{superdense}
 [qq]  +  [q \rightarrow q] \succeq  2[c \rightarrow c],   
\ee
showing that one can employ a shared ebit and a single bit of quantum communication to communicate 2 bits of classical information. Thus, complete classical information about two particles can be sent by direct manipulation of just one particle by the sender. This protocol starts with Alice and Bob sharing a Bell state. For example, they  share the state $\frac{|\uparrow \uparrow \rangle +  |\downarrow \downarrow\rangle}{\sqrt{2}}$.
 Alice then performs one of the following operations on her half of the Bell state - either the identity operation, $I$, or applies one of the three operators, i.e.  $\sigma_{x}, i\sigma_{y}$, or $\sigma_{z}$. The type of operation that Alice performs thus represents two bits of information which she wants to communicate with Bob.
This she does by sending over her half of the Bell state to Bob after performing the operation. Depending on Alice's operation (one of $I$, $\sigma_{x}, i\sigma_{y}$, or $\sigma_{z}$), Bob gets one of the four states: $\frac{|\uparrow \uparrow \rangle +  |\downarrow \downarrow\rangle}{\sqrt{2}}$, $\frac{|\downarrow \uparrow \rangle +  |\uparrow \downarrow\rangle}{\sqrt{2}}$,  $\frac{|\uparrow \downarrow \rangle -  |\downarrow \uparrow\rangle}{\sqrt{2}}$, or $\frac{|\uparrow \uparrow \rangle -  |\downarrow \downarrow\rangle}{\sqrt{2}}$. These states form what is known as the \textit{Bell basis}, which is an orthonormal basis. Since orthonormal quantum states can be distinguished by a suitable quantum measurement, Bob comes to know of the joint state he shared with Alice and hence the operation performed by Alice at her end. Thus, Alice succeeds in communicating two bits of classical information to Bob using a shared ebit and a single bit of quantum communication.
  
In both the protocols mentioned above, it is worth noting that it was the pre-existing entanglement that made them possible. Without entanglement, it is impossible to achieve both the teleportation and the super-dense coding.   
   
Thus we have seen a spectrum of quantum information processing protocols, from quantum computation to communications tasks and cryptography. Certain fundamental properties of quantum mechanics are exploited 
in accomplishing these tasks \cite{bennett95}. These include:
\begin{enumerate}
\item Superposition: A quantum state can exist in an arbitrary complex linear combination of classical logic states.
A classic example is the ``cat state", named after Schr$\ddot{\rm{o}}$dinger.
\begin{equation}
|\psi_{cat} \rangle = \frac{|Alive \rangle + | Dead \rangle}{\sqrt{2}}
\end{equation}
Both of the logic states evolve in ``parallel" according to a given unitary evolution. 
\item Interference: The quantum wave function undergoes interference, just like the classical waves. The quantum state can be in a superposition of different alternatives/paths, each of which is characterized by a complex probability amplitude, that has both real and imaginary components in general. If these paths are in principle indistinguishable, i.e. there is no information whatsoever in the universe that can help us distinguish between them, then in that case the corresponding probability amplitudes add up causing  interference. This feature is utilized in quantum computers, where different paths are explored in parallel in search of the solution and the probability amplitude of the path leading to the right solution gradually builds up.
\item Entanglement: As mentioned earlier, protocols like teleportation, superdense coding and pure state quantum computation are made possible due to entanglement.
\item Non-distinguishability of quantum states: Non-orthogonal quantum states cannot be unambiguously distinguished. Moreover, obtaining information about an unknown quantum state can cause disturbance 
and actually change it. This feature is exploited in designing cryptographic protocols.
\end{enumerate}

As we have seen, the theory of entanglement has been very helpful in understanding quantum mechanics from a fundamental point of view. In addition, entanglement has been shown to be a crucial resource in quantum information processing. However, this is not the whole story.
In recent years, there has been some progress in quantifying the quantum character of composite quantum systems that goes beyond entanglement. This is the focus of our next section.

 \section{Quantum Discord: A quantum resource?}
              
Quantum discord aims at generalizing the notion of quantum correlations in a quantum state, beyond entanglement~\cite{oz02,hv01,Glauber}. It aims to capture all the nonclassical correlations in a quantum system.
 Quantum measurements disturb a quantum system in a way that is unique to quantum theory. Quantum correlations in a bipartite system are precisely the ones that are destroyed by such disturbances. As we discuss below, this feature of quantum systems can be used to quantify the amount of purely quantum correlations present in a bipartite quantum system.
 
 Quantum mutual information is generally taken to be the measure of total correlations, classical and quantum, in a quantum state. For two systems, $A$ and $B$, it is defined as $ I(A:B) = H(A) + H(B) -H(A,B).$ Here $H(\cdot)$ denotes the von Neumann entropy of the appropriate distribution \cite{nielsen00a}. For a classical probability distribution, Bayes' rule leads to an equivalent definition of the mutual information as $I(A:B) = H(A)-H(A|B).$  This motivates a definition of classical correlation in a quantum state. 
 
 Suppose Alice and Bob share a quantum state $\rho_{AB} \in \mathcal{H}_A\otimes \mathcal{H}_B.$ If Bob performs a measurement specified by the POVM  set $\{\Pi_i\},$ the resulting state is given by the shared ensemble $\{p_i,\rho_{A|i}\},$ where
$$
 \rho_{A|i} = \tr_{\!\!B}(\Pi_i\rho_{AB})/p_i,\;\;\;p_i=\tr_{\!\!A,B}(\Pi_i\rho_{AB}).
$$
A quantum analogue of the conditional entropy can then be defined as $\tilde{S}_{\{\Pi_i\}}(A|B)\equiv\sum_ip_iS(\rho_{A|i}),$ and an alternative version of the quantum mutual information can now be defined as $\mathcal{J}_{\{\Pi_i\}}(\rho_{AB}) = S(\rho_A)-\tilde{S}_{\{\Pi_i\}}(A|B),$ where $S(\cdot)$ denotes the von Neumann entropy of the relevant state. The above quantity depends on the chosen set of measurements $\{\Pi_i\}.$ To capture all the classical correlations present in $\rho_{AB},$ we maximize $\mathcal{J}_{\{\Pi_i\}}(\rho_{AB})$ over all $\{\Pi_i\},$ arriving at a measurement independent quantity
\be
\label{eq:J}
\mathcal{J}(\rho_{AB}) = \max_{\{\Pi_i\}}(S(\rho_A)-\tilde{S}_{\{\Pi_i\}}(A|B)).
\ee
Then, quantum discord is defined as~\cite{oz02}
\ben
\label{discexp}
\mathcal{D}(\rho_{AB}) &=& I(\rho_{AB})-\mathcal{J}(\rho_{AB}) \\
                 &=& S(\rho_B)-S(\rho_{AB})+\min_{\{\Pi_i\}}\tilde{S}_{\{\Pi_i\}}(A|B)\nonumber
\een

\subsection{Operational interpretation of quantum discord}
The work in this dissertation seeks to answer very fundamental question about quantum discord: Is discord just a mathematical construct or does it have a definite physical role in information processing?
  We provide a link between quantum discord and an actual physical task involving communication between two parties~\cite{md10}. 
  
 The key insight to our findings is that quantum measurements and environmental decoherence disturb a quantum system in a way that is unique to quantum theory.
Quantum correlations in a bipartite system are precisely the ones that are destroyed by such disturbances, and therefore certain quantum communication protocols become overloaded by an amount exactly equal to quantum discord.
More specifically, we will show in Chapter 5 that discord is the markup in the cost of quantum communication in the process of quantum state merging~\cite{sm}, if the system undergoes measurement and/or decoherence. 

At this point, a natural question arises: Is there a role of quantum discord in quantum information theory as a whole beyond the lossy state merging protocol? We answer this question with a yes. We first observe that quantum state merging protocol is a derivative of the more general Fully Quantum Slepian Wolf (FQSW) protocol~\cite{sm,adhw09} and the closely related ``mother" protocol. The mother protocol is essentially a unification of all unidirectional, bipartite and memoryless quantum communication protocols like quantum teleportation, superdense coding and entanglement distillation.
A link between discord and state merging does indeed suggest a link between discord and the mother protocol and hence a possible role of discord in all bipartite, unidirectional and memoryless quantum communication protocols.

Before exploring the role of discord in quantum information theory, it is therefore necessary to review the mother protocol and other related concepts. 
\section{The mother protocol and quantum information's family tree}

So far, we have been discussing quantum communication tasks like teleportation, super-dense coding, quantum state merging as independent entities. Is there a unified way of looking at various quantum communication tasks? Do various quantum communication protocols described above have a common origin. Interestingly, the answer is yes. It was shown in \cite{adhw09} that essentially all unidirectional, bipartite and memoryless quantum communication protocols are actually siblings originating from one ``mother". The mother protocol can be seen to provide a hierarchical structure to the family of quantum protocols. We will describe the improved version of the mother protocol, which is the 
fully quantum Slepian-Wolf (FQSW) protocol. Throughout the thesis, we will use the names ``mother" and FQSW to describe essentially this same protocol.
\subsection{The mother protocol} 
At the start of the mother protocol, we have $n$ copies of a quantum state that can be expressed as the tensor power $ |(\psi^{ABR})^{\otimes n}\rangle$.  Alice holds the $A$ shares and Bob the $B$ shares. The reference system $R$ is ``purification" of the $AB$ system (which might be described by a mixed state) and does not actively participate in the protocol. The mother protocol can be viewed as an entanglement distillation between $A$ and $B$ when the only type of communication permitted is the ability to send qubits from Alice to Bob.
The transformation can be expressed in the resource inequality as
  
\begin{equation}
\label{mother}
\langle\psi^{AB}\rangle  + \frac{1}{2} I (A:R) [q \rightarrow q] \geq \frac{1}{2} I (A:B)[qq].
\end{equation}
Here, $|\psi^{AB}\rangle$ refers to the state shared between Alice and Bob whose purification is the state $|\psi^{ABR}\rangle$.
The above inequality states that $n$ copies of the state $|\psi^{AB}\rangle$ can be converted to $\frac{1}{2} I (A:B)$ EPR pairs per copy, provided Alice is allowed to communicate with Bob by sending him qubits at the rate $\frac{1}{2} I (A:R)$ per copy.

The mother protocol has a stronger version known as the FQSW protocol.
The FQSW protocol not only enables the two parties, Alice ($A$) and Bob ($B$), to distill  $\frac{1}{2} I (A:B)$ EPR pairs per copy, in addition Alice can ``merge" her state with Bob. By the state merging task, we mean that Alice is able to successfully transfer her entanglement with the reference system $R$ to Bob as shown in Fig 2.1.

 Writing the FQSW in terms of a resource inequality

\begin{equation}
\langle\psi^{AB}\rangle  + \frac{1}{2} I (A:R) [q \rightarrow q] \geq \frac{1}{2} I (A:B)[qq] +  $State Merging between A and B$
\end{equation}

 In a more rigorous mathematical notation, we write the above as
 \begin{equation}
 \label{fqsw}
\langle \mathcal{U}^{S \rightarrow AB} : \psi^{S}\rangle  + \frac{1}{2} I (A:R) [q \rightarrow q] \geq \frac{1}{2} I (A:B)[qq] + \langle id^{S \rightarrow \hat{B}} : \psi^{S}\rangle.
\end{equation}
It is important to explain the terminology used in the above inequality. When we have a noisy resource like a mixed state, $ \psi^{S}$, or a noisy channel, it is inserted between a ``$\langle   \rangle$". Thus a mixed state is represented by $\langle \psi^{S} \rangle $, and a noisy channel by $\langle \mathcal{N}\rangle$. A channel is a relative resource $\langle \mathcal{U}^{S \rightarrow AB} : \psi^{S}\rangle$ meaning that the protocol only works provided the input to the channel is the state $\psi^{S}$. On the LHS, $\mathcal{U}$ takes the state $ \psi^{S}$ and distributes it to Alice and Bob. On the RHS, the symbol ``$id$" is an identity channel taking the state $\psi^S$ to Bob alone. The state $\psi^{S}$ on the left-hand side of the inequality is distributed to Alice and Bob, while on the right-hand side, that same state is given to Bob alone.
This inequality states that starting from the
state  $|(\psi^{ABR})^{\otimes n}\rangle$, and using $\frac{1}{2} I (A:R)$ bits of quantum communication from Alice to Bob, they can distill
 $\frac{1}{2} I (A:B)$ EPR pairs per copy, and in addition Alice can accomplish merging her state with Bob. 
 Figure 2.1 \cite{adhw09} shows the action of the protocol.
\begin{figure}
\includegraphics[width=\linewidth]{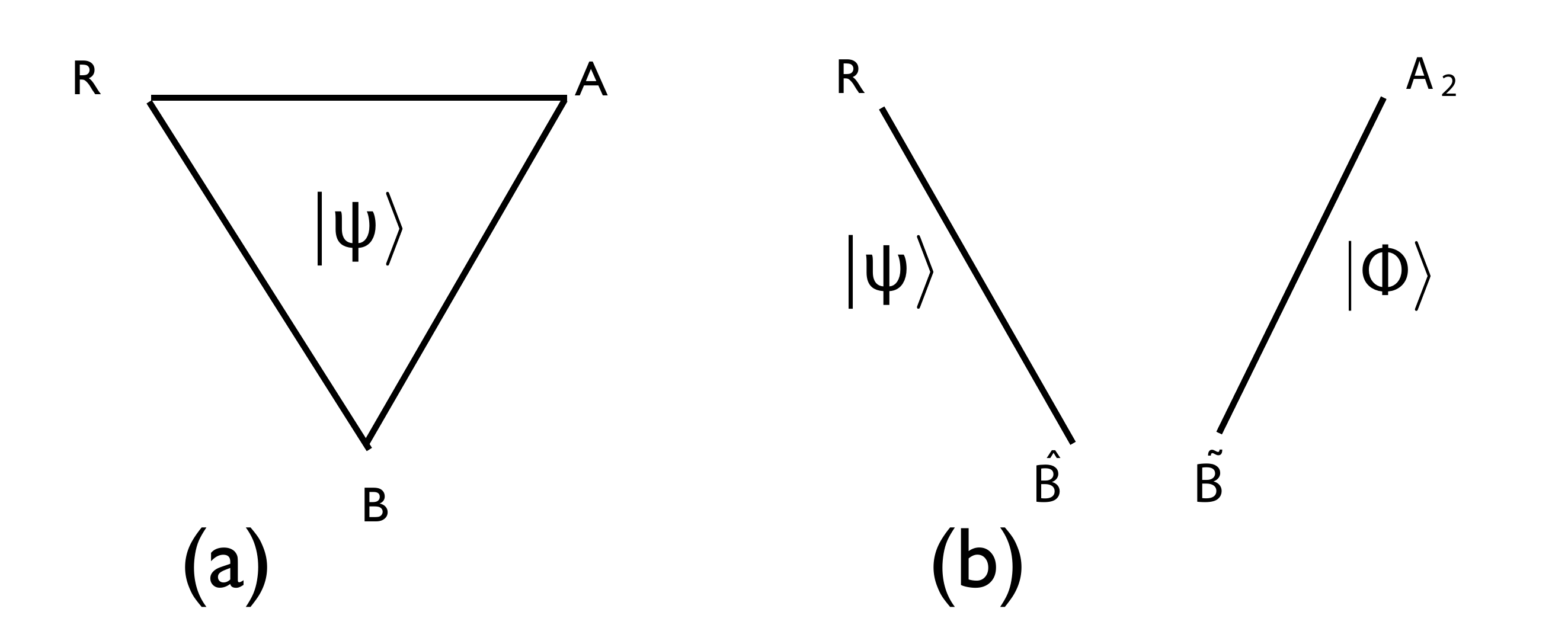}
\caption{a) The starting point of the FQSW protocol. A pure tripartite entangled 
state  $|(\psi^{ABR})^{\otimes n}\rangle$ shared between Alice and Bob. $R$ is the reference system purifying $AB$ system and does not participate actively in the protocol. b) After execution of the protocol. Alice's portion of the state has been transferred to Bob, and now Bob and $R$ hold the state $|(\psi^{R\hat{B}})^{\otimes n}\rangle$ where $\hat{B}$ is a register held with $B$.}
\label{mother}
\end{figure}
   
 In the state merging task, as described above, Alice is able to successfully transfer her entanglement with the reference system $R$ to Bob. This means that Alice transfers her portion of the state to Bob.
 In other words, they manage to create the state
 $|(\psi^{R\hat{B}})^{\otimes n}\rangle$, where $\hat{B}$ is a register held with $B$ and  $|(\psi^{R\hat{B}})^{\otimes n}\rangle = |(\psi^{ABR})^{\otimes n}\rangle$ in the limit $n \rightarrow \infty$.
  
Thus the FQSW protocol accomplishes state merging as well as entanglement distillation.
 The FQSW protocol is valid asymptotically in the limit of a large number of copies and this is denoted by the symbol $\geq$.

\chapter{Dynamically Generated Entanglement and Chaos}
  
\section{Introduction}

The connections between complexity, nonlinear dynamics, ergodicity, and entropy production, have long been at the heart of the foundations of statistical physics. A central goal of ``quantum chaos" has been to extend this foundation to the quantum world.  Classic works on the subject including level statistics \cite{Berry77}, properties of Wigner functions \cite{Berry77a}, and quantum scars in ergodic phase spaces \cite{HellerScar} have tended to focus on the properties of wave mechanics, e.g. the dynamics of single particle billiards \cite{McDonald} (also seen in the properties of classical waves, e.g. microwave cavities \cite{Stockmann}).  More recently, the tensor product structure of quantum mechanics, essential for understanding systems with multiple degrees of freedom, has come to the fore.  In that context, one is naturally led to consider how the dynamical generation of entanglement between quantum subsystems is connected with the chaotic dynamics of coupled classical degrees of freedom.  Such studies address fundamental issues of complexity in quantum systems and are potentially applicable in quantum information processing, where entanglement is considered to be an essential resource.  
The connection between chaos in the classical description of Hamiltonian dynamics and entanglement in the quantum description has been the subject of extensive study over the last decade.  The original motivation of Zurek and Paz  was to address the quantum-to-classical transition \cite{Zurek/Paz}.  By conjecturing that chaotic systems decohere exponentially fast through their entanglement with the environment, they hoped to resolve a paradox in which a macroscopic system would exhibit the effects of quantum coherence on a time scale logarithmic in $\hbar$.  

Work quickly following this turned to studies of the coupling of just two degrees of freedom, rather than system-environment coupling, as entanglement is most easily quantified for bipartite systems  \cite{Fuyura}.  In most cases, workers have considered systems described by a total Hamiltonian of the form
\begin{equation}
\label{joint}
H_{\text{total}}(t) = H_1(t)+H_2(t)+H_{\text{int}}(t),
\end{equation}
where $H_1(t)$ and $H_2(t)$ can exhibit chaos in the classical description of the dynamics and $H_{\text{int}}(t)$ couples the two degrees of freedom.  The most well-studied example has been two coupled kicked tops (a standard paradigm of quantum chaos \cite{Haake}).  Two separate questions have been addressed: (i) How does the rate of dynamical generation of entanglement correlate with the chaos in the subsystems 1 and 2? (ii) How does the entanglement content of the state, either in the eigenstates, or in the state that is dynamically generated in quasi-steady state, correlate with this chaos?  Miller and Sarkar \cite{Miller/Sarkar} were the first to study question (i) for this system, and through numerical studies, correlated the rate of generation of entanglement with the Lyapunov exponents associated with the mean positions of quantum wavepackets localized in a mixed phase space (weak chaos). This behavior was not found to be universal \cite{Fujisaki, Jacquod, Znidaric2003, Lakshminarayan/Bandyopadhyay2004}, but depended strongly on the degree of the chaos within subsystems when compared to the size of coupling between them.  In a seeming paradox, for strong chaos within the tops, the rate of entanglement generation decreased with coupling strength, asymptoting to a constant value.  Moreover, in a systematic study of the entangling power of the coupled kicked top system \cite{Kus04}, Demkowicz-Dobrzaski and Kus found additional anomalies, including a regime in which each top was described by highly regular dynamics, but exhibited the highest rate of generation of entanglement when compared with conditions where each top is highly chaotic. 

In relation to question (ii), Bandyopadhyay and Lakshminarayan \cite{Lakshminarayan/Bandyopadhyay2002,Lakshminarayan/Bandyopadhyay2004} explored the amount of entanglement that is associated with coupled kicked tops, with particular emphasis on the entanglement of the Floquet eigenstates \cite{Lakshminarayan}.  The entanglement of these eigenstates saturated to a value below the maximum possible value in a way that depended only on the Hilbert space dimension, not the chaoticity parameter.  The same was true of the dynamically generated entanglement.  (The relationship between the entanglement in the eigenstates and the dynamically generated entanglement is subtle \cite{Kus04}; we'll return to this point later). This work gave the first indication that the entanglement generated by the coupled tops was statistical in nature, and related to the theory of random states in Hilbert space. Using random matrix theory \cite{Mehta, Haake} they were able to determine the statistics of the Schmidt coefficients of a random bipartite pure state, and thus were able to predict the saturation value of the entanglement for the Floquet eigenstates. An extended analysis of the statistical properties of the Schmidt vectors of random states was carried out by Znidaric \cite{Znidaric2007}.  In other related work, dynamical generation of entanglement by chaotic maps and its relation to random matrices was also explored by Gorin and Seligman \cite{Seligman} as a way of modeling decoherence, by Scott and Caves \cite{Scott/Caves} and Abreu and Vallejos \cite{Abreu2006} as a way of comparing different quantizations of the Baker's map, and by Viola and coworkers \cite{Viola} as a means of quantifying complexity in quantum systems and its relationship to generalized entanglement.

In another approach to question (ii), Ghose and Sanders have shown that there are signatures of chaos in the entanglement dynamically generated by a single kicked top when the large angular momentum is thought of as a collection of symmetrically coupled qubits \cite{Ghose, Wang}.  They showed strong correlation between the classical Poincar\'{e} surface of section for a mixed phase space, and a contour plot of the dynamically generated entanglement as a function of the initial position of a localized coherent state.  Using the Floquet spectrum, they also explained the initial rise time and power spectrum in the entanglement history. 

While many of the elements connecting chaos and entanglement have been explored with a variety of successful numerical and analytic predictions, in some cases the key relations have been obscured.  In particular, in studies of systems of the form in Eq. \ref{joint}, entanglement is correlated with the chaos in the individual subsystems.  But, entanglement arises from coupling between subsystems, and is a global property of the state.  Likewise, chaos can also arise through the coupling of degrees of freedom when the overall dynamics are not integrable.  For this reason, we believe the key relations are best understood by correlating entanglement with chaos in the {\em joint system} (i.e., chaos in $H_{\text{total}}$), rather than chaos in the subsystems that one would see in the absence of coupling.  To do so, it is most natural to consider systems in which chaos and entanglement arise from the {\em same mechanism} -- the physical coupling between subsystems.  Moreover, by considering a total system that is chaotic only when the two parts are coupled, we focus on a classical phase space that describes the global system rather than a subsystem, and there is no ambiguity about the nature of the joint dynamics.  For this case, the distinction between weak and strong coupling cannot be made independently of weak and strong chaos, thereby sharpening our focus on the key relationships. 
 
  To address these issues, we consider a model system of kicked coupled-tops, rather then coupled kicked-tops, described in detail in this chapter.  This system is motivated by its connection to possible experimental realizations, our ability to easily visualize the classical phase space, and to analyze the Floquet map.  We use this system as a forum to explore question (ii) above -- how is chaos in the classical dynamics of the joint system correlated with the long-time averaged dynamically generated entanglement? 

The basic thesis of the work described in this chapter is as follows. Chaos can arise in classical dynamics when there are insufficient symmetries (integrals of motion) for a given number of degrees of freedom.  In the quantum analog, insufficient symmetries lead to the random matrix conjecture -- systems with global classical chaos have eigenvectors and eigenvalues that are statistically predicted by ensembles of random matrices \cite{Bohigas, Haake}.  Moreover, where classical chaos leads to ergodic dynamics and the generation of ``random" coarse-grained distributions on phase space, for times short compared to the Heisenberg time, but long compared to transient behavior, the quantum chaotic map generates a state with many properties that are statistically predicted by a random state in Hilbert space, picked according to the appropriate Haar measure \cite{Scott/Caves}.  The dynamically generated entanglement is then that of a random state (by this measure) in the relevant Hilbert state.  These predictions can be extended to mixed phase spaces with regular islands immersed in a chaotic sea.  With the help of Percival's conjecture \cite{Percival} that divides eigenstates into chaotic and regular classes, we can find the entanglement of a random state in a chaotic subspace and thus predict entanglement generation in a mixed phase space for chaotic initial conditions. Whereas in the globally chaotic case we can derive analytic results, for the mixed phase spaces we are relegated to numerical predictions, which nonetheless verify the connection between entanglement generation in chaotic dynamics and the creation of pseudo-random states in Hilbert space.

The remainder of this chapter is organized as follows.  In Sec. \ref{KCT} we introduce our model of kicked coupled-tops, studying the classical and quantum features. Section \ref{Ent}, the heart of this work, studies the entanglement in our system.  We perform numerical calculations of the entanglement of the system's eigenstates, the long-time averaged entanglement generated by the Floquet map, and its relationship to the classical phase space.  We then explain these results in terms of the properties of random states in Hilbert space.  Reviewing the essential ideas, we derive new analytic expressions for the typical entanglement of a random state when we are restricted to a subspace of the full tensor product space.  This is of relevance here given the symmetries of the system.  We also pay particular attention to the subtle distinctions between the eigenstates of random matrices and the random states generated from initially localized wavepackets.  In doing so we clarify previous works and make accurate predictions, especially for global chaos, but also extended to a more general mixed phase space scenario.  Our results are discussed and summarized in Sec. \ref{DS3}.

\section{Kicked Coupled-Tops}
\label{KCT}
\subsection{Quantum and classical descriptions }
We consider a bipartite system composed of two spins, $\mbf{I}$ and $\mbf{J}$,  isotropically coupled in a Heisenberg interaction, and subject to periodic kicks that act only on spin $\mbf{J}$. Choosing the direction of the kicks to be about the $z$-axis, the system evolves according to the Hamiltonian,
 \begin{equation}
 H=A \mbf{I} \cdot \mbf{J}+\sum_{n=-\infty}^\infty \delta (t - n \tau) B J_z .
 \label{Eq:HKCT}
 \end{equation}
Here $A$ gives the strength of the isotropic coupling, $B$ the strength of the kicking, and $\tau$ is the kicking period. Such a Hamiltonian describes the hyperfine interaction between nuclear spin $\mbf{I}$ and total electron angular momentum $\mbf{J}$, with a magnetic field that has negligible effect on the nucleus.  While this realization cannot reach deep into the semiclassical regime, for large atoms, with heavy nuclei and a large number of electrons in the valance shell, one can explore nontrival mesoscopic regimes.  The true semiclassical limit can potentially be attained in an atom-photon system where $\mbf{I}$ is the collective spin of an atomic ensemble coupled to the Stokes vector $\mbf{J}$ of a quantized electromagnetic field \cite{Mitchell}.  We will not consider here the feasibility of experimental realizations, instead focusing on the foundational theory.

\begin{figure}
\includegraphics[width=\linewidth]{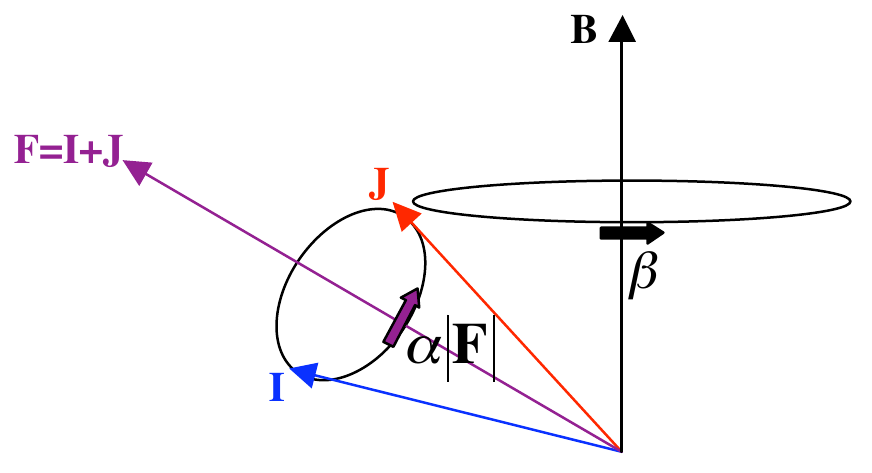}
\caption{The dynamics of the kicked coupled tops can be viewed as an alternating sequence of rotations. The two spins $\mbf{I}$ and $\mbf{J}$ precess around the total angular momentum $\mbf{F}$ by an angle $\alpha |\mbf{F}|$, and the spin $\mbf{J}$ is kicked around the space-fixed $z$-axis by $\beta$.}
\label{F1_1}
\end{figure}

Choosing the external field to act in delta kicks allows us to express the Floquet map (transformation after one period) in a simple form of sequential rotations,
\begin{equation}
\label{Eq:Floquet}
U_{\tau} = e^{-i\alpha \mbf{I} \cdot \mbf{J}}  e^{-i \beta J_z} \equiv e^{-i\alpha F^2 /2}  e^{-i \beta J_z},
\end{equation}
where $\alpha$ and $\beta$ are related to $A$ and $B$ in terms of the kicking period.  The Floquet map describes 
the evolution of the system for one period.
In the second form, we have expressed the rotation in terms of the total angular momentum $\mbf{F} = \mbf{I} + \mbf{J}$ and neglected irrelevant overall phases.  We can thus interpret the dynamics as alternating a rotation of $\mbf{J}$ about a space fixed $z$-axis by angle $\beta$, followed by a procession of $\mbf{I}$ and $\mbf{J}$ about $\mbf{F}$ by an angle $\alpha |\mbf{F}|$, as shown in Fig (\ref{F1_1}).  Such a simple transformation nonetheless leads to complex dynamics, including chaos in the classical limit as discussed below.  From the quantum perspective, since the two rotations don't commute, there are insufficient symmetries to specify Floquet eigenstates by a complete set of commuting operators; the system is not integrable.  Note, however, that the system is invariant under an overall rotation around the $z$-axis, so $F_z$ is a conserved quantity ($F^2$ is not conserved).

We treat the classical limit of quantum mechanical spin in the familiar way \cite{Haake}. Each of our spins has three components, but a fixed magnitude, and thus their orientations can be specified by two variables. The $z$-component of a spin and the angle $\phi$, denoting its orientation in the $x$-$y$ plane, are canonically conjugate, and thus each spin constitutes one canonical degree of freedom.  The classical dynamical map has the same physical action as described above in the quantum context -- rotation of $\mbf{J}$ by angle $\beta$ followed by precession of $\mbf{I}$ and $\mbf{J}$ about $\mbf{F}$ by angle $\alpha |\mbf{F}|$.  Here, the rotations are implemented by $3 \times 3$ SO(3) matrices.  The two spins, plus time-dependent Hamiltonian imply a five dimensional phase space.  Since $F_z$  is conserved, the dynamics is restricted to a four-dimensional hypersurface.  As there are no additional constraints, the dynamics are not integrable and can exhibit chaos.  Note, Eq. (\ref{Eq:HKCT}) is of the form of Eq. \ref{joint}, with $H_2=0$, but where chaos is only seen in the coupled dynamics, not the dynamics of the of $H_1$ alone.

To visualize the dynamics, we rewrite our system in terms of a new set of variables, $(F_z, \bar{\phi} \equiv \phi_I+\phi_J)$ and $(\delta F_z  \equiv I_z-J_z, \delta\phi \equiv\phi_I-\phi_J)$, 
 \begin{subequations}
 \begin{align}
J_z &= \frac{F_z-\delta F_z}{2},  I_z = \frac{F_z+\delta F_z}{2},\\
\mbf{I} \cdot \mbf{J} &= I_z J_z + \sqrt{I^2 - I_{z}^2} \sqrt {J^2 - J_{z}^2} \left( \sin \phi_I \sin \phi_J+\cos \phi_I \cos \phi_J \right) \nonumber \\
&=\left( \frac{F_z+\delta F_z}{2}\right) \left( \frac{F_z-\delta F_z}{2} \right) +\sqrt{I^2-\left({\frac{F_z+\delta F_z}{2}}^2\right)} \sqrt {J^2-\left({\frac{F_z-\delta F_z}{2}}^2\right)}\cos(\delta \phi).
 \end{align}
\end{subequations}
Because $F_z$ is a conserved quantity, $\bar{\phi}$ does not appear in our Hamiltonian. It is a cyclic coordinate, and thus we can ignore it without losing any information about the further evolution of the remaining variables. Neither do we require $\bar{\phi}$ to determine the Lyapunov exponent of a chaotic system. Thus, we need only consider the two difference variables,  $(\delta F_z , \delta\phi)$,  and time, taking us from a four to a three dimensional hypersurface. This allows us to visualize our system using a Poincar\'{e} surface of section as a stroboscopic plot.  We restrict our attention here to $F_z=0$ as this also leads to the largest subspace in the associated quantum problem.  The reduction of our system to essentially one degree of freedom is not generic, but simplifies the analysis without sacrificing our ability to study the essential relations between chaos and entanglement.
\begin{figure}[t]
\includegraphics[width=4.0in]{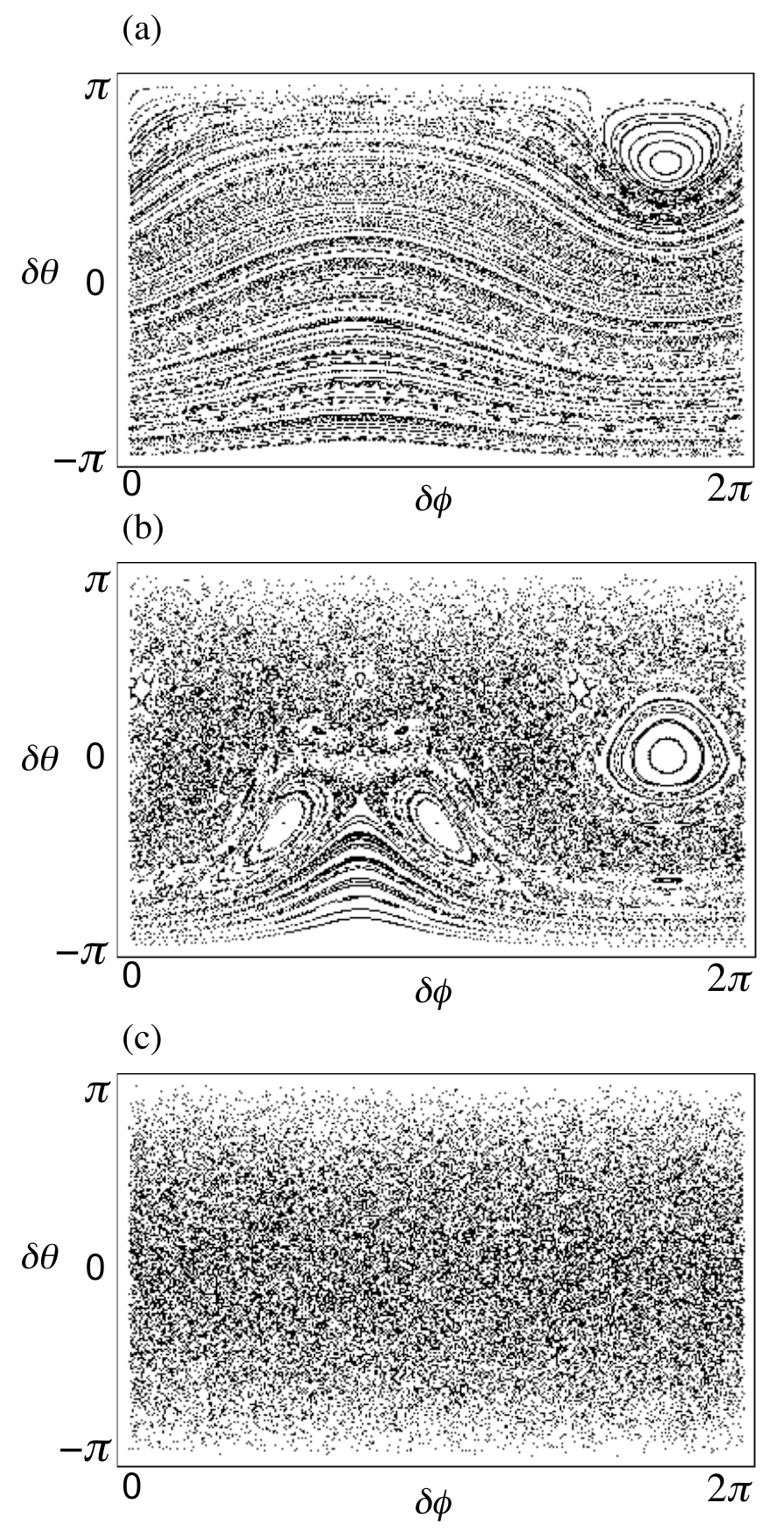}
\caption{Poincar\'{e} surface of section for the coupled kicked tops, with $F_z=0$ (a) Regular phase motion: $\alpha=1/2,\beta=\pi/2$, (b) Mixed phase space: $\alpha=3/2,\beta=\pi/2$, (c) Global chaos: $\alpha=6,\beta=\pi/2$.}
\label{F1_2}
\end{figure}
  \clearpage 
The classical equations of motion depend on the ratio $|\mbf{I}|/|\mbf{J}|$.  We focus here on equal spin magnitudes and fix $F_z=0$.  Thus, without loss of generality, since the SO(3) rotation matrices of classical dynamics are independent of spin magnitude, we take the spin vectors to be unit vectors. The basic structure of the phase space can be understood as follows. When the coupling is removed, our system has fixed points at the northern and southern ``poles". As the chaoticity parameter is turned up, chaos first forms around the unstable ``north pole" while regular behavior persists around the stable ``south pole". Further fixed points appear in the usual manner as bifurcations occur with increase of the chaoticity parameter. Figure (\ref{F1_2}) shows three different regimes of classical dynamics.  With the parameters $\alpha=1/2, \beta=\pi/2$ (Fig.  (\ref{F1_2}a), the dynamics are highly regular, with negligible stochastic motion.  When $\alpha=3/2, \beta=\pi/2$ (Fig.  (\ref{F1_2}b), we see a mixed space with chaotic and regular regions of comparable size. The parameters $\alpha=6, \beta=\pi/2$ (Fig.  (\ref{F1_2}c), give a completely chaotic phase space. 

We want to choose our quantum Hamiltonian so that we will recover our classical dynamics in the large spin limit. We would like to be able to vary the size of our spins, but we will keep the pair equal to each other in magnitude, $I=J$. Since the SU(2) rotation matrices depend on the spin magnitude, we must scale the Floquet operator.  By substituting $\alpha \rightarrow \tilde{\alpha}=\alpha/J$ we obtain the same Heisenberg equations of motion as the classical equations for equal magnitude spins.
\subsection{Quantum chaology}\label{S3}
In order to understand the dynamical generation of entanglement, we need to establish some basic understanding of the eigenstates of the system and their relationship to the classical dynamics.  As our system is time periodic, the states of interest are the eigenstates of the Floquet operator, Eq. (\ref{Eq:Floquet}).  It is useful to consider both the coupled and uncoupled representations of angular momentum connected by the usual Clebsch-Gordan expansion,
\begin{equation}
\ket{F,M_F} = \sum_{m_I,m_J}  \braket{I, m_I; J,m_J}{F,M_F} \ket{I, m_I}\ket{J,m_J}
\end{equation}
Conservation of $F_z$ implies that the operator is block diagonal for all states defined by quantum number $M_F$. The largest block, $M_F=0$, has dimension $2J+1$ as $F$ varies from 0 to $2J$. Using the uncoupled representation, denoting the product state by the single quantum number $m_J=-m_I$, the matrix,
\begin{eqnarray}
\bra{m'_J} U_{\tau} \ket{m_J}& =& \sum_{F} e^{-i \left( \alpha \frac{F(F+1)}{2J}+ \beta m_J \right) } \\
&  &\braket{F,0}{I, -m'_J; J,m'_J}  \braket{F,0}{I, -m_J; J,m_J} \nonumber 
\end{eqnarray}
can then be diagonalized to yield the Floquet eigenstates and eigenphases,
\begin{equation}
\label{Eq:Eigen}
\ket{k} = \sum_{m_J} c^{(k)}_{m_J}  \ket{I, -m_J}\ket{J,m_J}; \hspace{1 pc}
U_{\tau}\ket{k} = e^{i \phi_k} \ket{k}.
\end{equation}

A central result of quantum chaos is the connection with the theory of random matrices \cite{Haake}.  In the limit of large Hilbert space dimensions (small $\hbar$), for parameters such that the classical description of the dynamics shows global chaos, the eigenstates and eigenvalues of the quantum dynamics have the statistical properties of an ensemble of random matrices.  The appropriate ensemble depends on the properties of the quantum system under time-reversal \cite{Haake}.  We thus seek to determine whether there exists an anti-unitary (time reversal) operator $T$ that has the following action on the Floquet operator,
\begin{equation}
T U_{\tau} T^{-1} = U_{\tau}^{\dagger} = e^{i \beta J_z}e^{i \tilde{\alpha} \mbf{I}\cdot \mbf{J}}.
\end{equation}
Analogous to the case of the single kicked top, we consider the generalized time reversal operation,
\begin{equation}
\label{Eq:Reversal_1}
T=e^{i \beta J_z} K,
\end{equation}
where $K$ is complex conjugation in the uncoupled product representation.  Since {\em both} $I_y$ and $J_y$ change sign under conjugation, while the $x$ and $z$ components do not,
\begin{equation}
K J_z K = J_z; \hspace{1 pc} K\mbf{I} \cdot \mbf{J}K = \mbf{I} \cdot \mbf{J}.
\end{equation}
It then follows that 
\begin{eqnarray}
T U_{\tau} T^{-1} &=& \left( e^{i \beta J_z} K \right) \left( e^{-i\tilde{\alpha} \mbf{I} \cdot \mbf{J}}  e^{-i \beta J_z}  \right) \left( K e^{-i \beta J_z} \right)  \\
&=&  e^{i \beta J_z} \left( e^{i \tilde{\alpha} \mbf{I}\cdot \mbf{J}} e^{i \beta J_z}\right) e^{-i \beta J_z} \nonumber\\
&=& e^{i \beta J_z}e^{i \tilde{\alpha}\mbf{I}\cdot \mbf{J}}= U_{\tau}^{\dagger}, \nonumber
\end{eqnarray}
so the dynamics are time-reversal invariant. Moreover, $T^2=1$, so there is no Kramer's degeneracy. Given these facts, for parameters in which the classical dynamics are globally chaotic, we expect the Floquet operator to have the statistical properties of a random matrix chosen from the circular orthogonal ensemble (COE).

To further correlate the Floquet eigenstates with the classical phase space in the case of regular and mixed dynamics, it is useful to employ a Husimi representation. A spin coherent state has a minimum quantum uncertainty and is specified by polar orientation angles $\theta$ and $\phi$ on the sphere. In terms of the standard basis, a spin coherent state for a single spin is \cite{Puri}
\begin{equation}
\ket{\mu} = \sum_m \frac{\mu^{J-m}}{\left(1+|\mu|^2 \right)^J} \sqrt{\frac{(2J)!}{(J-m)!(J+m)!}}\space \ket{J,m},
\end{equation}
where $\mu =\tan(\theta/2) e^{i \phi}$.  For our system, because the subspaces in which the eigenstates live are not described by an irreducible representation of angular momentum, there are no such minimum uncertainty states for the difference angles.  Nonetheless, we obtain a useful set of states by projecting the product of spin coherent states associated with the two subsystems onto the subspace with a fixed value of $F_z$ (here $F_z=0$).  The result of the projection is
\begin{equation}
\label{Eq:Project}
\hat{P}_0 \ket{\mu_I} \ket{\mu_J} = \sum_m \left( \frac{\mu_I}{\mu_J} \right)^m \frac{(2J)!}{(J-m)!(J+m)!} \ket{m}_I \ket{-m}_J.
\end{equation}
Classically, in projecting onto the surface of section with $F_z=0$, we take $\theta_I + \theta_J = \pi$. Fixing this value in the quantum state one finds
\begin{equation}
\frac{\mu_I}{\mu_J}=e^{i(\phi_I-\phi_S)} \left[ \frac{1 + \sin\left( \frac{\theta_I - \theta_S }{2} \right)}{1 - \sin\left( \frac{\theta_I - \theta_S}{2}\right) }\right].
\end{equation}
The projected coherent state thus depends only on the difference of the angle variables, and allows us to consider localized quantum states correlated with the classical phase space of interest.  After normalizing, we arrive at an over-complete basis of states for the $F_Z=0$ subspace, parameterized by $\delta\theta$ and $\delta\phi$ . The Husimi distribution of a state $\ket{\psi}$ in this space,
   
\begin{equation}
\label{husimi_function}
Q(\delta\theta,\delta\phi) \equiv |\braket{\delta\theta,\delta\phi}{\psi}|^2
\end{equation}
then provides a visualization in phase space.
\begin{figure}
\includegraphics[width=\linewidth]{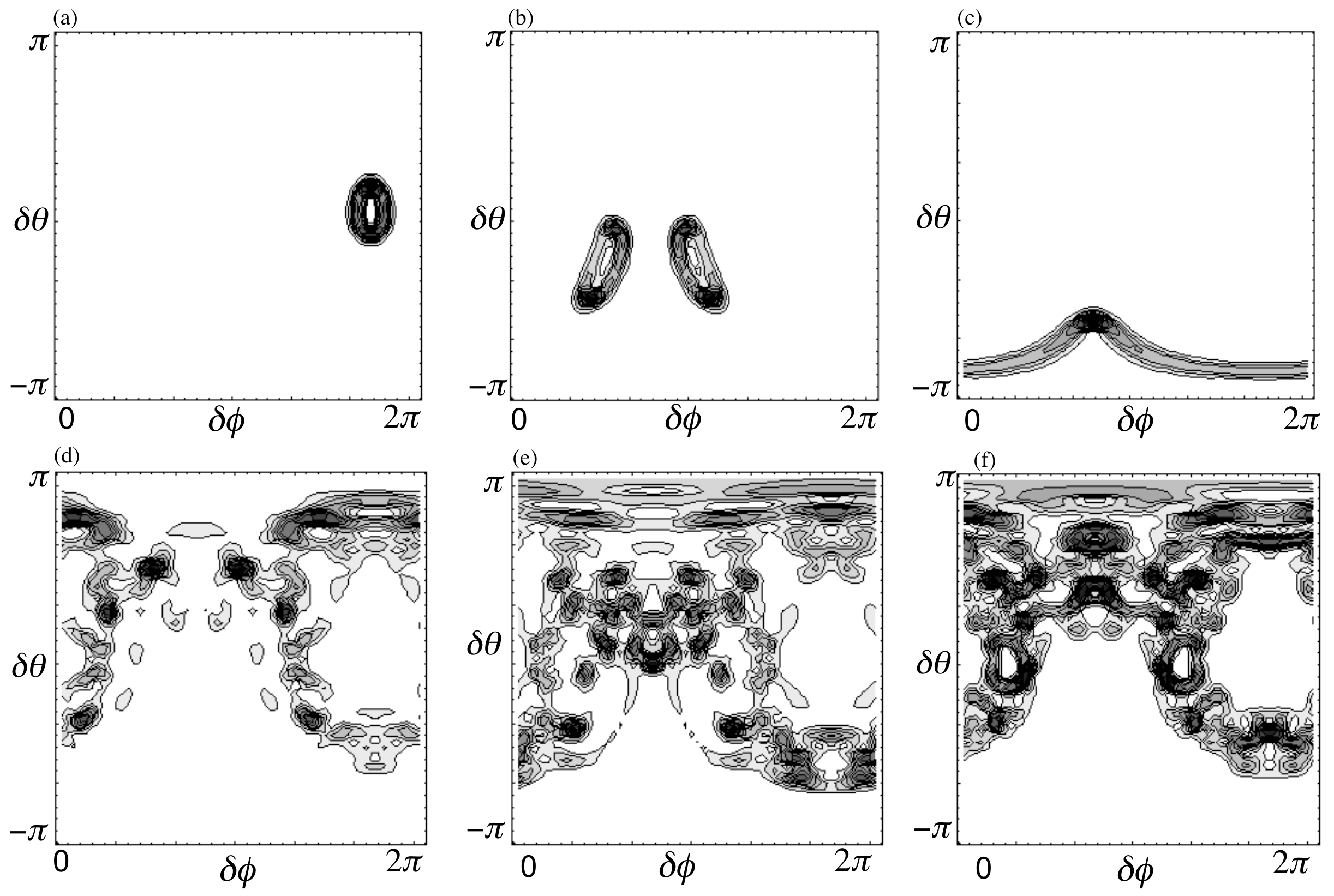}
\caption{ Husimi distributions of Floquet eigenstates associated with the parameters of a mixed phase space (Fig. \ref{F1_2}b). (a,b,c) Regular eigenstates around different fixed points. (d,e,f) Chaotic eigenstates, delocalized in the chaotic sea. For spin size, we have $I=J=150$.}
\label{F1_3}
\end{figure}
\clearpage
In order to explore the semiclassical limit, we choose $I=J=150$, corresponding to a $d=301$ dimensional Hilbert space in the $F_z=0$ subspace, or an ``effective $\hbar$" of $\hbar_{\text{eff}} = 1/301$. A $301$ dimensional Hilbert space is sufficiently large so that the minimum uncertainty spin coherent states can be effective considered 
as ``points" on the phase space. As the size of the Hilbert space increases, we expect a better quantum-classical correspondence.  
Figure \ref{F1_3} shows the Husimi plots of a few of the eigenstates for $\alpha/J=3/2, \beta=\pi/2$, for which the classical phase space is mixed (Fig. \ref{F1_2}b).  These plots exhibit the features expected according to Percival's conjecture.  The states roughly divide into regular and irregular sets, with regular eigenstates concentrated on invariant tori around stable fixed points, resembling harmonic oscillator eigenstates, and irregular ``chaotic" states randomly distributed within the chaotic sea.

\begin{figure}
\includegraphics[width=\linewidth]{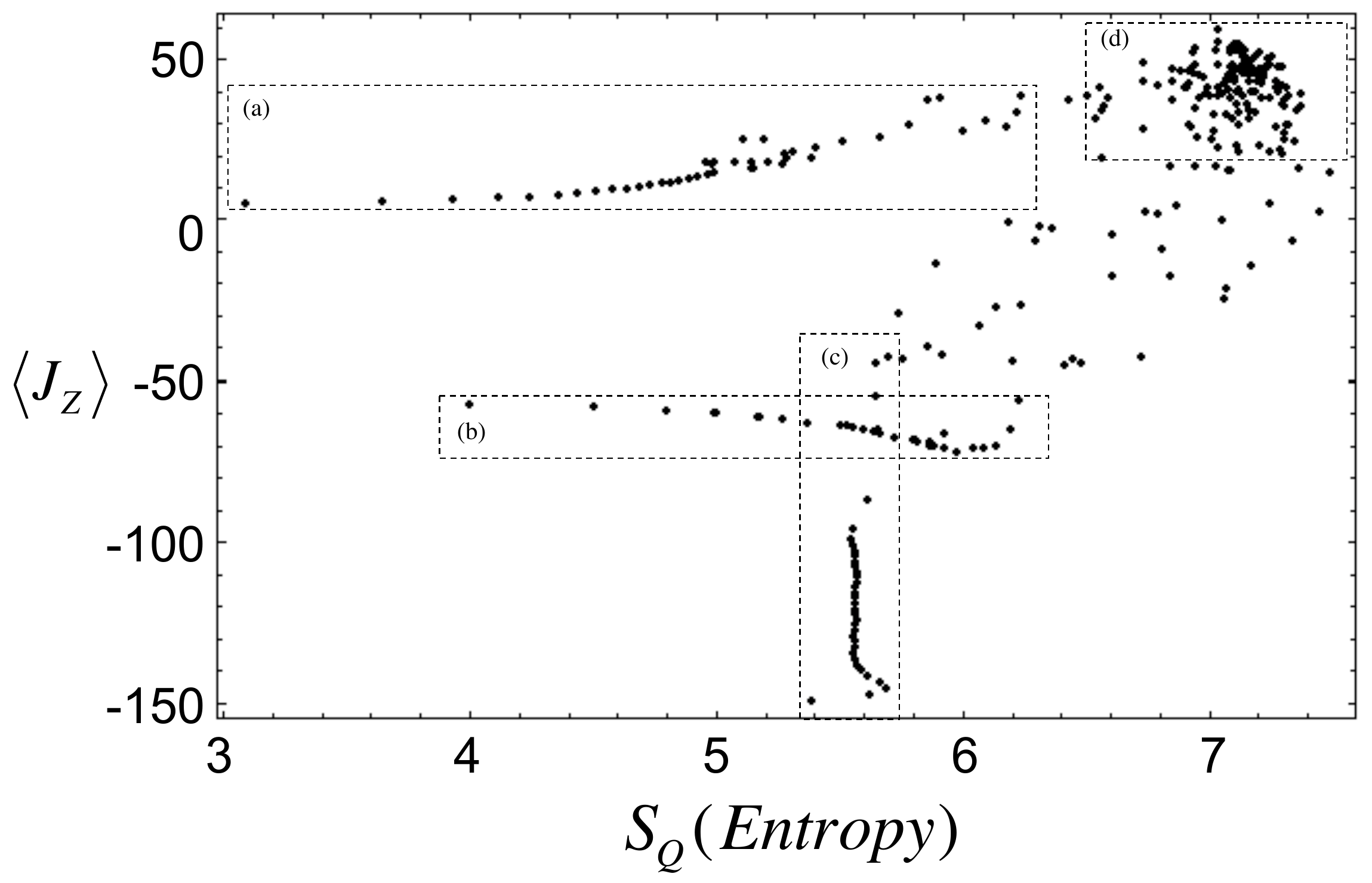}
\caption{Scatter plot of $\langle J_z \rangle$ vs. Husimi Entropy,  $S_Q$, for the Floquet eigenstates associated with the mixed phase space (Fig. (\ref{F1_2}b)). Boxed regions (a), (b), and (c) correspond to regular states centered around fixed points. States in region (d) are considered ``chaotic eigenstates".}
\label{F1_4}
\end{figure}
\clearpage
Though Percival's conjecture is largely born out in numerical analyses, it is not strictly true (especially in the finite $\hbar$ limit), nor is there a strict procedure for filtering the regular from chaotic eigenstates except for very special systems \cite{Tomsovic}. We can, nonetheless, create an approximate filter.  A useful measure for distinguishing states is the Shannon entropy of the Husimi distribution \cite{Wehrl}, 
\begin{equation}
S_Q= - \int d \mu \, Q(\delta \phi, \delta \theta) \log Q(\delta \phi, \delta \theta),
\end{equation} 
where $d\mu$ is the measure on the phase space of difference angles on the sphere and $Q$ is defined as in Eq. \ref{husimi_function}. We expect the states delocalized in the chaotic sea to have large entropy by this measure, while those states well-localized around fixed points have low entropy. This leaves some ambiguous situations, since highly excited states on regular tori also have high ``Husimi entropy''.  To improve the filter, we follow a procedure suggested by Korsch and coworkers \cite{Korsch}, which correlated the properties of the eigenstates to the classical phase space in order to distinguish the regular and irregular states for a nonlinear rotor.  In Fig. \ref{F1_4} we plot the values of $S_Q$ and $\langle J_z \rangle$.  The latter quantity correlates to the mean value of $\delta \theta$ in the semiclassical limit.  We see four distinct features in this plot.  Two lines of states with near constant $\langle J_z \rangle$ but increasing $S_Q$, boxed in Figs. (\ref{F1_4}a,b), correspond to the series of states localized around fixed points with increasing excitation (Figs \ref{F1_3}a,b).  The line of states with near constant $S_Q$ and increasing values of $\langle J_z \rangle$, boxed in Fig. \ref{F1_4}c,  correspond to the series of states localized around the stable ``south pole" (Fig. \ref{F1_3}c).  Finally, the cluster of states with high values of both $S_Q$ and $\langle J_z \rangle$, boxed in Fig. \ref{F1_4}d,  correspond to the states delocalized in the chaotic sea that are concentrated near the original unstable fixed point at the ``north pole" of the regular dynamics.  There is no clean division between this cluster and states clearly localized on invariant tori.  A qualitative examination, denoted in Fig \ref{F1_4}, nonetheless gives us an indication of the chaotic subspace for these mixed dynamics. As the size of the Hilbert space increases, we expect a sharper distinction between regular and chaotic eigenstates.
 Such an identification is useful for giving quantitative prediction of the dynamically generated entanglement, as we discuss in the next section. 

\section{Entanglement}
\label{Ent}
\subsection{Calculating Entanglement}
We consider only pure states of the bipartite system.  Entanglement is then uniquely determined by the coefficients in the Schmidt decomposition of the joint state of the system,
\begin{equation}
\ket{\Psi}_{IJ}=\sum_{i} \sqrt{\lambda_i} \ket{u_i}_I \ket{v_i}_J ,
\end{equation}
where $\lambda_i$ are the eigenvalues of the reduced density matrix of either subsystem, and the Schmidt basis vectors $\{\ket{u_i}_I, \ket{v_i}_J\}$ are their respective eigenvectors. The entanglement $E$ is the Shannon entropy of the Schmidt coefficients,
\begin{equation}
E = -\sum_i \lambda_i  \log( \lambda_i).
\end{equation}

Determination of the Schmidt decomposition is typically a nontrivial task, requiring partial trace and diagonalization of the reduced density operator\cite{nielsen00a}.  The Schmidt basis will generally depend on the state $\ket{\Psi}_{IJ}$.  For the system at hand, we have a unique situation -- within a subspace with a fixed value of $F_z$, the uncoupled basis of angular momentum is the Schmidt basis, {\em independent of the state}, as seen, e.g., Eq. (\ref{Eq:Eigen}).  Thus, for states within such subspaces, the entanglement is easily calculated as the Shannon entropy of the probability distribution of the state when expanded in the standard product basis. This not only simplifies calculations, but connects entanglement with the entropy of random states with respect to a fixed basis \cite{Wootters}. Thus, the choice of this particular metric of entanglement enables us to compare the entanglement properties of a state to the properties of random states in Hilbert space.

Throughout this section, we consider the $F_z=0$ subspace, and take $I=J=150$, corresponding to a Hilbert space of dimension $d=301$.  The maximum possible entanglement in this case is $E_{max} = \log d \approx 5.71$.

\subsection{Numerical Solutions}\label{S1}
\begin{figure}
\includegraphics[width=\linewidth]{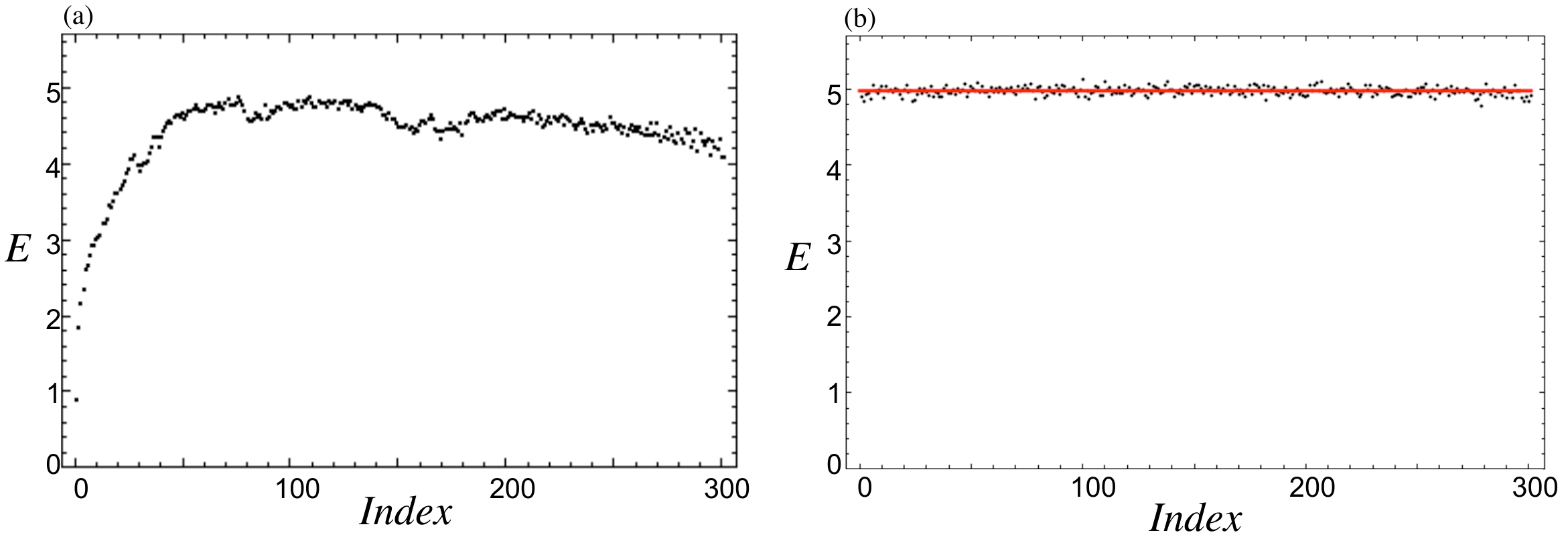}
\caption{Entanglement of the Floquet eigenstates. (a) Map corresponding to a mixed phase space: $\alpha=3/2,\beta=\pi/2$. (b) Map corresponding to global chaos: $\alpha=6,\beta=\pi/2$. The solid line gives the value expected from random matrix theory, Eq. (\ref{Eq:Ereal}).}
\label{F1_5}
\end{figure}

The entanglement of the Floquet eigenstates is easily calculated based on the discussion above.  Since the eigenstates reside in a subspace with fixed $F_z$, the uncoupled representation of angular momentum is the Schmidt basis, and the entanglement between spins in a given eigenstate $\ket{k}$ is the Shannon entropy of the probability distribution of the expansion $\lambda^{(k)}_{m_J}= |c^{(k)}_{m_J}|^2$ from Eq. (\ref{Eq:Eigen}).  Figure \ref{F1_5} shows a list plot of this entanglement for a mixed phase space (as shown in Fig. (\ref{F1_2}b)) and a completely chaotic space (as shown in Fig. (\ref{F1_2}c)).  In the latter case, the entanglement values are clustered around the value expected from random matrix theory, discussed below.

\clearpage
Our main interest is to study the dynamically generated entanglement and its correlation with the classical phase space.  We wish to associate quantum states with our classical initial conditions. The ``most classical" state of a quantum system is a coherent state, so it would be natural to associate a point in our four-dimensional classical phase space with a product of spin coherent states.  These states, however, have support on several subspaces with different values of $F_z$, and thus correspond to a distribution of classical surfaces of sections.  To avoid this complication, we project our coherent states into the $M_F=0$ subspace, and then renormalize them, as described in Eq. (\ref{Eq:Project}).  This gives us a pure state, which though no longer separable, typically has a low entanglement and is localized around a point in the classical phase space in the relevant difference angles.

\begin{figure}
\includegraphics[width=\linewidth]{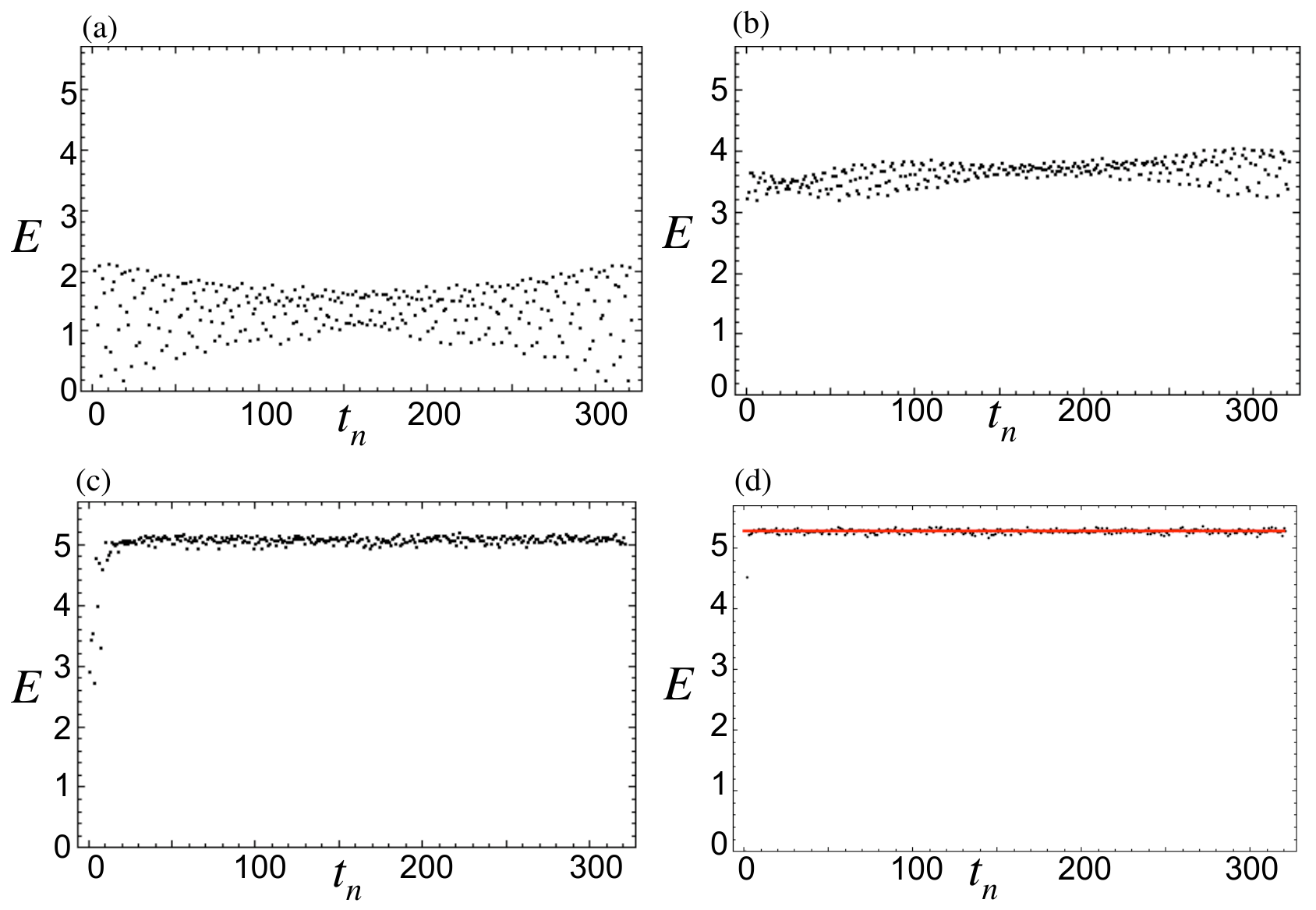}
\caption{ Dynamically generated entanglement as a function of the number of applications of the Floquet map. (a) Mixed phase space ($\alpha=3/2,\beta=\pi/2$), regular initial condition: $\ket{\psi_0}=\ket{I, I}\ket{J,-J}$. (b) Mixed phase space ($\alpha=3/2,\beta=\pi/2$), regular initial condition: $\ket{\psi_0}=\ket{\delta\theta=\pi/10,\delta\phi=53\pi/30}$. (c) Mixed phase space ($\alpha=3/2,\beta=\pi/2$), chaotic initial condition: $\ket{\psi_0}=\ket{I, -I}\ket{J,J}$ (d) Globally chaotic phase space ($\alpha=6,\beta=\pi/2$), chaotic initial condition:  $\ket{\psi_0}=\ket{\delta\theta=\pi/2,\delta\phi=\pi/3}$.  The solid line gives the value expected from random states in the Hilbert space, Eq. (\ref{Eq:Ecomplex}).}
\label{F1_6}
\end{figure}
\clearpage
The time-evolved state after $n$ applications of the Floquet operator to the projected coherent state is
\begin{equation}
\ket{\psi_n (\delta \theta, \delta \phi)}=U_{\tau}^n \ket{\delta \theta, \delta \phi} = \sum_k a_k e^{-i n\phi_k} \ket{k},
\end{equation}
expanded in the Floquet eigenstates, where $a_k = \braket{k}{\delta \theta, \delta \phi}$ is the initial spectral decomposition.  The Schmidt coefficients are the expansion of this state in the angular momentum product basis (the Schmidt basis) giving,
\begin{equation}
\label{Eq:lambda}
\lambda^{(n)}_{m_J}= \left|  \sum_k a_k   e^{-i n \phi_k} c^{(k)}_{m_J} \right|^2.
\end{equation}
according to Eqs. (\ref{Eq:Eigen}, \ref{Eq:lambda}). The Shannon entropy of these coefficients gives the dynamically evolved entanglement.  Figure \ref{F1_6} shows this quantum evolution for parameters such that the classical evolution is described by a mixed phase space.  For a coherent state initial condition chosen in the middle of a regular island ($\ket{\psi_0}=\ket{I, I}\ket{J,-J}=\ket{\delta\theta=-\pi,\delta\phi=0}$), the entanglement rises slowly and oscillates between high and low values.  For an initial condition in the chaotic sea ($\ket{\psi_0}=\ket{\delta\theta=\pi/2,\delta\phi=\pi/3}$), the entanglement rapidly rises and saturates to a near constant value, with small fluctuations about the quasi-steady state.

\begin{figure}
\includegraphics[width=\linewidth]{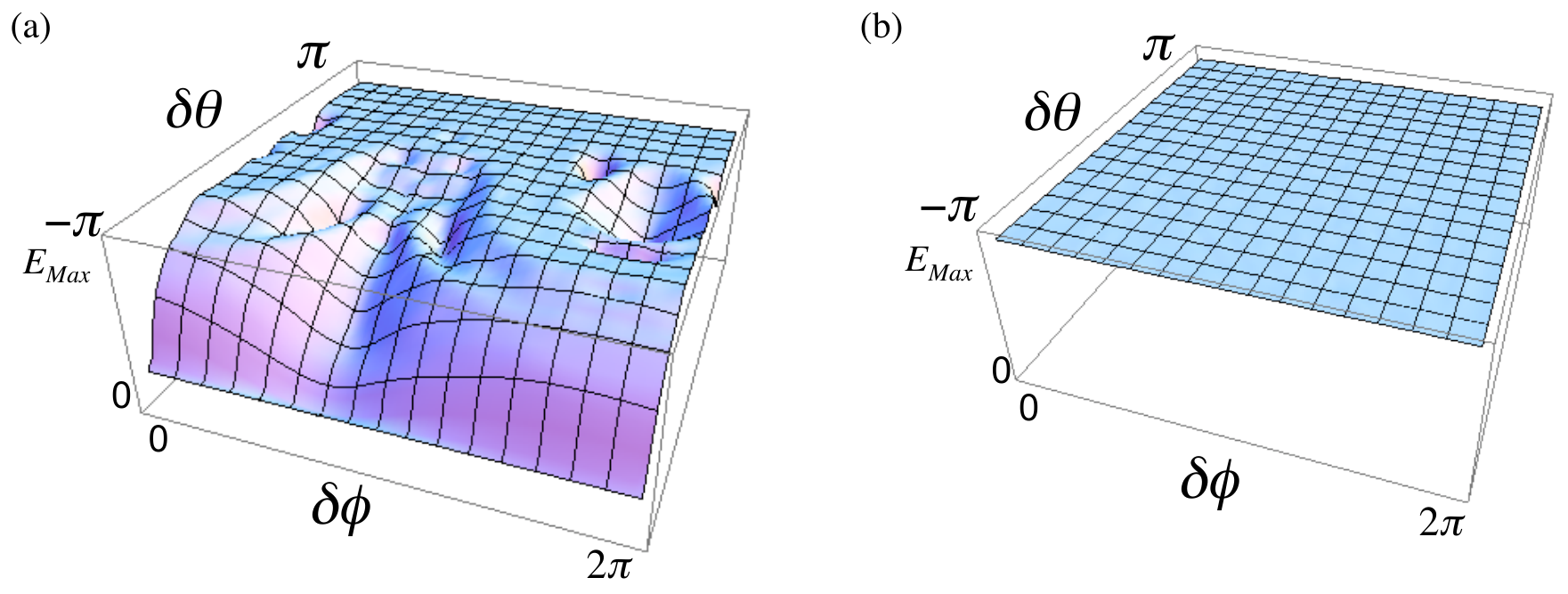}
\caption{Long-time average entanglement as a function of mean coordinate of the initial projected coherent state. (a) Mixed phase space: $\alpha=3/2,\beta=\pi/2$  (b) Globally chaotic phase space: $\alpha=6,\beta=\pi/2$)}
\label{F1_7}
\end{figure}

\begin{figure}
\includegraphics[width=\linewidth]{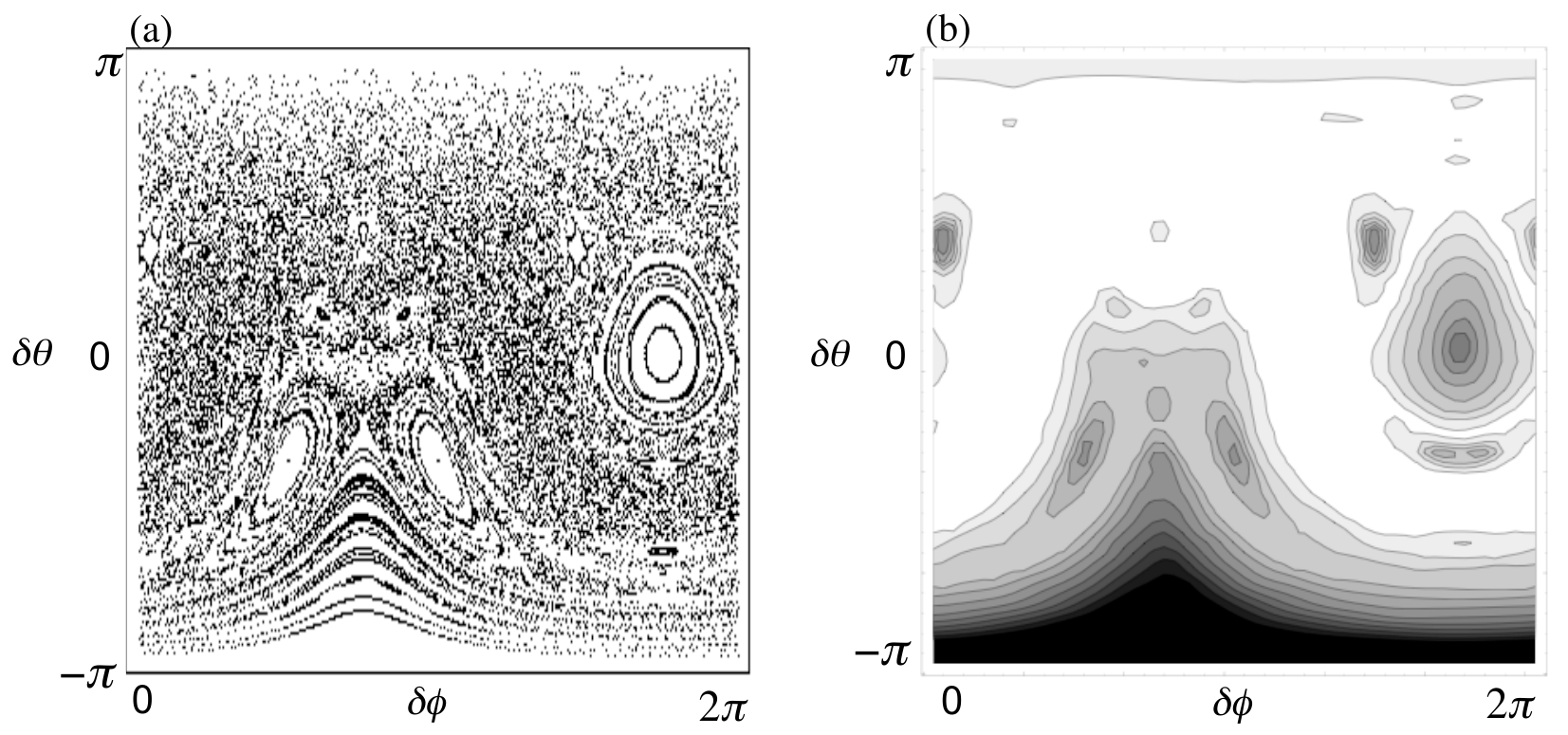}
\caption{Side-by-side comparison, showing dynamically generated entanglement as superb signature of classical chaos in a mixed phase space ($\alpha=3/2,\beta=\pi/2$). (a) Classical phase space, Poincar\'{e} section. (b) Long-time average entanglement as a function of mean coordinate of the initial projected coherent state}
\label{F1_8}
\end{figure}
\clearpage
In order to better explore how the entanglement evolution saturates to a particular value, we average over many time steps to find a long-time average of entanglement. We drop the first three hundred steps in order to remove transient effects and insure that the dynamics settle into a quasi-steady state, and then average over times steps 300-320.  By looking at a plot of this average, we can see how it correlates with initial conditions, a procedure initially carried out for the kicked top Hamiltonian by Wang {\em et al.} \cite{Wang}.  Figure \ref{F1_8} shows remarkably strong correlation between structures in the classical mixed phase space and the long-time entanglement average plot.  Chaotic initial conditions generally go to a higher average value than regular initial conditions, with the smallest values of entanglement generation near the classical fixed points.  Additionally, all initial conditions in the chaotic sea saturate to nearly the same average entanglement. 

For parameters corresponding to global chaos, we can see that the surface plot is very flat (see Fig. \ref{F1_8}a), with all initial conditions converging to nearly the same long-time entanglement average.  For the parameters at hand, averaging over all initial conditions, the dynamically generated entanglement is $\bar{E}_{dynam}=5.28$, as compared to the value $\bar{E}_{eigens}=4.97$ found for the average entanglement of the eigenstates of the Floquet map.   For the mixed phase space, the value of long-time entanglement is flat for initial conditions that correlate with the classical chaotic sea.  To find the entanglement characteristic of the chaotic initial conditions, we take a grid of coherent states across the phase space.  Each point in the grid is determined as ``regular" or ``chaotic" by the local Lyapunov exponent of the classical dynamics.  For those states with positive Lyapunov exponent we evolve according to the Floquet operator and calculate the long-time entanglement average, as described above.  Weighting these values according to the measure on phase space gives us an average entanglement of $\bar{E}_{dynam}=5.08$ in the chaotic sea, significantly lower than that for the globally chaotic phase space. Below, we interpret these results with statistics of random states in Hilbert space and their connection to quantum chaos.

\subsection{Entanglement and random states in Hilbert space}
The numerical studies in Sec. \ref{S1} reveal some empirical facts.  When the Floquet map corresponds to a fully chaotic phase space, the entanglement of the eigenstates are all nearly equal, with an average value independent of the coupling strength and below the maximum possible entanglement for the bipartite system.  Moreover, the dynamically generated entanglement when starting from a projected coherent state localized in a chaotic sea saturates to a nearly constant value after a few applications of the Floquet map.  In a mixed phase space, the amount of entanglement increases as the size of the chaotic sea increases.  For a completely chaotic space, the value no longer changes with coupling strength.  This saturation value is {\em different} from the entanglement seen in the eigenstates.  These facts leads us to conclude that the value of entanglement generation for chaotic maps is {\em statistical} in nature, as emphasized by Bandyopadhyay and Lakshminarayan \cite{Lakshminarayan/Bandyopadhyay2002,Lakshminarayan/Bandyopadhyay2004}, and Scott and Caves \cite{Scott/Caves}. The predicted values follow from the theory of random matrices and random states in Hilbert space, which we briefly review.

The random matrix conjecture of quantum chaos states that when the Floquet map (in a periodically driven system) classically generates global chaos, the quantum operators have many of the statistical properties of a random matrix drawn from an appropriate ensemble depending on fundamental symmetries \cite{Haake}.  Systems with time-reversal symmetry (and no Kramer's degeneracy), have Floquet maps with many of the statistical properties of random matrices chosen from the Circular Orthogonal Ensemble (COE) -- the set of symmetric unitary matrices with a probability distribution defined by the orthogonally invariant Haar measure \cite{Dyson}.  Without time-reversal symmetry, the Floquet maps have many of the statistical properties of random matrices chosen from the  Circular Unitary Ensemble (CUE)  -- the set of general unitary matrices with a probability distribution defined by the unitarily invariant Haar measure \cite{Dyson}.  When expressed in the basis of their eigenvectors, both such matrices have the form $U=\sum_k \exp{i \phi_k} \ket{k}\bra{k}$, where the phases are randomly distributed from $0$ to $2 \pi$ with a uniform probability distribution.  In the case of COE, the eigenvectors are invariant with respect to an anti-unitary operator T; for CUE the eigenvectors have no time-reversal invariance. 

We seek to predict the entanglement of state vectors based on statistical arguments. We can do this by averaging over an appropriate distribution of random states in Hilbert space \cite{Wootters}. To construct the probability measure for sampling random states, we employ a parameterization equivalent to the Hurwitz parameterization of random unitaries \cite{Haar}.  Such a measure can be constructed by connecting the vector space with a manifold upon which there is a known geometric measure. A normalized state in a $d$-dimensional complex Hilbert space can be visualized as a point on the surface of a hypersphere in a $2d$ dimensional real space, where for each of the $d$ basis vectors in Hilbert space we assign a pair of orthogonal directions that project out the real and imaginary parts of the state's probability amplitude. The surface area of a differential patch on a hypersphere is then the probability measure for picking uniformly distributed random states.  The coordinates of a state, parameterized by angles on the hypersphere, and the corresponding measure over the space are
\begin{subequations}
\label{Eq:complex}
\begin{gather}
c_{1,r}=\cos\theta_1 ,\\
c_{1,i}=\sin\theta_1\cos\theta_2 ,\\
c_{n,r}=\sin\theta_1\dots\sin\theta_{2n-2}\cos\theta_{2n-1} ,\\ 
c_{n,i}=\sin\theta_1\dots\sin\theta_{2n-1}\cos\theta_{2n} ,\\ 
c_{d,r}=\sin\theta_1\dots\sin\theta_{2d-2}\cos\theta_{2d-1} ,\\
c_{d,i}=\sin\theta_1\dots\sin\theta_{2d-1} ,\\
d\lambda=N \sin^{2d-2}\theta_1\sin^{2d-3}\theta_2\dots \nonumber ,\\
\sin\theta_{2d-2} d\theta_1 d\theta_2\dots d\theta_{2d-1},
\end{gather}
\end{subequations} 
where $c_{n,r}$, $c_{n,i}$ are the real and imaginary expansion coefficients in the $n^{th}$ basis state, $d\lambda$ is the surface element, and $N$ is a normalization constant. The angles all range from $(0,\pi)$ except for the last angle which varies from $(0,2\pi)$. This defines the measure for random vectors over the field of complex numbers.

For random states in a real vector space, the probability measure is the area element on a $d$ dimensional hypersphere, with each direction corresponding to a basis vector. In this case the coordinates of the state and measure over the space are
\begin{subequations}
\label{Eq:real}
\begin{gather}
c_{1,r}=\cos\theta_1 ,\\ 
c_{n,r}=\sin\theta_1\dots\sin(\theta_{n-1})\cos(\theta_{n}) ,\\ 
c_{d,r}=\sin\theta_1\dots\sin(\theta_{d-1}) ,\\
d\lambda=N \sin^{d-2}\theta_1\sin^{d-3}\theta_2\dots \nonumber ,\\
\sin\theta_{d-2}d \theta_1 d \theta_2\dots d \theta_{2d-1}.
\end{gather}
\end{subequations} 
This defines the measure for random vectors over the field of real numbers.

With these measures in hand, we can calculate expected values of entanglement of random states in an appropriate ensemble and compare them to the numerically predicted results.  For large $d$-dimensional spaces, the variance scales as $1/\sqrt{d}$  \cite{Wootters}, so when the states in question are well-described by the statistics above, we anticipate the expectation value to give good predictive power.  A well known example is the entanglement of a ``typical state" picked at random from a $d_1 \otimes d_2$ tensor product Hilbert space, with no other restrictions of symmetry.  The Haar measure average of the entanglement over the whole space gives \cite{Page,Scott/Caves, Hayden}
\begin{equation}
\bar{E}_{d_1 \otimes d_2}=\sum^{d_1d_2}_{k=d_1+1} \frac{1}{k}-\frac{d_1-1}{2d_2}, d_2\geq d_1.
\end{equation}
For large dimensions, $\bar{E}_{d_1 \otimes d_2} \approx \log d_1 - d_1/(2 d_2)$, which is close to the maximum possible value of entanglement, but saturates slightly below.  Typical pure states in an unconstrained bipartite Hilbert space are highly entangled \cite{Hayden}.  For the case at hand, symmetries constrain the accessible Hilbert space.  We thus turn to study the typical entanglement expected under these conditions.

\subsubsection{Typical entanglement in a subspace}\label{S2}
Our system has an additional symmetry, its rotational invariance around the $z$-axis.  This restricts our system so that eigenstates and dynamics take place in subspaces with fixed values of $F_z$.  Calculation of entanglement within a subspace is generally a nontrivial task as there is no simple expression for the entanglement in terms of variables that we can average over the Haar measure \cite{Georgeot}. In our case, there is a happy accident -- the uncoupled basis of angular momentum, $\vert J, m_J \rangle \otimes \vert I, M_F-m_J \rangle$, is also the Schmidt basis for {\em all} states in the subspace. This implies that we can take the {\em fixed} Schmidt vectors as the directions that define the space on a hypersphere, and thereby employ the same parameterization of the Haar measures as in Eqs. (\ref{Eq:complex},\ref{Eq:real}), where now $d$ is the dimension of the subspace.  Note, this would not in general be possible for an arbitrary subspace because the entanglement is not a simple function of the expansion coefficients in a fixed basis.  The key question we must address is whether, with respect to the Schmidt basis, the state vector is random over the field of real or complex numbers, since the statistical properties of these two vector spaces differ, as discussed by Wootters \cite{Wootters}.  Once that question is answered, one can predict the entanglement based on the expected entropy in the Schmidt basis. 

For a state in a fixed $F_z$ subspace, expanded in the uncoupled basis, $\ket{\Psi}=\sum c_{m_J} \ket{m_J} \ket{-m_J}$, the entanglement is 
\begin{equation}
\label{Eq:Entropy}
E = - \sum_{m_J} \left| c_{m_J} \right|^2 \log \left(  \left| c_{m_J} \right|^2 \right ).
\end{equation}
Note that, the Schmidt basis is T-invariant according to time-reversal operator Eq. (\ref{Eq:Reversal_1}).  Thus, any other vector which is T-invariant will have real expansion coefficients $c_{m_J}$.  If the vector is random with respect to this basis, then these real coefficients are distributed on the hypersphere according to Eq. (\ref{Eq:real}). The contribution of each term in the expression for the entanglement given above should be equal, so we can shortcut by integrating only the first term, and multiplying by the number of terms, $d$. We normalize by an integral over the measure for that variable.  The result for T-invariant vectors is
\begin{eqnarray}
\label{Eq:Ereal}
\bar{E}_{R} &=&d\frac{ - \int \left| \cos\theta_1 \right|^2 \log \left(  \left| \cos\theta_1 \right|^2 \right ) \sin^{d-2} \theta_1 d \theta_1}{ \int \sin^{d-2}\theta_1 d \theta_1} \nonumber \\
&=&\mathcal{H}_{d/2}+\log 4-2,
\end{eqnarray}
where
\begin{equation}
\mathcal{H}_D = 1 + 1/2 + 1/3 + \dots + 1/D
\end{equation}
is the harmonic series. 

When the state is not T-invariant, its expansion coefficients in the Schmidt basis will be complex.  For random complex states states, it is useful to first simplify our parameterization by specifying the magnitudes of the expansion coefficients in terms of the angles on the hypersphere, rather than the real and imaginary parts of the expansion coefficients. Our new parameterization and the associated surface element are as follows:
\begin{subequations}
\begin{gather}
\left|c_{1}\right|=\cos\theta_1 \\
\left|c_{m}\right|=\sin\theta_1\dots\sin\theta_{m-1}\cos\theta_m \\
\left|c_{d}\right|=\sin\theta_1\dots\sin\theta_{d-1} \\
d\lambda=N \sin^{2d-3}\theta_1\sin^{2d-5}\theta_2\dots\nonumber \\
\sin\theta_{d-1}\cos\theta_1\dots\cos\theta_{d-1}d\theta_1 d\theta_2...d\theta_{2d-1},
\end{gather}
\end{subequations} 
where $\theta_m$ now ranges from $(0,\pi/2)$. Since the entanglement for a state in the subspace depends only on the the magnitudes $\{|c_m|\}$, Eq. (\ref{Eq:Entropy}) can be expressed in terms of this parameterization of the manifold.  Performing the average, the typical entanglement for a complex state, restricted to a $F_z$ subspace, is
\begin{eqnarray}
\label{Eq:Ecomplex}
\bar{E}_{C} &=&d\frac{ -  \int \left| \cos\theta_1 \right|^2 \log \left(  \left| \cos\theta_1 \right|^2 \right ) \sin^{2d-3} \theta_1 \cos\theta_1 d \theta_1}{ \int \sin^{2d-3}\theta_1 \cos\theta_1 d \theta_1} \nonumber \\
&=& \mathcal{H}_d-1.
\end{eqnarray}
These averages hold regardless of dimension of the space, though the variance of the distribution rapidly narrows as $d$ increases.

In the limit of large dimensional spaces, we recover the results of  Wootters \cite{Wootters} and Zyczkowski \cite{Zyczkowski} for the entropy of a random state in a real or complex vector space,
\begin{subequations}
\label{Eq:Log}
\begin{gather}
\bar{E}_{\text{R}} \rightarrow \log d - 2 + \log 2 + \gamma ,\\
\bar{E}_{\text{C}} \rightarrow \log d - 1 + \gamma,
\end{gather}
\end{subequations}
where $\gamma \approx 0.577 $ is Euler's constant.  Whereas these expressions give the entanglement of our state in the $F_z$ subspace, in general the entropy of a random state with respect to a fixed basis is not equal to its entanglement.  For example, for the full tensor product space, for large dimensional Hilbert spaces with $d_1=d_2$, $\bar{E}_{d_1 \otimes d_2} \rightarrow \log d_1 -1/2$, which differs from the Wooters/Zyczkowski entropy, Eq. (\ref{Eq:Log}), taking $d=d_1^2$.

As an aside, we can repeat our calculations for the linear entropy, an entanglement monotone. The linear entropy is determined by the purity of the reduced density operator of one subsystem,
\begin{equation}
S_{\text{L}}(\rho)=1-Tr(\rho_{\text{red}}^2) = 1- \sum_m \lambda_m^2  =1-\sum_m \left| c_m \right|^4
\end{equation}
where $\lambda_m = \left| c_m \right|^2$ are the Schmidt coefficients for a state in the subspace.  We repeat our integrals over the appropriate manifolds and find 
\begin{equation}
\bar{S}_{\text{L,R}}=1-\frac{3}{d+2}, \hspace{1 pc} \bar{S}_{\text{L,C}}=1-\frac{2}{d+1},
\end{equation}
the same results found by Brown and Viola by different methods \cite{Brown/Viola}.

\subsubsection{Typical entanglement prediction for the kicked coupled-tops}
With the results of Sec. \ref{S2} in hand, we can compare the predictions of the typical entanglement of random states to the entanglement found numerically in Sec. \ref{S1}.  Since the system is time reversal invariant without Kramer's degeneracy as shown in Sec. \ref{S3}, under the random matrix conjecture of quantum chaos, we expect the eigenstates of the Floquet operator for globally chaotic classical dynamics to be random real states \cite{Kus88}.  The eigenstates are restricted to a subspace with fixed value of $F_z$, so Eq. (\ref{Eq:Ereal}) applies.  We consider the $F_z=0$ subspace with dimension $d=2J+1$.  For spin $J=150$, one finds $\bar{E}_{\text{R}} = 4.98$, in excellent agreement with the mean entanglement of the eigenstates for the globally chaotic case, $\bar{E}_{\text{eigens}}=4.97$.  

Next we consider the dynamically generated entanglement, starting from a spin coherent product state projected into the $F_z=0$ subspace.  The key conjecture, seen numerically in prior studies, is that chaotic maps acting on a fiducial state generate states with the statistics of random states in Hilbert space, chosen according to the appropriate ensemble.  However, contrary to prior claims \cite{Lakshminarayan/Bandyopadhyay2002}, though the Floquet operator is a member of the COE, the dynamically generated state is {\em not} a random real vector in the Schmidt basis.  To see this, first note that since the Floquet operator is a member of the COE, we know the eigenstates are time-reversal invariant, $T\ket{k}=\ket{k}$.  However, according to Eq. (\ref{Eq:Reversal_1}), time reversal acting on the dynamically evolved state gives
\begin{equation}
T\ket{\psi_n (\delta \theta, \delta \phi)}=\sum_k a_k^* e^{+i n\phi_k} \ket{k} \neq \ket{\psi_{n} (\delta \theta, \delta \phi)}.
\end{equation}
Thus, the dynamically evolved state is {\em not} an eigenstate of the time reversal operator.  This is true even when the initial state itself is a time-reversal eigenstate (e.g., the coherent state at the pole), in which case $T\ket{\psi_n}=\ket{\psi_{-n}}$.  

To further put a point on this, consider the state expanded in the Schmidt basis.  Simplifying our notation, let $\ket{m} = \ket{I, m_I=m}\ket{J,m_J=-m}$ be a Schmidt vector.   After $n$ applications of the Floquet operator, $\ket{\psi_n} = \sum_{m} c^{(n)}_m  \ket{m}$.  In the transformation from the initial to final vector in this basis, $c^{(n)}_m = \sum_{m'}M^{(n)}_{m,m'} c^{(0)}_{m'}$, the matrix $M^{(n)}_{m,m'} = \sum_k e^{i n \phi_k} \braket{m}{k}\braket{k}{m'} $ is not an orthogonal matrix.  It is a random unitary matrix with complex entries with respect to the basis of interest -- the Schmidt basis.  The vector $c^{(n)}_m$ is thus a random vector in complex vector space.

Given the observations above, we expect the dynamically generated entanglement to be predicted by the statistics of random complex vectors.  This is indeed born out in the numerics.  For the globally chaotic map, we evolve and average to find the quasi-steady state value, as discussed in Sec. \ref{S1}. The long-time entanglement average is almost independent of the initial coherent state, projected in the $F_z=0$ subspace. For these initial condition Eq. (\ref{Eq:Ecomplex}) predicts $\bar{E}_{C}=5.28$ in good agreement with the long-time average value of 5.28. 

In the case of a mixed phase space, we saw that the long-time entanglement average was almost constant for initial states localized in the chaotic sea.  Clearly, this value of entanglement is a statistical property of Hilbert space.  Just as the quantum dynamics lead to a random state in the entire $F_z=0$ subspace when the classical dynamics are globally chaotic, for a classically mixed phase space, based on Percival's conjecture, the quantum dynamics generate a random state in the {\em chaotic subspace}. The structure of the chaotic sea cannot be described by a simple symmetry, so we cannot determine the entanglement of a typical state analytically. However, we can filter the eigenstates to determine which are in the chaotic subspace, as discussed in Sec. \ref{S3}, and sample randomly from a unitarily invariant measure over this subspace in order to find the typical entanglement value.  In this case there is no simple expression for the entanglement as a function of the states, so we cannot analytically take the average over the appropriate measure as before. Instead, we generate a large number of random states in the chaotic subspace, and find their entanglements. We do this by picking the real and imaginary parts of the expansion coefficients with respect to the chaotic eigenstates according to a Gaussian distribution. After normalizing, the entanglement is calculated for this state, and the process is repeated 100 times. The results are averaged to find an estimate of the average entanglement of a random state in the chaotic sea. We find that the average entanglement of a random state in the chaotic subspace picked according to the measure for complex random vectors is 5.13, in good agreement with the numerically determined value of $\bar{E}_{dynam}=5.08$ found in Sec. \ref{S1}.   Part of this discrepancy is likely due to the greater degree of variation of entanglement across the chaotic sea in the mixed phase space compared to the relatively flat completely chaotic phase space.   In addition our filter for determining the members of the chaotic subspace was somewhat crude with an ambiguous ``grey zone".  We would expect this to improve deeper in the semiclassical regime, where Percival's conjecture applies better.  

\section{Discussion and Summary}
\label{DS3}
Classical chaotic dynamics lead to ergodic mixing in phase space. Quantum analogs of ergodicity have long been considered, including ergodicity of eigenfunctions \cite{Zelditch}, ``spectral chaos" \cite{Heller84}, and increase in entropy associated with the wave function when expanded in a fixed (non-stationary-state) basis \cite{Peres}.  Recent numerical studies indicate that quantum dynamics generated by nonintegrable Hamiltonians generates pseudo-random states in a Hilbert space \cite{Scott/Caves, Emerson}.  In that sense, quantum chaotic dynamics is the classical analog of ergodic mixing in quasi-steady state, for times sufficiently long compared to the transient behavior, but short compared to the Heisenberg time or the time when correlations in the pseudo-random matrix appear \cite{Abreu2007}.  Such a result is not new, having its roots in the random matrix theory conjecture of quantum chaos \cite{Bohigas} -- the typical Hamiltonian of a nonintegrable system has the statistical properties of random matrices of an ensemble picked according to the symmetries of the system under time reversal.  The classic works on the subject, however, focus on the properties of the stationary states and spectra -- Berry's ``quantum chaology" \cite{Berry87}.  

The dynamical generation of random quantum states has implications for the dynamical generation of entanglement.  It is well known that for large dimensional bipartite Hilbert spaces, a random state is highly entangled with almost the maximum entanglement allowed by the dimension \cite{Hayden}.  As the large dimensional limit is equivalent to the $\hbar \rightarrow 0$ semiclassical limit, and to the degree that the quantum analogs of chaotic Hamiltonians generate random states, one expects near maximal dynamical generation of entanglement in quantum chaos, to a value that is predicted by the statistics at hand.  This is not to say that regular dynamics (quantum analogs of integrable motion) cannot lead to highly entangled states.  Indeed, such behavior is seen, and has been previously noted in \cite{Kus04}.  Regular dynamics, however, shows oscillatory behavior, including in the generation of entanglement.  Chaotic dynamics, by contrast, lead to quasi-steady state behavior, and typically lead to higher values of time-averaged entanglement than regular motion.  Taken together, these facts imply that the long-time average entanglement in a bipartite system should be a strong signature of classical chaos, closely associated with ergodicity in the two dynamical descriptions. An initial coherent state in the 
chaotic sea has a support over a large number of Hamiltonian eigenstates. This is in contrast to an initial coherent state in the regular islands that have a support over relatively fewer number of eigenstates. This causes the dynamically generated entanglement, when starting from a projected coherent state localized in a chaotic sea,  to saturate to a nearly constant value after a few applications of the Floquet map \cite{Ghose}. The entanglement dynamics observed are thus closely tied to the support of the initial coherent state in the basis of the Hamiltonian eigenstates. The entanglement for a coherent state initially placed in a regular island shows a quasiperiodic behaviour due to a support over a relatively fewer eigenstates. 
 
Since entanglement is a global property of the total system, it is critical to study the chaos in the {\em joint} system dynamics rather than chaos in the separate degrees of freedom.  It is the joint-system dynamics that mixes the two subsystems and leads to random states of the bipartite system with statistically predictable entanglement.  This perspective helps us to understand some previous results, which though predicted analytically and/or numerically, appear to be paradoxical or raise questions about the connection between chaos and the dynamical generation of entanglement.  For example, the results of \cite{Fujisaki} show that in the case of coupled kicked-tops, when the individual subsystems are strongly chaotic but weakly coupled, the rate of generation of entanglement decreases with increased chaoticity.  This can be understood by the fact that in this regime the chaotic mixing {\em between subsystems} is suppressed due to the increasing mismatch between the time scale governing individual top dynamics and the time scale governing coupling between them.  Indeed, Tanaka {\em et al.}\cite{Fujisaki} explained this in terms of rapid ``dynamical averaging" that washes out the correlations that determine the rate of entanglement generation.  

In another example mentioned in the introduction, Demkowicz-Dobrzaski and Kus noted that in a highly-regular regime of kicked tops ($k=0.01$ chaoticity parameter in the standard notation), the rate of entanglement generation was anomalously large.  This result, however, is understood by noting that in addition to weak nonlinearity $k$ for individual tops, the coupling between tops was weak and equal to the nonlinear strength, $\epsilon = 0.01$.  In that case, all time scales in Eq. \ref{joint} are of the same order, and given the lack of integrability of the {\em total Hamiltonian}, $H_{\text{total}}$, we expect the global dynamics to be highly chaotic.  Indeed, the fact the entanglement saturated to the same value seen for chaotic tops without oscillation indicates that we reach the entanglement level of a random state in the joint Hilbert space.  Our central conclusion is thus that chaos in the subsystem is not a strong indicator of the dynamical generation of entanglement, but rather chaos in the joint dynamics of the coupled degrees of freedom.  The amount of entanglement generated in quasi-steady state is statistically predicted by the typical entanglement of a random state in the chaotic sea. 

We have studied the relationship between entanglement and chaos for a system of isotropically coupled tops in which one of the tops receives a periodic kick around a fixed axis.  Here the chaos and entanglement arise from the same coupling mechanism removing any ambiguity between chaos in the subsystem vs. chaos in the joint-system dynamics.  The results reported here give further evidence of the fact that chaotic systems take quantum initial conditions to pseudo-random quantum states, and that the high long-time entanglement average of states undergoing quantum chaotic dynamics is just that of a typical state in the Hilbert space. We see the confirmation of this picture in the excellent agreement between the properties of ensembles of quantum states and the numerical results for the eigenvector statistics and long-time entanglement average for the completely chaotic system. This approach was also found to be highly flexible, applying to subspaces and mixed phase spaces.

\chapter{Quantum Signatures of Chaos in Quantum Tomography} 
\label{QSCQT}
 At a fundamental level chaos represents “unpredictability”, so this seems at odds with the goal of gaining information in order to estimate an unknown quantum state. However, on the flip side, this unpredictability represents the potential “information” to be gained in an estimation process. If everything is predicted and known, we learn nothing new.  The missing information in deterministic chaos is the \textit{initial condition}. A time history of a trajectory at discrete times is an archive of information about the initial conditions given perfect knowledge about the dynamics.   Moreover, if the dynamics is chaotic the rate at which we learn information increases due to the rapid Lyapunov divergence of distinguishable trajectories and we expect unbiased information because of the ergodic mixing of phase space.  That is, if the information is generated by chaotic dynamics, the trajectory is random, and all initial conditions are equally likely until we invert the data and discover the initial state.  
 
  As in the case of deterministic chaos, the missing information in quantum tomography is the \textit{initial condition}. We try to accurately model all of the quantum dynamics occurring in the system, and then use the measurement time history to give us information about the initial quantum state.  The dynamics is “informationally complete” if the time history contains information about an arbitrary initial condition. Our goal is to characterize and quantify the performance of tomography, when the dynamics driving the system is chaotic in the classical limit. We use this to draw comparisons between the role played by regular and chaotic dynamics in the information gain in the tomography procedure.
      
  The remainder of this chapter is organized as follows.  In Sec. \ref{Bck}, we review the protocol for quantum tomography via continuous measurement \cite{sjd05}. We also review the essential features of the ``quantum chaology" associated with the kicked top \cite{Haake}, which is the dynamics employed to drive the system for the purpose of tomography. Section \ref{FQT}, the heart of this chapter, studies the performance of tomography as the system dynamics becomes increasingly chaotic.  We perform numerical calculations of the average fidelity obtained for an ensemble of random states as a function of the chaoticity of the driving dynamics.  We then explain these results in terms of the information gain in tomography. In section \ref{RMTT}, we make predictions for the information gain using random matrix theory in the fully chaotic regime and show a strong agreement between the two. Our results are discussed and summarized in Sec. \ref{DS2}.

\section{Background} 
\label{Bck}
\subsection{Tomography via weak continuous measurement}

The protocol for quantum tomography via continuous measurement \cite{sjd05, rjd11} is as follows: We are given an ensemble of $N$, noninteracting, simultaneously prepared quantum systems in an identical but unknown state given by the density matrix 
$\rho_0$. Our goal is to determine $\rho_0$ by continuously measuring an observable $\mathcal{O}_0$. The ensemble is collectively controlled and probed in a time-dependent manner to obtain an ``informationally complete" continuous measurement record. In order to achieve informational completeness, when viewing in the Heisenberg picture, the set of measured observables should span an operator basis for $\rho_0$. For a Hilbert space of finite dimension d, and fixing the normalization of $\rho_0$, the set of Hermitian operators must form a basis of $su(d)$.
The measurement record is inverted to get an estimate of the unknown state. Laboratory Realization of such a record is intimately tied to \textit{controllability}, i.e., designing the system evolution is such a way as to generate arbitrary unitary maps. While it is desirable to obtain an informationally complete measurement record, we shall see that we can obtain high fidelity in tomography in some cases even when this is not the case ~\cite{mrfd10}.

 In an idealized form, the probe performs a QND measurement that couples uniformly to the collective variable across the ensemble and measures $\sum_{n=j}^ N \mathcal{O}_0^{(j)}$. For a strong QND measurement, quantum backaction will result in substantial entanglement between the particles. For a sufficiently weak measurement, the noise on the detector (shot noise of a laser probe) dominates the quantum fluctuation intrinsic to the measurement outcomes of the state (projection noise). In this case, we can neglect the backaction on the quantum state, and the ensemble remains factorized. Then we can write the measurement record obtained as,
\begin{equation}
\label{m_record}
\mathcal{M} = Tr(\mathcal{O}_0 \rho(t)) + \sigma W(t),
\end{equation}
amplified by the total number of atoms. Here $\sigma W(t)$ is a Gaussian-random variable with zero mean and  variance $\sigma^{2}$, which accounts for the noise on the detector.
   
In order to obtain a measurement record that can be inverted to reconstruct an estimate of the initial state, one must drive the system by a carefully designed dynamical evolution that continually maps new information onto the measured observable. In order to do so, the system is manipulated by external fields. The Hamiltonian of the system, $H(t) = H[\phi_{i}(t)]$, is a functional of a set of time dependent control functions, $\phi_i(t)$, so that the dynamics produces an informationally complete measurement record $\mathcal{M}$. Since our goal is to estimate the initial state from the measurement record and the system dynamics, we will work in the Heisenberg picture.
Rewriting eq. (\ref{m_record}) in the Heisenberg picture, we get 
\begin{equation}
\label{m_record_heisenberg}
\mathcal{M} = Tr(\mathcal{O}(t) \rho_{0}) + \sigma W_i.
\end{equation}
We sample the measurement record at discreet times so that
\begin{equation}
\label{m_record_heisenberg_d}
\mathcal{M}_i = Tr(\mathcal{O}_i \rho_{0}) + \sigma W_i.
\end{equation}
    Thus, the problem of state estimation is reduced to a linear stochastic estimation problem.
    
     The goal is to determine $\rho_0$ given $\{\mathcal{M}_i\}$ for a well chosen $\{\mathcal{O}_i\}$ in the presence of noise $\{W_i\}$.
 We use a simple linear parametrization of the density matrix 
 
\begin{equation}
\label{density}
\rho_0 = \frac{I}{d} + \sum_{\alpha =1}^ {d^2-1} r_{\alpha} E_{\alpha},
\end{equation}
where $d$ is the dimension of the Hilbert space, $r_{\alpha}$ are $d^2-1$ real numbers (the components of a generalized Bloch vector), and $\{E_{\alpha}\}$ is an orthonormal Hermitian basis of traceless operators. We can then write Eq. \ref{m_record_heisenberg_d} as
  \begin{equation}
\label{m_record_heisenberg_d_parameter}
\mathcal{M}_i = \sum_{\alpha =1}^ {d^2-1} r_{\alpha} Tr(\mathcal{O}_i  E_{\alpha}) + \sigma W_i, 
\end{equation}
or, in the matrix form as,
 \begin{equation}
\label{m_record_heisenberg_d_parameter_matrix}
\textbf{M} = \tilde{\mathcal{O}}\textbf{r} + \sigma \textbf{W},
\end{equation}    
which in general is an overdetermined set of linear equations with $d^2-1$ unknowns $\textbf{r} = (r_1, ..., r_{d^2-1})      $. 

The conditional probablity distribution for the random variable $\textbf{M}$, given the state $\textbf{r}$, is the Gaussian distribution
 \begin{equation}
\label{probability_M}
\mathcal{P}(\textbf{M}|r) \propto \text{exp} ( - \frac{1}{2 \sigma^2} (\textbf{M} - \tilde{\mathcal{O}}\textbf{r})^{T}(\textbf{M} - \tilde{\mathcal{O}}\textbf{r}))
\end{equation}    

We can use the fact that the argument of the exponent in Eq. \ref{probability_M} is a quadratic function of $\textbf{r}$ to write the likelihood function (ignoring any priors)
\begin{equation}
\label{probability_ML}
\mathcal{P}(r|\textbf{M}) \propto \text{exp} ( - \frac{1}{2 \sigma^2} (\textbf{r} - \textbf{r}_{ML} )^{T} (\textbf{r} - \textbf{r}_{ML} ))
\end{equation}  
is a Gaussian function over the possible states $\textbf{r}$ centered around the most likely state, $\textbf{r}_{ML}$, with the covariance matrix given by $C = \sigma^{2} (\tilde{\mathcal{O}}^{T}\tilde{\mathcal{O}})^{-1}$. The uncontrained maximum liklihood solution is given by
 \begin{equation}
\label{ML}
\textbf{r}_{ML} = (\tilde{\mathcal{O}}^{T}\tilde{\mathcal{O}})^{-1} \tilde{\mathcal{O}}^{T} \textbf{M}.
\end{equation}  
  The measurement record is “informationally complete” when the covariance matrix has full rank, $d^{2}-1$.
 If the measurement record is incomplete and the covariance matrix is not full rank,  we replace the inverse
 in Eq. \ref{ML} with the Moore-Penrose pseudo inverse \cite{ig03}.  The eigenvectors of $\textbf{C}^{-1}$ represent the orthogonal directions in operator space that we have measured up to the final time, and the eigenvalues  determine the uncertainty, or signal-to-noise associated with those measurement directions.  
 
 When we have an incomplete measurement record, or in the presence of noise, the unconstrained maximum likelihood 
 procedure does not give a density matrix that corresponds to a physical state. The estimated density matrix might have negative eigenvalues. We fix this by finding a valid density matrix that is ``closest" to $\rho_{ML}$, the density matrix obtained by the unconstrained maximum likelihood procedure. We use the inverse of the covariance matrix, $\textbf{C}^{-1}$, as the cost function or metric to measure the distance between $\rho_{ML}$ and $\bar{\rho}$, the new estimate. The metric is defined as 
 \begin{equation}
\label{norm}
\parallel \textbf{r}_{ML} - \bar{\textbf{r}} \parallel ^2= (\textbf{r}_{ML} - \bar{\textbf{r}}) \textbf{C}^{-1} (\textbf{r}_{ML} - \bar{\textbf{r}}).
\end{equation} 
This metric can be justified in the following way. The inverse of the covariance matrix, $\textbf{C}^{-1}$, encodes the 
nature of information obtained by our procedure. A small eigenvalue of $\textbf{C}^{-1}$, implies that we have a low 
signal to noise ratio associated with the measurement of the eigen-operator or more uncertainty in that particular direction in the operator space. By defining the cost function as in Eq. \ref{norm}, we take into account the fact that different directions in operator space have been measured unequally. The metric in Eq. \ref{norm} makes sure that during the numerical optimization while finding a positive matrix, $\bar{\rho}$, the uncertain components can be adjusted with more freedom that the more certain ones thereby maintaining faithfulness with the measurement record.

  Thus we see, the key quantity that quantifies the information gain about the state is the inverse of the covariance matrix, $\textbf{C}^{-1}$ the multivariate Gaussian associated with the measurement record. Using $\textbf{C}^{-1}$,   we will be constructing metrics for information gain in tomography and see how these metrics behave as we change the control dynamics of our experiment from regular to chaotic.
     
\subsection{The kicked top}
How does the presence of chaos in the control dynamics influence our ability to perform tomography? In order to address this question, we chose the ``kicked top" dynamics \cite{Haake} as the paradigm to explore quantum chaos in tomography. The Hamiltonian for the kicked top (after setting $\hbar$ = 1) is given by 

  \begin{equation}
\label{kicked_top}
H(t) = \frac{1}{\tau}p  J_{x} + \frac{1}{2j} \kappa  J_{z}^{2} \sum_{n=-\infty}^\infty \delta (t - n \tau)
\end{equation} 
Here, the operators, $J_{x}$, $J_{y}$ and $J_{z}$ are the angular momentum operators obeying the commutation relation $[J_{i}, J_{j}] = i \epsilon_{ijk} J_{k}$
The first term in the Hamiltonian describes a precession around the \textit{x} axis with an angular frequency $\frac{p}{\tau}$ and the second term described a periodic sequence of kicks separated by time period $\tau$. Each kick is an impulsive rotation about the \textit{z} axis by an amount proportional to $J_z$.
Choosing the external field to act in delta kicks allows us to express the Floquet map (transformation after one period) in a simple form of sequential rotations,
\begin{equation}
\label{Floquet}
U_{\tau} = e^{\frac{-i \lambda J_z^2}{2j}} e^{-i\alpha J_{x}},     
\end{equation}
where $\alpha$ and $\lambda$ are related to $p$ and $\kappa$ in terms of the kicking period.
The classical map can be obtained by considering the Heisenberg evolution of the expectation values of the angular momentum operators in a familiar way \cite{Haake}. 
 
The classical dynamics consists of the motion of a unit spin vector
on the surface of the sphere.  The $z$-component of a spin and the angle $\phi$, denoting its orientation in the $x$-$y$ plane, are canonically conjugate, and thus the spin constitutes one canonical degree of freedom.  The classical dynamical map has the same physical action as described above in the quantum context -- precession of the spin around the \textit{x} axis with an angular frequency $\alpha$ followed by  an impulsive rotation around the \textit{z} axis by an amount proportional to $J_{z}$ and a proportionality constant $\lambda$.
In our analysis, we fix $\alpha = 1.4$ and choose $\lambda$ to be our chaoticity parameter. As we vary $\lambda$ from $0$ to $7$, the dynamics change from highly regular to completely chaotic.  
Since the total magnitude of the spin is a constant of motion, our classical map is two dimensional.  We visualize the phase by plotting the \textit{z} and \textit{y} components of motion after every application of the dynamical map. 

\begin{figure}
\includegraphics[width=5.0in]{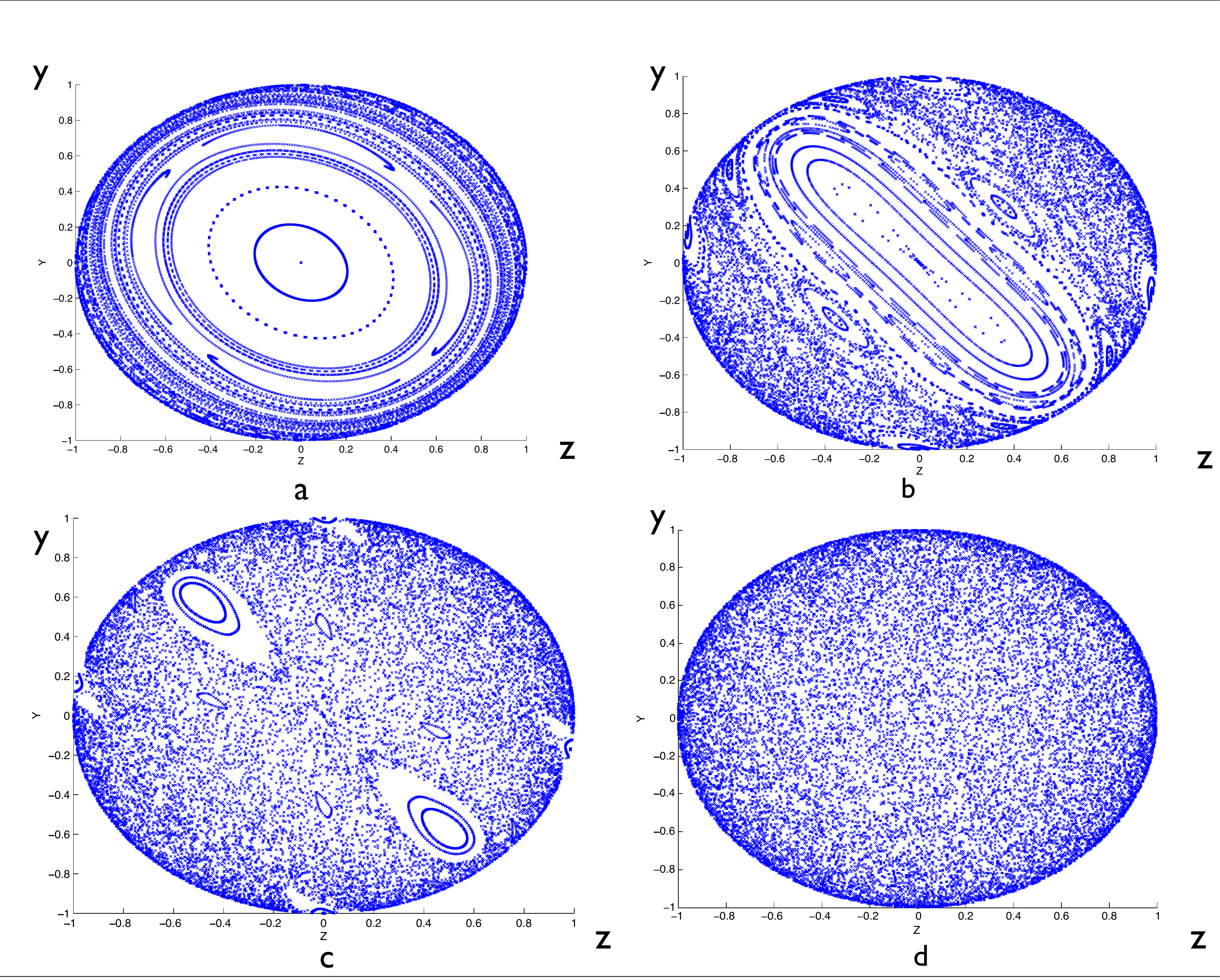}
\caption{Phase space plots for the kicked top in four regimes. (a) Regular phase space: $\alpha=1.4,\lambda=0.5$, (b) Mixed phase space: $\alpha=1.4,\lambda=2.5$, (c) Mostly chaotic: $\alpha=1.4,\lambda=3.0$, (d) Fully chaotic phase space: $\alpha=1.4,\lambda=7.0$. The figures depict trajectories on the southern hemisphere ($\textit{x} < 0$) of the unit sphere where $X = \frac{J_x}{j}$, $Y = \frac{J_y}{j}$ and $Z = \frac{J_z}{j}$ and we take the limit $j \rightarrow \infty $ to get the classical limit as in \cite{Haake}.}
\label{F1}
\end{figure}

In order to explore the connection between the degree of chaos and the information gain in tomography, we will consider the kicked top dynamics in four different regimes, with varying degrees of chaos.
Figure \ref{F1} shows four different regimes of classical dynamics.  With the parameters $\alpha=1.4, \lambda= 0.5$ (Fig. \ref{F1}a), the dynamics are highly regular.  When $\alpha=1.4, \lambda=2.5$ (Fig. \ref{F1}b), we see a mixed space with chaotic and regular regions of comparable size. The parameters $\alpha=1.4, \lambda = 3.0$ (Fig. \ref{F1}c), give a phase space that has mostly chaotic regions and finally, $\alpha = 1.4, \lambda = 7.0$ gives a completely chaotic phase space (Fig. \ref{F1}d).

\clearpage

\subsection{Metrics to quantify information gain}

Our protocol for quantum tomography via continuous measurement of a driven system \cite{sjd05} gives us a window into the complexity of quantum dynamics and its relationship to chaos. The experimental implementation of tomography by continuous measurement provides a nice platform for exploring these ideas in the laboratory \cite{jessen}. Quantum tomography deals with extraction of information about an unknown quantum state through
measurements. In our attempt to study dynamical chaos under this paradigm, we define metrics to quantify this information gain. These metrics allow us to study the ability of our control dynamics to generate a sufficiently high signal-to-noise ratio in different directions of the operator space. As we shall see, these metrics elucidate the connection between the degree of chaos and the fidelities obtained in tomography.
We can quantify the information gain in a number of ways. 

1) If we normalize the eigenvalues of the inverse of the covariance matrix, then as a probability distribution, its Shannon entropy, $E$ is a measure of how evenly we have sampled all the directions in the operator space.
 We reach the maximum entropy when we have measured all directions in the space of matrices equally, $E_{max}  = \text{log} (d^2 -1) $. This is the most unbiased measurement we can implement that will lead to the highest fidelities, on average, for a random state.    

2) The Fisher information sets the Cramer-Rao bound on our uncertainty about the parameters that define the density operator. In general, for a multivariate parameter estimation problem, the Cramer-Rao bound gives
\begin{equation}
\label{Cramer-rao}
\textbf{Cov}_{\theta}(T(X)) \geq \textbf{F}^{-1}
\end{equation}
 where, matrix inequality $A \geq B$ is understood to mean that the matrix $A - B$ is positive semidefinite. $\textbf{T(X)}$  is the unbiased estimator of the multivariate parameter,  $\theta = [{\theta_{1}, \theta_{2}, ..., \theta_{d}}]$ and $\textbf{Cov}_{\theta}(T(X)) $ is the covariance matrix of a set of unbiased estimators for the parameters $\theta$. It quantifies the error in our estimation process. $\textbf{F} $ is the multivariate generalization of the Fisher information \cite{ct}
\begin{equation}
\label{Multi-fisher}
F_{mn} = \textbf{E}(\frac{\partial}{\partial \theta_{m}} \text{log} f(x;\theta) \frac{\partial}{\partial \theta_{n}} \text{log} f(x; \theta)).
\end{equation}
For a multivariate Gaussian, the Fisher information matrix is equal to the inverse of the covariance matrix
 \begin{equation}
\label{fisher-gaussian}
\textbf{F} = \textbf{C}^{-1}
\end{equation}
and thus the Cramer-Rao bound reads
\begin{equation}
\label{CR-gaussian}
\textbf{Cov}_{\theta}(T(X)) \geq \textbf{C}
\end{equation}
We consider the basis in which $F$ and hence $C^{-1}$ is diagonal.
\begin{equation}
\label{Fisher-diagonal}
F^{'} = U F U^{T}.
\end{equation}
Such a transformation is provided by $U$ composed from the eigenvectors of $C$. In this representation, the estimate of the newly transformed parameters fluctuate independently of each other. This suggests the possibility to form a single number quantifying the performance of the tomography scheme as a whole by adding those independent errors
\begin{equation}
\label{error}
\epsilon \geq Tr(C)   
\end{equation}
   Thus, $\frac{1}{Tr(C)}$, which is the collective Fisher information (F. I.), serves as a measure of the amount of information about the parameter $\theta$ that is present in the data. 

 


\section{ Fidelity in quantum tomography as a signature of chaos}
\label{FQT}
How does the presence of chaos in the control dynamics influence our ability to perform tomography? In order to address this question, we chose the ``kicked top" dynamics \cite{Haake} as the paradigm to explore quantum chaos in tomography. While one can show that repeated application of such a single unitary, like the kicked top,  does not generate an informationally complete record \cite{mrfd10}, one can still obtain sufficiently high fidelities taking into account the positivity constraint in the numerical optimization. In \cite{mrfd10}, it was shown that such a repeated application of the ``kicked top" dynamics for the purpose of tomography is equivalent to the repeated application of a single random unitary as far as the final fidelities of reconstruction are concerned. The arguments made in \cite{mrfd10}
were more ``kinematic" in nature as they mostly dealt with the properties of random matrices and how close such a dynamics driven by a single random unitary comes to obtaining an informationally complete measurement record.
In this chapter we explore, even more thoroughly, the role of chaos in information gain in tomography. Our approach is more ``dynamical" in the sense that we are not only interested in the final fidelities obtained, but also how the rate of information gain is influenced by the degree of chaos in the dynamics.

\begin{figure}
\includegraphics[width=\linewidth]{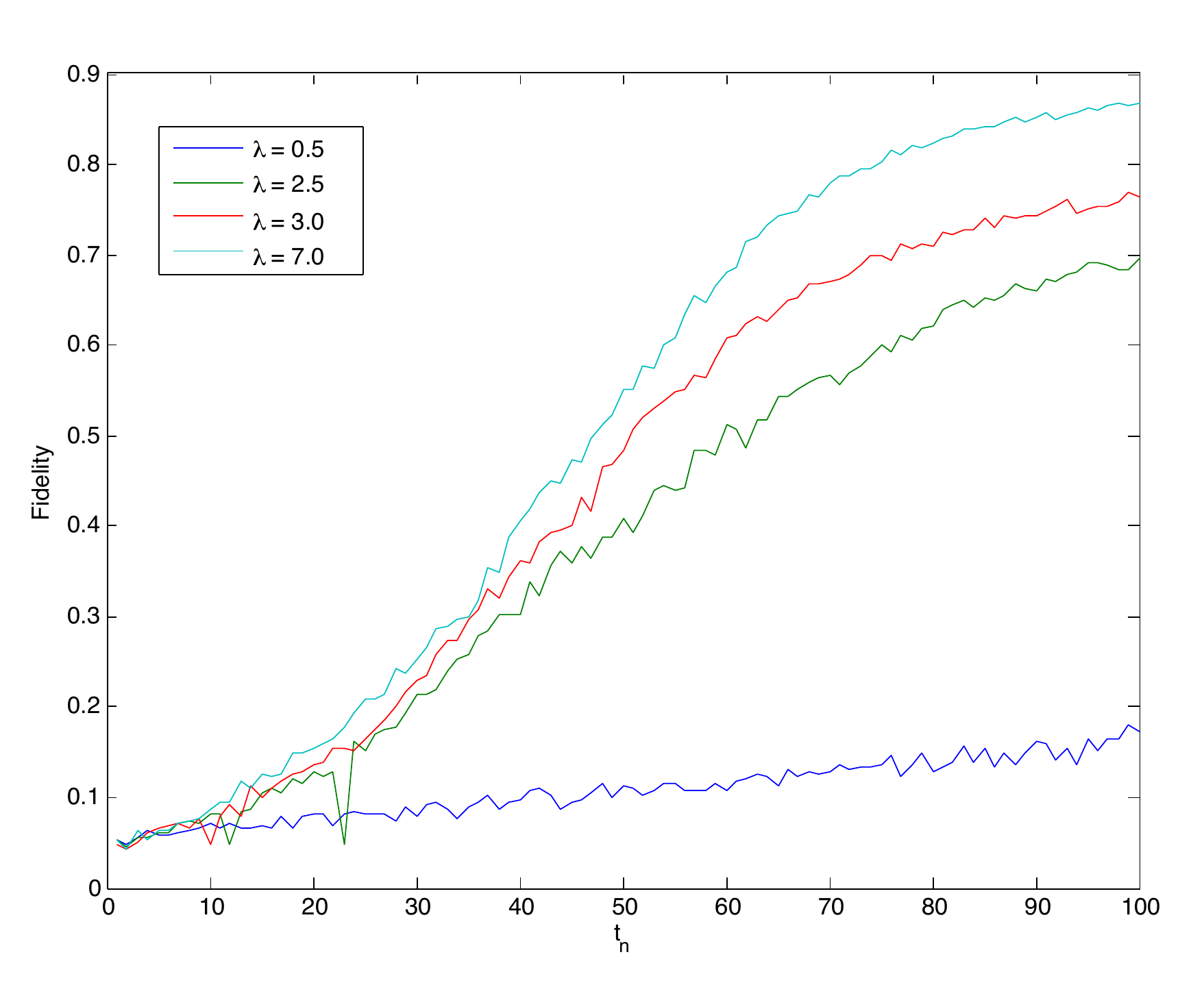}
\caption{Fidelity of reconstruction as a function of the number of applications of the kicked top map. The fidelity calculated as the average fidelity of reconstruction of 100 states picked at random according to the Haar measure. The parameters of the kicked top are as described in the text, with $\alpha$ fixed. We show the fidelity for different choices of the chaoticity parameter. Both the rate of growth and the final value of the fidelities are increased with higher values of $\lambda$.}
\label{F2}
\end{figure}

We are now ready to explore the role of chaos in the performance of tomography. Throughout this section, we consider spin $J=10$, which is sufficiently large that a minimum uncertainty spin coherent state can resolve features in the classical phase space.
Figure \ref{F2} shows the average fidelity of reconstruction of 100 states picked at random according to the Haar measure as a function of the number of applications of the kicked top map, and for different values of the chaoticity parameter. We see that the rate of increase in fidelity increases with the degree of chaos. The final fidelity achieved after a fixed number of kicks is also correlated with the degree of chaos. 

\begin{figure}
\includegraphics[width=\linewidth]{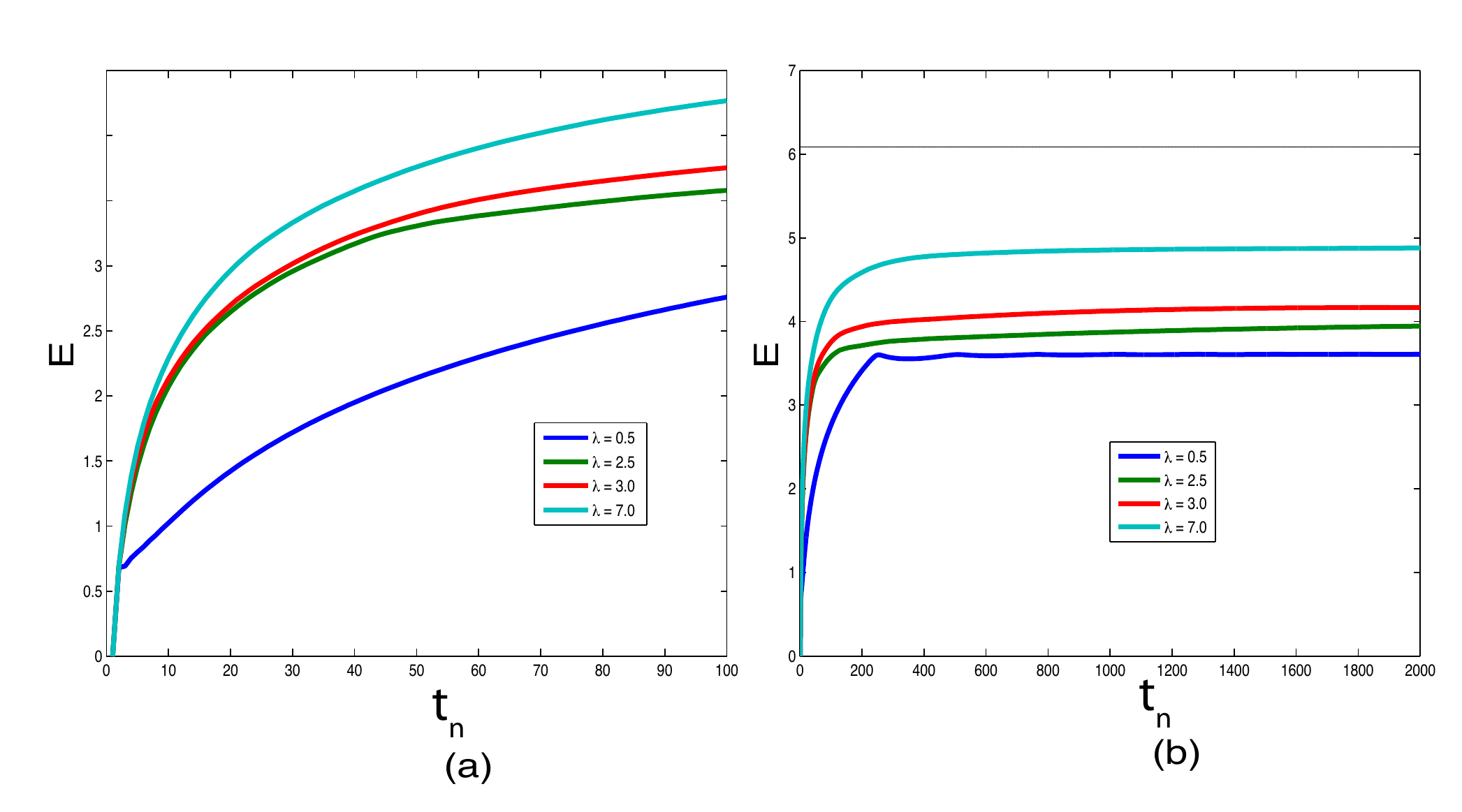}
\caption{The Shannon entropy of the normalized eigenvalues of the inverse of covariance matrix as a function of the number of applications of the kicked top map: (a) Short time behavior (b) long time/asymptotic behavior. The parameters are as described in the text.}
\label{F3}
\end{figure}

\begin{figure}
\includegraphics[width=\linewidth]{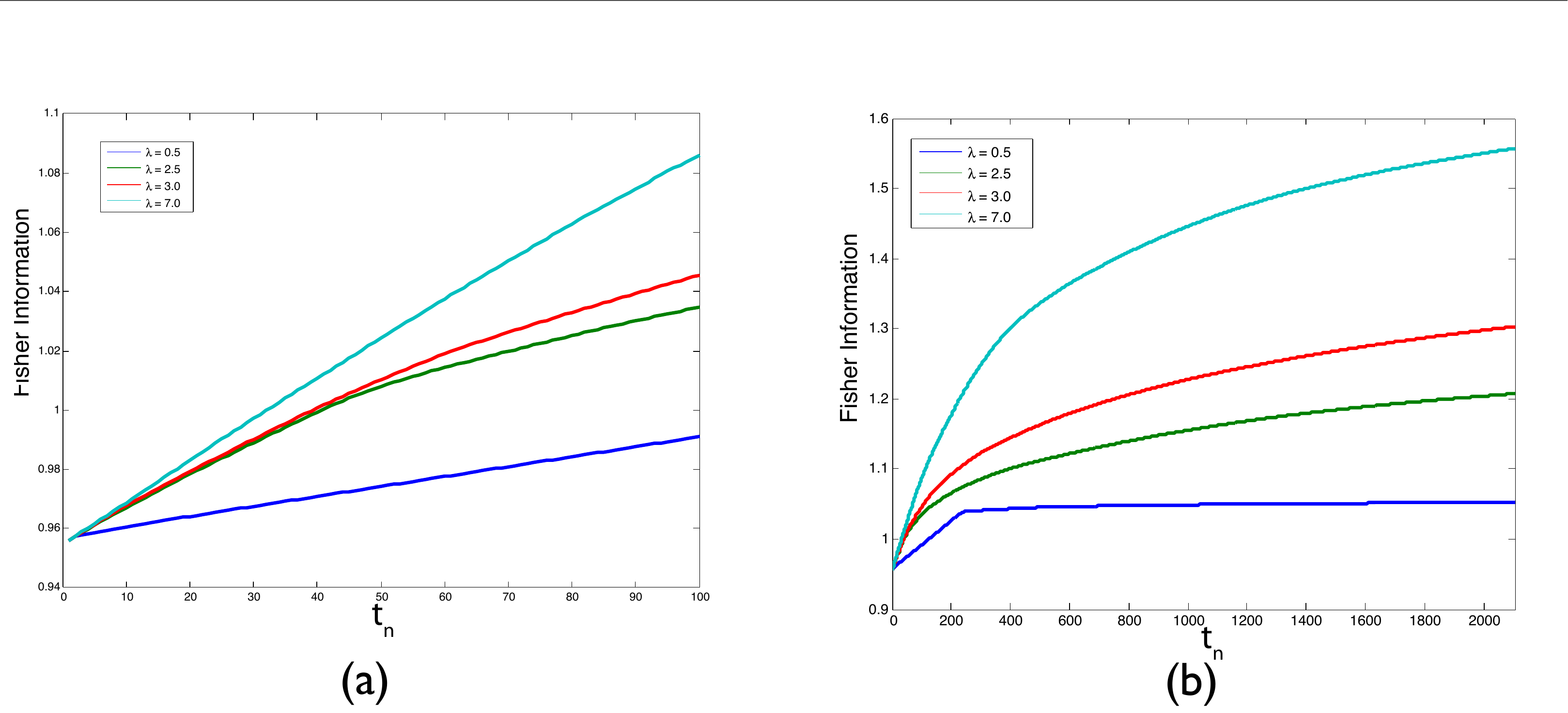}
\caption{The Fisher information of the parameter estimation in tomography as a function of the number of applications of the kicked top map: (a) Short time behavior (b) long time/asymptotic behavior. The parameters are as described in the text.}
\label{F4}
\end{figure}

\clearpage
We can understand the above results by studying the information gain in tomography as a function of the degree of chaos in the control dynamics. Figure \ref{F3} shows the behavior of the entropy $E$ of the covariance matrix, as defined above, as a function of the number of applications of the kicked top map, and for different values of the chaoticity parameter. We see that the rate of entropy increase for short times, Fig. \ref{F3}a, is correlated with the degree of chaos present in the control dynamics. The asymptotic value of the entropy reached also increases with the chaoticity parameter. Chaotic dynamics provides a measurement record with a large signal-to-noise ratio in all the directions in the operator space.
Increase in the chaoticity parameter results in an increasingly unbiased measurement process that will yield 
high fidelities for estimating random quantum states. Figure \ref{F3}a shows the behavior of the entropy at short time scales, while we see asymptotic behavior in Fig. \ref{F3}b.
 The collective Fisher information, $\frac{1}{Tr(C)}$, tells us about the amount of information our measurement record contains about the parameters that define the density matrix. Figure \ref{F4} shows the behavior of the Fisher information as a function of the number of applications of the kicked top map, and for different values of the chaoticity parameter. We see that the rate of increase of the Fisher information is correlated with the degree of chaos present in the control dynamics. As our dynamics become increasingly chaotic, we obtain higher values for the Fisher information at a given time. We expect the Fisher information to be correlated with the average fidelities of estimation for an ensemble of random states.
\clearpage
\section{Signatures of chaos : Information gain in the fully chaotic regime and random matrix theory}
\label{RMTT}
As discussed in the previous chapters, a central result of quantum chaos is the connection with the theory of random matrices \cite{Haake}.  In the limit of large Hilbert space dimensions (small $\hbar$), for parameters such that the classical description of the dynamics shows global chaos, the eigenstates and eigenvalues of the quantum dynamics have the statistical properties of an ensemble of random matrices.  The appropriate ensemble depends on the properties of the quantum system under time-reversal symmetry\cite{Haake}.  We thus seek to determine whether there exists an anti-unitary (time reversal) operator $T$ that has the following action on the Floquet operator,
\begin{equation}
T U_{\tau} T^{-1} = U_{\tau}^{\dagger} = e^{i\alpha J_{x}}e^{\frac{i \lambda J_z^2}{2j}}.
\end{equation}
 Considering the generalized time reversal operation,
\begin{equation}
\label{Eq:Reversal}
T=e^{i \alpha J_x} K,
\end{equation}
where $K$ is the complex conjugation operator.  
It then follows that 
\begin{eqnarray}
T U_{\tau} T^{-1} &=& \left( e^{i \alpha J_x} K \right) \left( e^{\frac{-i \lambda J_z^2}{2j}} e^{-i\alpha J_{x}} \right) \left( K e^{-i \alpha J_x} \right)  \\
&=&  e^{i \alpha J_x} \left( e^{\frac{+i \lambda J_z^2}{2j}} e^{i \alpha J_x}\right) e^{-i \alpha J_x} \nonumber\\
&=& e^{i \alpha J_x} e^{\frac{+i \lambda J_z^2}{2j}} = U_{\tau}^{\dagger}, \nonumber
\end{eqnarray}
so the dynamics is time-reversal invariant.  Moreover, $T^2=1$, so there is no Kramer's degeneracy. Given these facts, for parameters in which the classical dynamics is globally chaotic, we expect the Floquet operator to have the statistical properties of a random matrix chosen from the circular orthogonal ensemble (COE).

\begin{figure}
\includegraphics[width=\linewidth]{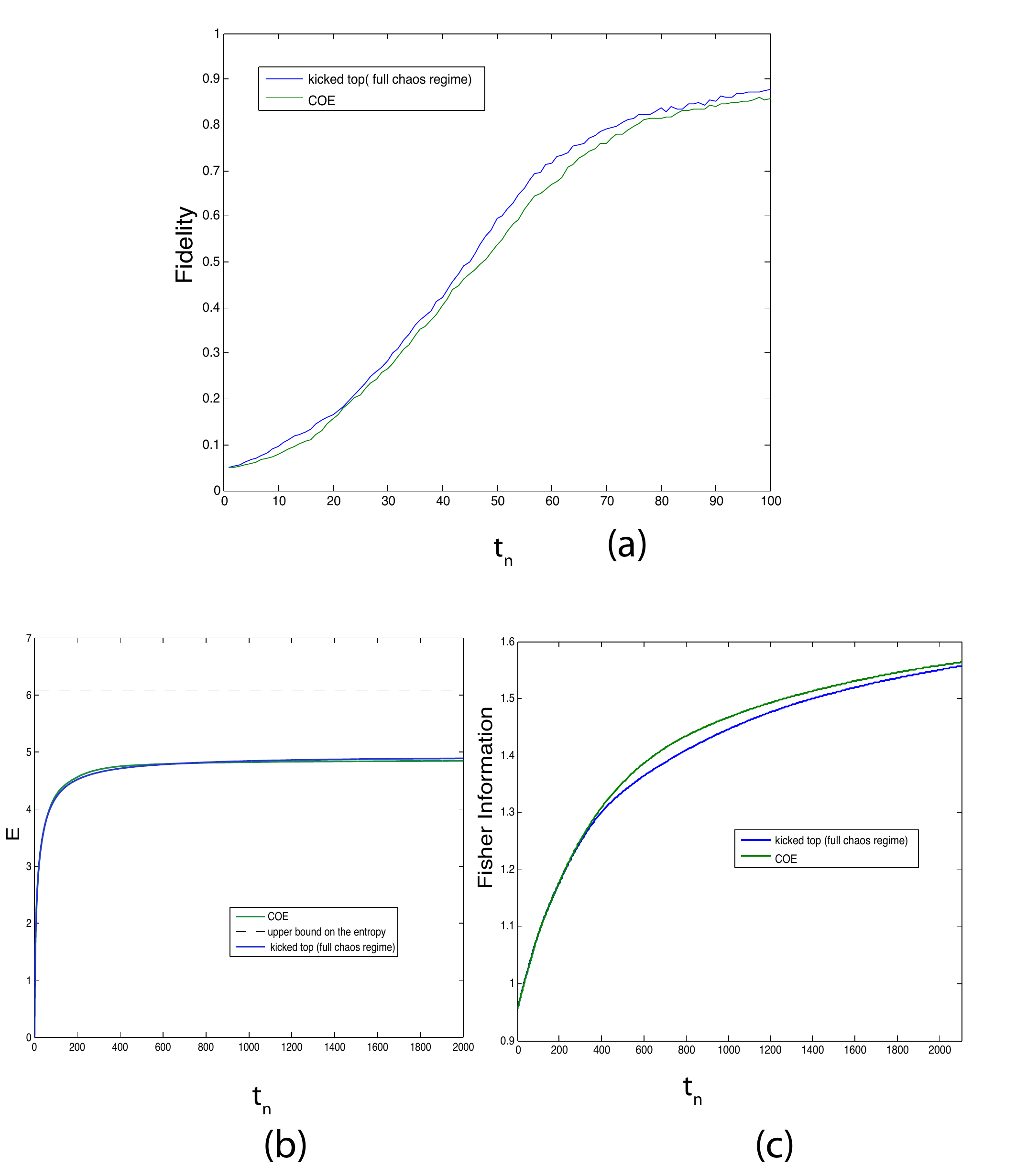} 
\caption{Comparison between the tomography performed by the repeated application of kicked top in the fully chaotic regime (the blue line) and that by a typical random unitary picked from the COE (the green line). (a) The average fidelity of reconstruction of 100 states picked at random according to the Haar measure. (b) The Shannon entropy of the normalized eigenvalues of the inverse of covariance matrix as a function of the number of applications of the map.  The dotted line gives the upper bound on the entropy, $E_{max} = \log (d^2 -1 )$.}
\label{F5}
\end{figure}

When the system is driven by dynamics that are completely chaotic, we expect the information gain and the fidelity to follow the predictions from random matrix theory.
Figure \ref{F5} shows the behavior of the fidelity, Shannon entropy and the Fisher information of the inverse of the covariance matrix as a function of the number of applications of the kicked top map (the blue line) and compares it with the Shannon entropy a typical random unitary picked from the COE (the green line). We see a remarkable agreement between our predictions from random matrix theory and the entropy calculation for the evolution by a completely chaotic map.
 
\begin{figure}[!htp]
\includegraphics[width=\linewidth]{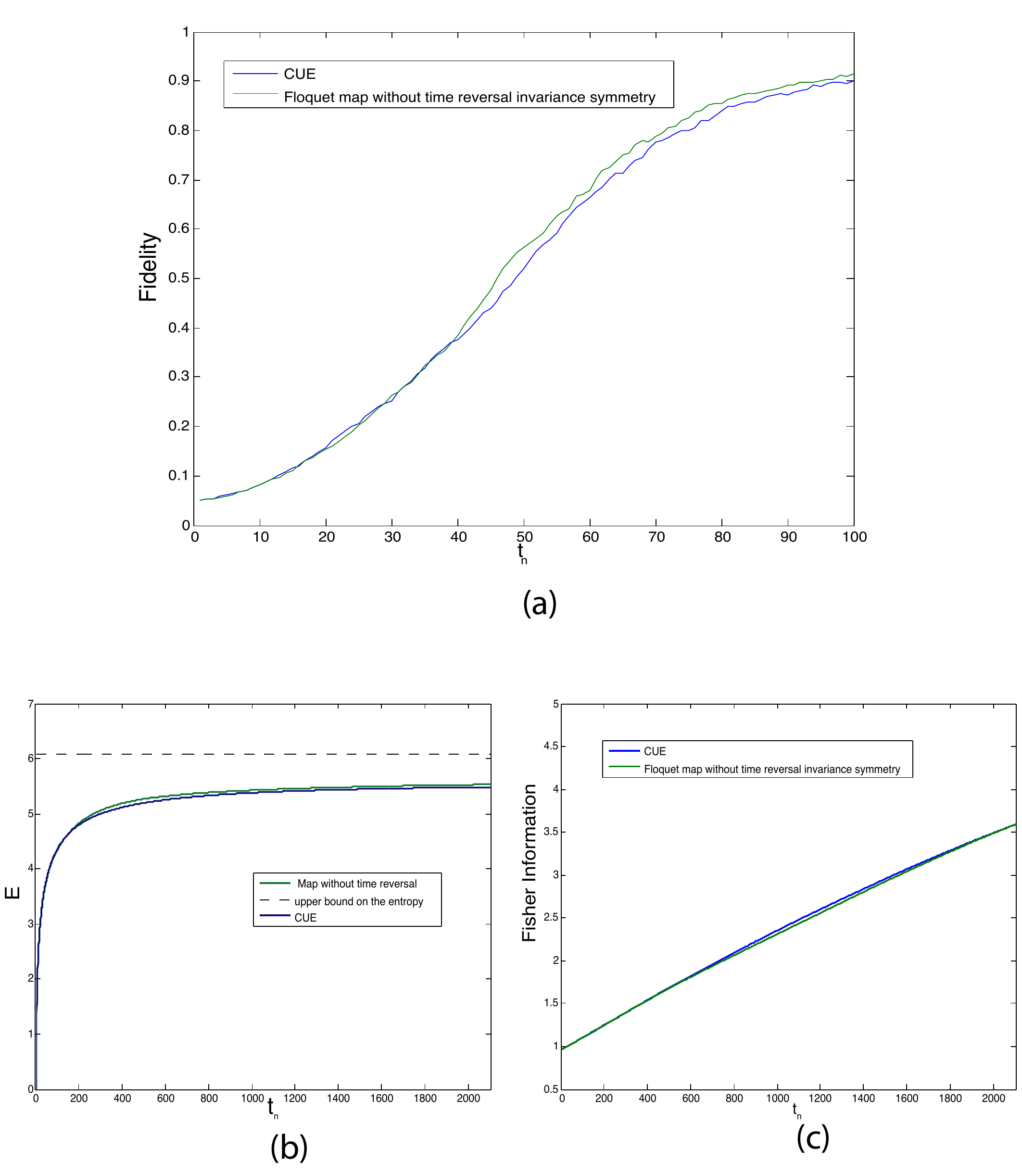}
\caption{Same as Figure 4.5 but for a kicked top without time reversal invariance (Eq. \ref{Floquet_noTR}). In this case the results are well predicted by modeling the dynamics by random matrices sampled from the CUE.}
\label{F7}
\end{figure}

\clearpage
We test our predictions from  the random matrix theory for chaotic maps without a time reversal symmetry.
For example, another type of the ``kicked top" map without time reversal symmetry is given by
\begin{equation}
\label{Floquet_noTR}
U_{\tau}= e^{-i \lambda_1 - J_x^2-i\alpha_1 J_{x}} e^{-i \lambda_2 -  J_y^2-i\alpha_2 J_y} e^{-i \lambda_3 J_z^2-i\alpha_3 J_{z}}.
\end{equation}
In Fig. \ref{F7}, we repeat the above calculations for this map. In this case, the appropriate random matrix ensemble is the CUE. We see an excellent agreement between the behavior of the Shannon entropy as predicted by the random matrix theory and that for the evolution by a completely chaotic map without the time reversal symmetry \cite{kmh88}.

\clearpage

When all the eigenvalues of the inverse of the covariance matrix are equal, we have an upper bound on the entropy,
 $E_{max}  = \text{log} (d^2 -1) $. Figures \ref{F5} - \ref{F7} compare the entropy values achieved 
 by the repeated application of the same unitary (time reversal invariant or otherwise) to $E_{max}$. We see that we fall significantly short of $E_{max}$ by such a procedure. 
 
\begin{figure}[!htp]
\includegraphics[width=\linewidth]{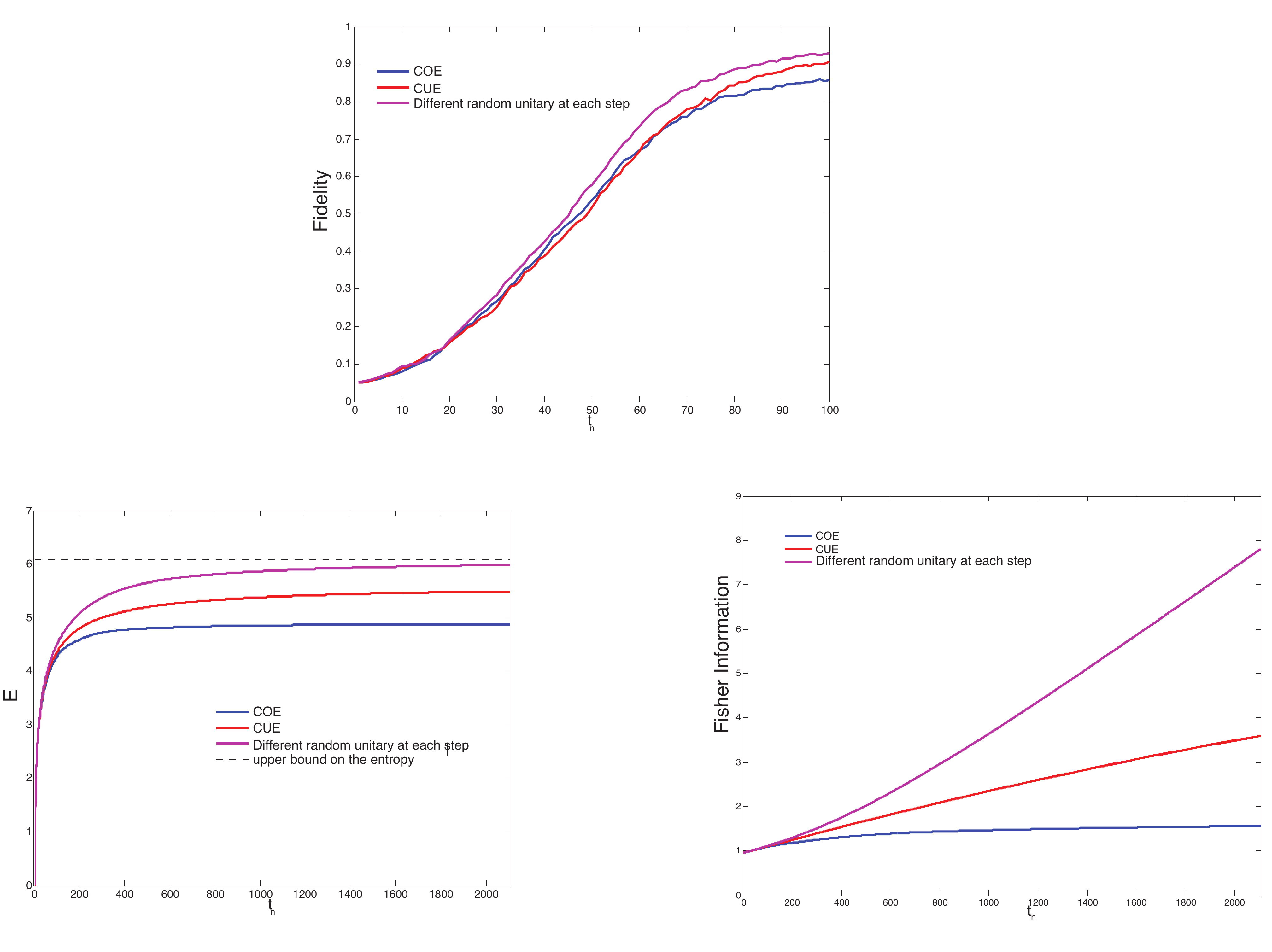}
\caption{Comparison between tomography performed by applying a \textit{different} random unitary at each time step, picked from the unitarily invariant Haar measure (magenta line) and that by a repeated appplication of a random unitary picked from the COE (blue line) and the CUE (red line) . The dotted line gives the upper bound on the entropy, $E_{max} = \log (d^2 -1 )$}
\label{F9}
\end{figure}

 So far, we have considered the application of the same unitary matrix over and over again to obtain the measurement record.  However, this does alone give us an informationally complete measurement record and high fidelities are reached only when we make use of the positivity constraint.  On the other hand, we can consider application of a series of \textit{different} random unitaries, randomly chosen.  In that case we expect to rapidly reach an informationally complete set and thus rapidly gain information about tomography. In Fig. \ref{F9}, we plot the fidelities, Shannon entropy and Fisher information achieved by applying a \textit{different} random unitary, picked from the unitarily invariant Haar measure, and compare it with the results obtained by the repeated application of the same unitary (picked from the COE and CUE). We also see that we reach the upper bound, $E_{max}$, asymptotically, by this method. Indeed, an application of a different random unitary is the most unbiased estimation we can hope to perform.   
                               
\clearpage

\section{Discussions and Summary}

\label{DS2}

 Dynamics sensitive to the initial conditions will reveal more information about them as one observes the system trajectory in the course of time. 
  Classically chaotic dynamics, being sensitive to the initial conditions, generates this unpredictability or information to be gained about the initial coordinates of the trajectory. 
  Similarly, we found that the rate at which one obtains information about an initially unknown quantum state in quantum tomography is correlated with the extent of chaos in the system. This is a new quantum signature of classical chaos that we have found. In fact, our results can be regarded as signatures of chaos in closed quantum systems undergoing unitary evolution. We have been able to quantify the information gain using the Fisher information associated with estimating the parameters of the unknown quantum state. When the system is fully chaotic, the rate of information gain agrees with the predictions of random matrix theory. 

 At its core, our approach is akin to the Kolmogorov-Sinai (KS) entropy measure of chaos \cite{s59}.  Incomplete information about the initial condition leads to unpredictability of a time history.  In the presence of classical chaos, in order to predict which coarse-grained cell in phase space a trajectory will land at a later time, we require an exponentially increasing fine-grained knowledge of the initial condition.  The KS entropy is the rate of increase, and is related to the positive Lyapunov exponents of the system. Is there a meaningful quantum definition of KS entropy?  In order to predict the measurement record with a fixed uncertainty we learn more and more about the initial condition.  Is the rate at which we obtain this information exponentially fast when the system is quantum chaotic?  Does this converge to the classical Lyapunov exponents in the  limit of large action (small $\hbar$)? There are many important subtleties in these questions.  As we gain more and more information, eventually quantum backaction becomes important in the measurement history.  The number of copies we have and the shot noise on the probe limits the ultimate resolution with which we can deduce the quantum state.  Unlike classical dynamics, we can never consider infinite resolution, even in principle.  The quantum resolution is limited by the size of $\hbar$. As the dimension of the Hilbert space increases, and hence the effective $\hbar$ decreases, we expect 
 to see an even sharper difference in the information gain as a function of chaoticity. In the limit when $d$, the dimension of the Hilbert space, becomes infinity, we expect the rate of information gain to be intimately related to the classical Lyapunov exponents.
 How does all this translates into a quantum definition of KS entropy is an important subject of further investigation. 
 
 Still further, in principle we never have perfect knowledge of the dynamics.  Another hallmark of chaos is hypersensitivity to perturbations\cite{Scott/Caves}.  How does this fundamentally limit our ability to perform quantum state reconstruction when the system is sufficiently complex, and the equivalent dynamics is chaotic. Though quantum systems show no sensitivity to initial conditions, due to unitarity, they do show a sensitivity to parameters in the Hamiltonian~\cite{per00}. This poses interesting questions for quantum tomography and, more interestingly, for quantum simulations. Under what conditions are the system dynamics sensitive to perturbations and how does this effect our ability to perform quantum tomography? Under what conditions does the underlying quantum chaos affect our ability to accomplish quantum simulations in general? We hope to answer these questions in our future work.

\chapter{Interpreting Quantum Discord through Quantum State Merging}

\section{Introduction}

The characterization of the resources behind the enhancements and speedups provided by quantum mechanics over best known classical procedures is one of the most fundamental questions in quantum information science. This has generally been expressed in terms of quantum entanglement~\cite{hhhh07}. There are, however, quantum processes which provide an exponential advantage in the presence of little or no entanglement~\cite{kl98, dqc1}. In the realm of mixed-state quantum computation, quantum discord~\cite{oz02,hv01} has been proposed as a resource~\cite{dsc08} and there has been continued progress~\cite{adv}. It has also been shown to be a resource in quantum state discrimination~\cite{rra11, lfwf11} and quantum locking~\cite{lock}. 

Quantum discord was originally suggested as a measure of quantumness of correlations~\cite{oz02}, and has since been studied in variety of systems and settings~\cite{belld,disc,ad10}. The initial motivation for its definition arose in the context of pointer states and environment-induced decoherence~\cite{darwinism}. It has since been related to quantum phase transitions~\cite{s09} and the performance to quantum and classical Maxwell's demons~\cite{demon}. Though satisfactory from a physical perspective, a benchmark for accepting some quantity as a resource in quantum information science is that it appears as the solution to appropriate asymptotic information processing tasks. It is this \textit{information theoretic} interpretation that has been lacking for quantum discord, and which we now provide in this dissertation. This also addresses a more fundamental dichotomy in quantum information science, where resources and their manipulations can have both a thermodynamic and an information theoretic interpretations independently, which are not intuitively or mathematically reconciled.

\section{Quantum State Merging And Discord}
Consider a party Bob having access to some incomplete information $Y,$ and another party Alice having the missing part $X.$  We can think of $X$ and $Y$ as random variables. If Bob wishes to learn $X$ fully, how much information must Alice send to him? Evidently, she can send $H(X)$ bits to satisfy Bob. Here $H(.)$ is the Shannon entropy associates with the random variable under consideration.
 However, Slepian and Wolf showed that she can do better, by merely sending $H(X|Y) = H(X,Y)-H(Y),$ the conditional information~\cite{thomascover}.  Since $H(X|Y) \leq H(X)$, Alice can take advantage of correlations between $X$ and $Y$ to reduce the communication cost needed to accomplish the given task. 

The quantum state merging protocol is the extension of the classical Slepian-Wolf protocol into the quantum domain where Alice and Bob share n copies of a quantum state, expressed in the form of a tensor power as $\rho_{AB}^{\otimes n}$, with each party having the marginal density operators $\rho_{A}^{\otimes n}$ and $\rho_{B}^{\otimes n}$ respectively. Let $\ket{\Psi_{ABC}}$ be a purification of $\rho_{AB}.$ Assume, without loss of generality, that Bob holds $C.$ The quantum state merging protocol quantifies the minimum amount of quantum information that Alice must send to Bob so that he ends up with a state arbitrarily close to $\ket{\Psi}_{B'BC}^{\otimes n},$ $B'$ being a register at Bob's end to store the qubits received from Alice. It was shown that in the limit of $n\rightarrow \infty$, and asymptotically vanishing errors, the answer is given by the quantum conditional entropy~\cite{how05,how07}, $S(A|B) = S(A,B) - S(B)$. When $S(A|B)$ is negative, Bob obtains the full state with just local operations and classical communication, and distills $-S(A|B)$ ebits with Alice, that can be used to transfer additional quantum information in the future.

It is to be noted that quantum discord can be nonzero even for
separable quantum states. The state merging protocol does in fact work in such a scenario as was shown
by Horodecki et. al \cite{sm}. In such a case, the entanglement needed for state merging is provided by sharing
the right amount of EPR pairs between Alice and Bob. The cost of sharing these EPR
pairs is properly taken into account when one computes the net cost of quantum
communication needed to accomplish state merging.
 We quote from Horodecki et. al. \cite{sm}:\\
Theorem 2 (Quantum State Merging). For a state $\rho_{AB}$ shared by Alice and
Bob, the entanglement cost of merging is equal to the quantum conditional
entropy $S(A|B) = S(AB) - S(B)$ in the following sense. When the $S(A|B)$ is
positive, then merging is possible if and only if $R>S(A|B)$ ebits per input copy
are provided. When $S(A|B)$ is negative, then merging is possible by local
operations and classical communication, and moreover $R<-S( A|B)$ maximally
entangled states are obtained per input copy.

Although it is true that one needs entanglement for state merging (and thus one needs to
share extra EPR pairs) in the case when we have a separable state having nonzero discord, we demonstrate that even in such a scenario, discord is equal to
the mark up in the amount of entanglement needed for state merging after one of the parties
(B) makes a measurement.

In both scenarios (independent of whether $S(A|B)$ is positive or negative), the
cost of state merging is more expensive when one destroys the quantum part of correlations
(which are quantified by discord), by making measurements on one subsystem B.
Thus we see, state merging and hence our interpretation is valid in general including the
cases when we have a separable state with nonzero discord.

%
%
\subsection{Quantum operations increase the cost of state merging}
%

An intuitive argument for our interpretation of quantum discord begins with the strong subadditivity theorem \cite{nc00}, which states that~\cite{how07}
\be
\label{eq:ssa}
S(A|B,C) \leq S(A|B).
\ee
From the point of view of the state merging protocol, the above has a very clear interpretation: having more prior information makes state merging cheaper. Or in other words, throwing away information will make state merging more expensive. Thus, if Bob discards system C, it will increase the cost of quantum communication needed by Alice in order to merge her state with Bob. Our intent here shall be to relate this increase in the cost of state merging to quantum discord between A and B.


To that end, we expand the size of the Hilbert space so that an arbitrary measurement (with forgetting) can be modeled by coupling to the auxiliary subsystem and then discarding it. This permits us to apply strong subadditivity to the problem in question. We assume $C$ is initially in a pure state $\ket{{0}}$, and a unitary interaction $U$ between $B$ and $C$. Letting primes denote the state of the system after $U$ has acted we have $S(A,B) = S(A,BC)$ as $C$ starts out in a product state with $AB$. We also have $I(A : BC) = I(A' : B'C')$. As discarding quantum systems cannot increase the mutual information, we get $I(A' : B') \leq I(A' : B'C')$. Now consider the state merging protocol between A and B in the presence of $C$. We have $S(A|B) = S(A) - I(A : B) = S(A) - I(A : BC) = S(A|BC).$ After the application of the unitary $U$, but before discarding the subsystem $C$, the cost of merging is still given by $S(A'|B'C') = S(A|B)$. This implies that one can always view the cost of merging the state of system $A$ with $B$,  as the cost of merging $A$ with the system $BC$, where $C$ is some ancilla (initially in a pure state) with which $B$ interacts coherently through a unitary $U$. Such a scheme does not change the cost of state merging, as shown, but helps us in counting resources. Discarding system $C$ yields
\be
\label{eq:Idiff}
I(A' : B') \leq I(A':B'C') = I(A:BC) = I(A : B),
\ee
or alternatively,
\be
\label{eq:Sdiff}
S(A'|B') \geq S(A'|B'C') = S(A|B).
\ee
Computing the mark up in the price in the state merging on discarding information gives $D = I(A:B) - I(A':B').$ This quantity $D$ is equal to quantum discord when our quantum operations are quantum measurements maximizing $I(A' : B')$. Thus, discord is the minimum possible increase in the cost of quantum communication in performing state merging, with a measurement on the party receiving the final state. This also addresses the asymmetry that is inherent is quantum discord. This is exhibited operationally in our interpretation since the state merging protocol is not invariant under exchanging the parties.


We now show that the minimum of $D$ over all possible measurements is the quantum discord. The state $\rho_{AB},$ under measurement of subsystem $B$, changes to $\rho_{AB} ' = \sum_{j} p_{j} \rho_{A|j} \otimes \pi_j,$ where $ \{\pi_j\} $ are orthogonal projectors resulting from a Neumark extension of the POVM elements \cite{neumark40}. The unconditioned post-measurement states of $A$ and $B$ are
$$
\rho_{A}' =\sum_{j} p_{j} \rho_{A|j} = \rho_{A},~~~\rho_{B}' = \sum_{j} p_{j} \pi_j.
$$ Computing the value of $I(A':B')$, we get
\ben
I(A':B') &=&  S(A') +S(B') - S(A',B'),\nonumber  \\
        &=& S(A') + H(p) -\big\{H(p) + \sum_{j} p_{j} S(\rho_{A|j})\big\}, \nonumber\\
            &=& S(A) - \sum_{j} p_{j} S(\rho_{A|j}).
\een
After maximization, it reduces to $\mathcal{J}(\rho_{AB})$, as in Eq. (\ref{eq:J}). The reduction to rank 1 POVMs follows as stated earlier.

We can also rewrite the expression for $D$ using Eq. (\ref{eq:Sdiff}) instead of Eq. (\ref{eq:Idiff}) as the increase of the conditional entropy $D =  S(A'|B') - S(A|B).$ The above expression makes our interpretation even more transparent. Quantum measurements on $B$ destroys quantum correlations between $A$ and $B$. This increases the average cost of quantum communication needed by $A$, to merge her post measurement state with $B$.  Since, $S(A'|B') = \sum_{j} p_{j} S( \rho_{A|j}) \geq S(A|B)$, there is always a mark up in the cost of state merging. 

\section{An Example}

 Consider the separable state $\rho_{AB}=\left(\proja{0}\otimes\projb{0} + \proja{1}\otimes\projb{+}\right)/2,$ where $\ket{+}=(\ket{0}+\ket{1})/\sqrt{2}.$ The cost incurred by $A$ to merge her state with $B$ is $S(A|B)=0.399124$ ebits, while that after measuring B using the projectors $(I \pm \frac{\sigma_x-\sigma_z} {\sqrt{2}})/2$ is  $S(A'|B')=0.600876$ ebits. The markup in the cost of state merging is $S(A'|B')-S(A|B) = 0.201752$ ebits, which is exactly the quantum discord of the state $\rho_{AB}.$ Hence, any information lost through the measurement results in making the quantum state merging more expensive by exactly the same amount.

\subsection{Properties of Discord}

We can now use our quantum state merging perspective to derive the various properties of discord. Since measurements on system $B$ will always result in either discarding some information or at best preserving the original correlations, we will always get a price hike in state merging or at best we can hope to just break even. Hence, discord, which is the mark up, will always be greater than or equal to zero~\cite{oz02,datta10}.

Quantum discord of a state is zero if and only if the density matrix is of the form $\rho_{AB} = \sum_{i} p_{i} \rho_{A|i} \otimes \proj{\lambda_i},$ in the basis which diagonalizes $\rho_{B}$. Measuring the projectors $\proj{\lambda_i}$ and discarding the measurement results on such a state yields $\rho^{M}_{AB} = \sum_{i} P_{j} \rho_{AB} P_{j} = \rho_{AB}.$ Thus, we have a measurement which causes no loss of information, and retains all the correlations between $A$ and $B$. Hence there is no mark up in the cost of merging a zero discord state.

The converse can be seen through the application of strong subadditivity in Eq.~(\ref{eq:ssa}). The equality of mutual information, $I(A:B)$, of the initial state and that of the state after quantum operations on $B$, $I(A':B') $ coincides with the equality condition for strong subadditivity. But this is exactly the condition for quantum discord~\cite{datta10} to vanish. Thus a zero mark up in the cost of state merging implies zero discord.

An upper bound on discord is decided by an upper bound on the mark-up we can obtain. Since Bob cannot lose more information than there is at his disposal, the entropy of the state at Bob's end, $S(B),$ is an upper bound on quantum discord.



Finally, for pure states, quantum discord reduces to entanglement, and $S(A|B)$ = $S(A) - I(A:B)  = -S(A) \leq 0$. From our perspective, measurement destroys all the correlations present between $A$ and $B$. Though the post-measurement state merging of the state of $A$ with that of $B$ occurs at zero cost, they lose the $-S(A|B)$ potential Bell pairs that could have been put to some use. This provides us a novel way of measuring entanglement, as the markup in merging a pure state, when $B$ is measured.

In conclusion, this study places quantum discord squarely in the midst of quantum informational concepts and opens up the way for its manipulation as a resource in quantum information processing. We also hope that our work will serve as a stepping stone for a more comprehensive and unified understanding of quantum physics, thermodynamics and information theory.


\chapter{Quantum discord as a resource in quantum protocols}
   
Quantum correlations lie at the heart of the mystery of quantum mechanics and also serve as a resource for the potential benefits quantum information science might provide. However, recent developments suggest that entanglement may not fully capture the complete quantum character of a system~\cite{oz02}. Quantum discord aims to fill this gap and quantify essentially all the quantum correlations in a quantum state.

In the previous chapter, we addressed the important question: Does quantum discord have a definite physical role in information processing?  We provided an affirmative answer by linking quantum discord to the yield of physical tasks in quantum information theory. At this point, a natural question arises: Is there a role of quantum discord in quantum information theory as a whole beyond the state merging protocol? This chapter answers precisely this question.
   
We show quantum discord to be a resource in quantum information processing. This is accomplished by proving a relationship between quantum discord and the yield of the quantum protocols. Our results are derived by studying the fully quantum Slepian-Wolf protocol\cite{dhw04} -- a unification of essentially all bipartite, unidirectional and memoryless quantum communication protocols -- in the presence of environmental decoherence. As examples, we elucidate the significance of quantum discord in quantum teleportation, superdense coding, and entanglement distillation. Finally,  we provide the first quantitative relation between quantum discord and the query complexity of quantum computations.

The key insight to our findings is that quantum measurements and environmental decoherence disturb a quantum system in a way that is unique to quantum theory.
Quantum correlations in a bipartite system are precisely the ones that are destroyed by such disturbances, and therefore certain quantum communication protocols become overloaded by an amount exactly equal to quantum discord.
More specifically, we showed that discord is the markup in the cost of quantum communication in the process of quantum state merging~\cite{sm}, if the system undergoes measurement and/or decoherence. At this point, a natural question arises: Is there a role of quantum discord in quantum information theory as a whole beyond the lossy state merging protocol? We answered this question with a yes. We first observed that quantum state merging protocol is a derivative of the more general Fully Quantum Slepian Wolf (FQSW) protocol~\cite{sm,adhw09} and the closely related ``mother" protocol. The mother protocol is essentially a unification of essentially all unidirectional, bipartite and memoryless quantum communication protocols like quantum teleportation, superdense coding and entanglement distillation.
A link between discord and state merging does indeed suggest a link between discord and the mother protocol and hence a possible role of discord in essentially all bipartite, unidirectional and memoryless quantum communication protocols.

This is made possible by comparing the performance of the fully quantum Slepian-Wolf (FQSW) protocol \cite{dhw04} in the presence and absence of decoherence and linking it to the discord of the state involved. While decoherence is expected to diminish the gain provided by a quantum protocol, we provide, for the first time, a general lower bound on the amount of this deterioration. Our bound is only dependent on the state involved, independent of the details of the protocol as well as the nature of the decoherence. Within the resource framework of quantum Shannon theory~\cite{dhw04,dhw08}, we couple the performance of the FQSW protocol to the most general environmental decoherence to show that quantum discord of the state participating in the protocol is the lower bound to the depreciation of the protocol's performance. The FQSW protocol - a quantum communication-assisted entanglement distillation protocol -  is the parent protocol from which all information processing protocols emanate~\cite{adhw09}. The generality of the FQSW protocol allows us to establish the role of quantum discord in the performance of noisy versions of quantum teleportation, super-dense coding, and distillation. Finally, interpreting quantum computation as the classical capacity of a quantum communication channel~\cite{brv00}, we also elucidate the role of quantum discord in quantum computational processes.

Our result connecting quantum discord to quantum protocols is subtle as well as intriguing. Although it is known that entanglement is often necessary for the success of quantum protocols, and that the presence of decoherence affects its performance, we have now provided a quantitative result of the amount of such a depreciation. We have shown that the amount by which a protocol suffers in the presence of decoherence is an inherent property of the quantum states involved. It suggests that the choice of the best state for any noisy quantum protocol must be a tradeoff between the entanglement and discord of the state involved.  Given the non-monotonic relation between quantum discord and entanglement in quantum states~\cite{mono}, choosing the optimal state for a quantum task is a non-trivial one, though for which our work identifies the proper certificate.  
   

\section{The FQSW protocol}
\subsection{The mother protocol and the quantum information's family tree}

So far, we have been discussing quantum communication tasks like teleportation, super-dense coding, quantum state merging as independent entities. Is there a unified way of looking at various quantum communication tasks? Do various quantum communication protocols described above have a common origin. Interestingly, the answer is yes. It was shown in \cite{adhw09} that essentially all unidirectional, bipartite and memoryless quantum communication protocols are actually siblings originating from one ``mother". The mother protocol can be seen to provide a hierarchical structure to the family of quantum protocols. We will describe the improved version of the mother protocol, which is the 
fully quantum Slepian-Wolf (FQSW) protocol. Throughout this chapter, we will use the names ``mother" and FQSW to describe essentially this same protocol. We have described the FQSW protocol in detail in Chapter 2 (section 2.4). In the next section, we examine the FQSW protocol in the presence of environmental decoherence.

\section{Quantum discord as a measure of coherence in the FQSW protocol}
\label{sec:Main}
This section contains our main result. The FQSW is essentially a non-dissipative protocol in that no information is leaked to the environment in each step of the protocol, but any practical implementation of a quantum information protocol will be affected by loss and noise. In particular, we will consider loss of information and coherence at Bob's end. This can be studied by considering a unitary coupling between Bob's system $B$ and an ancillary environment system, say $C,$ and then tracing $C$ out. Physically, such a quantum operation will emulate environmental decoherence.


We begin by expanding the size of the Hilbert space so that an arbitrary measurement (or any other quantum operation) can be modeled by coupling to the auxiliary subsystem and then discarding it. We assume the ancilla $C$ to initially be in a pure state $\ket{{0}}$, and a unitary interaction $U$ between $B$ and $C$. Letting primes denote the state of the system after $U$ has acted we have $H(A,B) = H(A',B'C')$ as $C$ starts out in a product state with $AB$. We also have $I(A : BC) = I(A' : B'C')$. As discarding quantum systems cannot increase the mutual information, we get $I(A' : B') \leq I(A' : B'C')$. 

Now consider the FQSW protocol between $A$ and $B$ in the presence of $C$. We can always view the yield of the FQSW protocol on the system $AB$ to be the same as that of performing the protocol between systems $A$ and $BC$, where $C$ is some ancilla (initially in a pure state) with which $B$ interacts coherently through a unitary $U$. Such an operation does not change the cost or yield of the FQSW protocol, as shown, but helps us in counting resources. Discarding system $C$ yields
\be
\label{eq:Idiff1}
I(A' : B') \leq I(A':B'C') = I(A:BC) = I(A : B).
\ee
Now consider a protocol which we call as $FQSWD_{B}$  (FQSW after decoherence), where the subscript refers to the decoherence at $B$. The resource inequality for $FQSWD_{B}$ is
 \begin{equation}
\avg{\mathcal{U}^{S' \rightarrow A'B'} : \psi^{S'}}  + \frac{1}{2} I (A':R') [q \rightarrow q] \geq \frac{1}{2} I (A':B')[qq] + \avg{\mathbb{I}^{S' \rightarrow \hat{B}} : \psi^{S'}}.
\end{equation}
It is important to explain the terminology used in the above inequality. When we have a noisy resource like a mixed state, $ \psi^{S'}$, or a noisy channel, it is inserted between a ``$\langle   \rangle$". Thus a mixed state is represented by $\langle \psi^{S'} \rangle $, and a noisy channel by $\langle \mathcal{N}\rangle$. A channel is a relative resource $\langle \mathcal{U}^{S' \rightarrow AB} : \psi^{S'}\rangle$ meaning that the protocol only works provided the input to the channel is the state $\psi^{S'}$. On the LHS, $\mathcal{U}$ takes the state $ \psi^{S'}$ and distributes it to Alice and Bob. On the RHS, the symbol ``$\mathbb{I}$" is an identity channel taking the state $\psi^{S'}$ to Bob alone. The state $\psi^{S'}$ on the left-hand side of the inequality is distributed to Alice and Bob, while on the right-hand side, that state is given to Bob alone. The primed letters, $A'$, $B'$ etc., indicate that the protocol is taking place in the presence of decoherence at Bob's end.

As in the fully coherent version, Alice is able to transfer her entanglement with the reference system $R'$, and is able to distill $\frac{1}{2} I (A':B')$ EPR pairs ($[qq]$) with Bob. The net quantum gain for the fully coherent protocol is $G=\frac{1}{2} I (A:B) -\frac{1}{2} I (A:R) = -H(A|B)$ ebits. This is the difference between the yield obtained and the cost of quantum communication incurred. Likewise, the net gain for the protocol suffering decoherence at $B$ is $G_{D}=\frac{1}{2}I(A':B') - \frac{1}{2}I(A':R') = -H(A'|B')$. Therefore, the net advantage of the coherent protocol over the decohered one is given by  $D = G-G_{D} = H(A'|B') - H(A|B)$ ebits. Evidently, this quantum advantage depends on the exact nature of the environment and the system's interaction with it via $U.$ Employing the original definition of quantum discord due to Zurek~\cite{z00}( Zurek's original definition of discord did not consider optimizing over all measurements), $D$ quantifies the loss in the yield of a quantum protocol due to environmental decoherence. Our results therefore provide a standard way of quantifying, in entropy units, the damage to the performance of quantum process and protocols in the presence of any decoherence process in any experimental scenario.

The strength of our result, however, comes from the next step of minimizing $D$ over all environmental operations performing measurements. Using the measurement model of quantum operations~\cite{nc00}, the state $\rho_{AB}$ under measurement of subsystem $B$, changes to $\rho_{AB}' = \sum_{j} p_{j} \rho_{A|j} \otimes \pi_j,$ where $ \{\pi_j\} $ are orthogonal projectors resulting from a Neumark extension of the POVM elements \cite{neumark40}. The unconditioned post-measurement states of $A$ and $B$ are
\be  
\label{eq:cond}
\rho_{A}' =\sum_{j} p_{j} \rho_{A|j} = \rho_{A},~~~\rho_{B}' = \sum_{j} p_{j} \pi_j.
\ee
Invoking these relations, we get
\be
H(A'|B') = \sum_{j} p_{j} H(\rho_{A|j}).
\ee
After minimization over all POVMs, $D$ reduces to $\mathcal{D}(A:B)$ as defined in Eq.~(\ref{discexp}). \textit{Quantum discord thus quantifies the minimum loss in yield of the FQSW protocol due to decoherence.} This is our main result, and shows that the performance of all the protocols in the quantum information family tree must be judged by the quantum discord. The connection between quantum discord and the FQSW protocol provides a metric for studying the advantage of coherence in accomplishing any of the children protocols that can be derived from the FQSW protocol. For example, we look at the noisy versions of quantum teleportation, superdense coding, and entanglement distillation. We then apply our general result to quantum computation.

The connection we have made here is subtle. Although it is known that entanglement is necessary for the success of the protocol, and that the presence of decoherence affects its performance, we have now provided a quantitative result of the amount of such a depreciation. We have shown that the amount by which a protocol suffers in the presence of decoherence is an inherent property of the quantum states involved. It suggests that the best state to be employed in a noisy quantum communication protocol is a tradeoff between the entanglement and discord of the state involved. The variation of discord and entanglement in quantum states is not monotonic~\cite{mono}, and a detailed study of this tradeoff will be presented in future work. In the next section, however, we demonstrate the power of our result by applying it to some well-known quantum information protocols.

\section{Quantum discord in the children protocols}
\label{sec:child}
The connection between quantum discord and the FQSW protocol provides a metric for studying the effect of coherence in accomplishing any of the so called ``children protocols" that can be derived from the FQSW protocol. In this section, we show that by connecting quantum discord with the FQSW protocol, we can interpret discord as the advantage of quantum coherence in noisy versions of teleportation, super-dense coding, and entanglement distillation. Finally, we reproduce an earlier result on the connection of quantum discord and quantum state merging.

\subsection{Noisy teleportation}

The noisy teleportation resource inequality can be expressed as
\be
\label{nt}
\langle\Psi^{AB}\rangle  + I (A:B) [c \rightarrow c] \geq I (A\rangle B)[q \rightarrow q],
\ee
obtained by combining the mother protocol with teleportation~\cite{dhw08}.  Here, $I(A\rangle B) = -H(A|B)$ is also known as the coherent information~\cite{ydh08}. When Bob undergoes decoherence, we get,
\begin{equation}
\label{ntd}
\langle\Psi^{A'B'}\rangle  + I (A':B') [c \rightarrow c] \geq I (A'\rangle B')[q \rightarrow q].
\end{equation}
The above can be interpreted as following: The net loss in the number of qubits that can be teleported when comparing the coherent teleportation (the one without any decoherence),
Eq.~(\ref{nt}), and the one which suffers decoherence, Eq.~(\ref{ntd}), is given by  $I(A\rangle B) - I(A'\rangle B') = H(A'|B') - H(A|B)$. We assume the classical communication to be free in this case, as long as we are teleporting unknown quantum states. We have $H(A|B) = H(A) - I(A : B) = H(A) - I(A : BC) = H(A|BC).$ As in Sec.~(\ref{sec:Main}), the application of the unitary $U$, but before discarding the subsystem $C$, the cost of teleportation is still given by $H(A'|B'C') = H(A|B)$. From Eq.~(\ref{eq:Idiff1}),
\be
\label{eq:Sdiff1}
H(A'|B') \geq H(A'|B'C') = H(A|B).
\ee
Therefore, we see that the advantage of the coherent protocol over the noisy version in teleporting unknown quantum states is equal to the quantum discord of the original state.

\subsection{Noisy super-dense coding}

Noisy super-dense coding can be derived by combining the mother protocol with super dense coding~\cite{dhw08}
\be
\label{superdense}
 [qq]  +  [q \rightarrow q] \succeq  2[c \rightarrow c],
\ee
showing that one can employ a shared ebit and a single bit of quantum communication to communicate 2 bits of classical information. Here, $[q \rightarrow q]$ represents one qubit of communication between two parties and $[qq]$ represents one shared ebit between two parties. Similarly, $[c \rightarrow c]$ represents one classical bit of communication between the parties.
The symbol $\succeq $ is used to denote exact attainability as compared to $\geq$ which is to denote asymptotic attainability. Combining these, the noisy super-dense coding protocol can be expressed as,
\be
\langle\Psi^{AB}\rangle  + H (A) [q \rightarrow q] \geq I (A:B)[c \rightarrow c].
\ee
When the party $B$ is undergoing decoherence, the noisy superdense coding can be expressed as,
\be
\langle\Psi^{A'B'}\rangle  + H (A') [q \rightarrow q] \geq I (A':B')[c \rightarrow c].
\ee
We note that $ H(A) =H(A')$. Thus, due to decoherence, the number of classical bits communicated through this protocol gets reduced by the amount $I (A:B) -  I (A':B')$, which is equal to the discord of the original state.

\subsection{Entanglement distillation}

The one-way entanglement distillation can be expressed as
\be
\langle\Psi^{AB}\rangle  + I (A:R) [c \rightarrow c] \geq I (A\rangle B)[qq].
\ee
This inequality can be derived by combining the FQSW protocol Eq.~(\ref{fqsw}) and recycling the $\frac{1}{2} I(A:R)$ ebits out of the total $\frac{1}{2} I(A:B)$ produced for teleportation, as shown in~\cite{dhw08}. Decoherence at Bob's end $B$ provides
\be
\langle\Psi^{A'B'}\rangle  + I (A':R') [c \rightarrow c] \geq I (A'\rangle B')[qq].
\ee
The net change in entanglement distillation is $I(A' \rangle B') - I(A\rangle B) = H(A|B) - H(A'|B'),$ which is the negative of the quantum discord of the original state. As is well known, classical communication between parties cannot enhance entanglement, and we can neglect the overhead of $I(A:R)-I(A':R')$ classical bits.

\subsection{Quantum state merging}

We showed in Chapter 5 that quantum state merging provides an operational interpretation for quantum discord~\cite{md10,cabmpw10}. It is the markup in the cost of quantum communication in the process of quantum state merging, if one discards relevant prior information. An intuitive argument for the above interpretation of quantum discord can be made through strong subadditivity~\cite{sm}
\be
\label{eq:ssa}
H(A|B,C) \leq H(A|B).
\ee
From the point of view of the state merging protocol, the above has a very clear interpretation. Having more prior information makes state merging cheaper. In other words, throwing away information will make state merging more expensive. Thus, if Bob discards system $C$, it will increase the cost of quantum communication needed by Alice in order to merge her state with Bob. 

We can easily derive the results of ~\cite{md10} starting from the FQSW protocol. We start by expressing the quantum state merging as a resource inequality
\ben
\label{sm}
\langle\Psi^{AB}\rangle  &+& H(A|B) [q \rightarrow q]  \nonumber \\
&+& I (A:B)_{\Psi} [c \rightarrow c] \geq \langle \mathbb{I}^{S \rightarrow \hat{B}} : \Psi^{S}\rangle.
\een
This accomplishes state merging from Alice to Bob at the cost of $H( A|B)$ bits of quantum communication. When $H(A|B)$ is negative, Alice and Bob can distill this amount of entanglement in the form of Bell pairs. Thus, quantum state merging provides an operational interpretation of $H(A|B).$

We can derive quantum state merging from the FQSW if the entanglement produced at the end of the FQSW protocol can be used to perform teleportation. We start by first noting the teleportation protocol~\cite{dhw08}
\be
\label{teleport}
 [qq]  + 2 [c \rightarrow c] \succeq  [q \rightarrow q].
\ee
From the FQSW protocol Eq.~(\ref{fqsw}), using the entanglement produced at the end for quantum communication Eq.~(\ref{teleport}), one gets the quantum state merging primitive Eq.~(\ref{sm}).

As in Sec.~(\ref{sec:Main}), the resource inequality for the noisy version of the quantum state merging protocol
\ben
\langle\Psi^{A'B'}\rangle  &+& H(A'|B') [q \rightarrow q] \nonumber \\
&+& I (A':B')_{\Psi} [c \rightarrow c] \geq \langle \mathbb{I}^{S \rightarrow \hat{B}} : \Psi^{S}\rangle.
\een
The cost of quantum communication in this case is $S(A'|B'),$ and the mark up in this cost is $S(A'|B') - S(A|B)$, which is equal to the quantum discord of the original state.

\section{Quantum discord in quantum computation}
\label{sec:comp}
Our approach also allows us to quantify the effect of decoherence on the efficiency of quantum computations as a communication protocol. While our work is restricted to a particular model of quantum computation (where computation is viewed as a communication process), it provides insights into the role of discord in quantum computation.
 In this model, a quantum computation can be described using two registers, the memory ($M$) and the computational register ($C$), and two programmers, the sender and the receiver~\cite{brv00}.  At the beginning of the computation, the sender encodes the problem to be solved (the message) as a quantum state in the memory register.  The initial state is operated upon by a sequence of unitaries (the channel), and the receiver collects the answer at a later time by looking at the computational register $C$, which contains the output of this communication process when the computation is finished. Consequently, the classical capacity of a quantum communication channel can be connected to the efficiency of quantum computation. Initially, the two registers are uncorrelated. As the computation/communication proceeds, the channel, which are the sequence of unitaries that implement the the black box oracle queries generates correlations between the two registers.

Let the memory register of size $N$ be in the state $\sum_{j=1}^N p_{j} | j \rangle_{M}$. The computational register is in the state $\rho_{C}^{0}$ giving the combined initial state $\sum_{j=1}^N p_{j} | j \rangle_{M} \otimes \rho_{C}^{0}$. As the computation proceeds, the two registers become correlated $\left(\sum_{j=1}^N p_{j} | j \rangle_{M}\right) \otimes \rho_{C}^{0} \rightarrow \sum_{j=1}^N p_{j} | j \rangle_{M} \otimes \rho_{C}(j).$
The mutual information between $M$ and $C$ is given by $ I(M:C)= H(M)-H(M|C)= H(C) - \sum_{j=1}^N p_{j} H\left(C(j)\right),$
where $\rho_{M}$ and $\rho_{C}$ are the reduced density matrices of registers $M$ and $C$ respectively and $\rho_{MC}$ is the density matrix for the entire system. The mutual information thus calculated is also equal to the Holevo bound for the classical capacity of a quantum channel ~\cite{hol73}. The goal of quantum computation is to maximize this mutual information. Now, any quantum operations on $C$ will act to reduce the mutual information between $M$ and $C$, i.e. $I(M' : C') \leq I(M : C).$ Thus, to the degree that $I (M : C)$ is a measure of the efficiency of the quantum computation, it is reduced by the amount $I (M:C) -  I (M':C')$, which is equal to the discord of state $\rho_{MC}$.

In a query complexity model of quantum computation, the change of mutual information in a single step sets a limit to the scaling of the computation with the problem size. If the variation in mutual information $I(M:C)$ in a single step, say $t,$ of a computation is $|I(M_{t+1}:C_{t+1})-I(M_t:C_t)|=f(N)$ bits, then the query complexity for the given computation is at least $\log N/f(N),$ since $\log N$ bits is the maximum mutual information for a system of size $N.$ If the corresponding variation in $I(M':C')$ is $|I(M'_{t+1}:C'_{t+1})-I(M'_t:C'_t)|=g(N)$ bits, then
\ben
|g(N)-f(N)| &\leq& \Big| I(M'_{t+1}:C'_{t+1})-I(M_{t+1}:C_{t+1}) \nonumber \\
    &&  -(I(M'_{t}:C'_{t})-I(M_{t}:C_{t}) ) \Big| \nonumber \\
    &=& \left| \mathcal{D}(M_{t+1}:C_{t+1})- \mathcal{D}(M_{t}:C_{t})\right| \nonumber \\
    &\leq&  \overline{\Delta}(M:C)
\een
where $ \overline{\Delta}(M:C)$ is the maximum possible change in quantum discord in a single step of the computation. As the query complexity of the computation is now bounded by $\log N/g(N),$ which in turn is bounded by the consumption of quantum discord in the computational process, our result provides a quantitative characterization of quantum discord as a resource in quantum computation.

\section{Discussions}
\label{sec:conc}

The FQSW protocol is the unification of a large class of quantum information theory protocols. We developed a vital link between the FQSW protocol and quantum discord, providing a unified picture illuminating the role of quantum discord and hence quantum correlations in essentially all quantum communication protocols.

While all our results are derived for finite-dimensional cases, gaussian quantum discord \cite{ad10} has been related to a generalisation of quantum dense coding for continuous-variable states, when all the states and operations involved are gaussian. The problem was cast as the advantage that can be harnessed by using nonlocal quantum interactions. Our results can be extended to continuous variable systems.

Our work elucidates the role non-classical correlations, as captured by quantum discord, play in quantum communication tasks. For an important and large class of protocols, quantum discord serves as a metric of how coherently the protocol performs. We have been able to quantify the loss quantum communication protocols suffer due to decoherence and show that this is aptly captured by quantum discord.

\chapter{ Summary and Outlook}

\section{Summary of the main results}

This thesis addresses some fundamental questions in quantum mechanics as seen in the light of quantum information theory. We also explore the unique properties of quantum systems, like quantum correlations in bipartite quantum systems, that serve as a resource and therefore can be harnessed for building future technology and engineering applications based upon quantum theory.

We have studied the relationship between entanglement and chaos for a system of isotropically coupled tops in which one of the tops receives a periodic kick around a fixed axis.  Here the chaos and entanglement arise from the same coupling mechanism removing any ambiguity between chaos in the subsystem vs. chaos in the joint-system dynamics.  The results reported here give further evidence of the fact that chaotic systems take quantum initial conditions to pseudo-random quantum states, and that the large long-time entanglement average of states undergoing quantum chaotic dynamics is just that of a typical state in the Hilbert space. We see the confirmation of this picture in the excellent agreement between the properties of ensembles of quantum states and the numerical results for the eigenvector statistics and long-time entanglement average for the completely chaotic system. This approach was also found to be highly flexible, applying to subspaces and mixed phase spaces.

The dynamical generation of random quantum states has implications for the dynamical generation of entanglement.  It is well known that for large dimensional bipartite Hilbert spaces, a random state is highly entangled with almost the maximum entanglement allowed by the dimension \cite{Hayden}.  As the large dimensional limit is equivalent to the $\hbar \rightarrow 0$ semiclassical limit, and to the degree that the quantum analogs of chaotic Hamiltonians generate random states, one expects near maximal dynamical generation of entanglement in quantum chaos, to a value that is predicted by the statistics at hand.  This is not to say that regular dynamics (quantum analogs of integrable motion) cannot lead to highly entangled states.  Indeed, such behavior is seen, and has been previously noted in \cite{Kus04}.  Regular dynamics, however, show oscillatory behavior, including in the generation of entanglement.  Chaotic dynamics, by contrast, lead to quasi-steady state behavior, and typically lead to higher values of time-averaged entanglement than regular motion.  Taken together, these facts imply that the long-time average entanglement in a bipartite system should be a strong signature of classical chaos, closely associated with ergodicity in the two dynamical descriptions.


In chapter \ref{QSCQT}, we found that the rate at which one obtains information about an initially unknown quantum state in quantum tomography is correlated with the extent of chaos in the system. This is a new quantum signature of classical chaos that we have found. In fact, our results can be regarded as signatures of chaos in closed quantum systems undergoing unitary evolution. We have been able to quantify the information gain using the Fisher information associated with estimating the parameters of the unknown quantum state. When the system is fully chaotic, the rate of information gain agrees with the predictions of the random matrix theory. 

In the last two chapters we gave an operational interpretation of quantum discord through quantum state merging 
and established the role of quantum discord in quantum communication protocols. Our work elucidates the role non-classical correlations, as captured by quantum discord, play in such quantum information processing tasks. For an important and large class of protocols, quantum discord serves as a metric of how coherently the protocol performs. We have been able to quantify the loss quantum communication protocols suffer due to decoherence and show that this is aptly captured by quantum discord. In conclusion, this study places quantum discord squarely in the midst of quantum informational concepts and opens up the way for its manipulation as a resource in quantum information processing. We also hope that our work will serve as a stepping stone for a more comprehensive and unified understanding of quantum physics, thermodynamics and information theory. Our work can be summarized by one slogan: \textbf{``Quantum discord is the advantage of quantum coherence in quantum information theory".}

\section{Outlook}

One of the indicators of good science is the questions it raises as much as the ones it solves.
This work can be extended in several interesting directions. I outline a few of them.
 Though quantum systems show no sensitivity to initial conditions due to unitarity, they do show a sensitivity to parameters in the Hamiltonian~\cite{per00}. This poses interesting questions for quantum tomography and, more interestingly, for quantum simulations. Under what conditions are the system dynamics sensitive to perturbations and how does this affect our ability to perform quantum tomography? Under what conditions does the underlying quantum chaos effect our ability to accomplish quantum simulations in general? For example, Shepelyansky has done extensive work on the issue of many-body quantum chaos in the quantum computer hardware and its effect on the accuracy of quantum computation \cite{Shepelyansky}. Recently classical simulations of quantum dynamics have been connected to intergrability and chaos \cite{pz07}.
 
  
 Information theoretic characterization of quantum chaos has thrown light on the connections between entanglement and chaos~\cite{tmd08}. As already mentioned, quantum chaotic dynamics will drive the system into arbitrary superposition of quantum states. For a bipartite system and an initially pure product state of minimum uncertainty wave packet, this is reflected in the generation of highly entangled states in the Hilbert space. What happens when the initial state is a mixed state? I conjecture that in general, the generation of correlations between subsystems of a chaotic system can be explored using quantum discord. Moreover, discord can serve as a useful quantity to quantify the amount of quantum information generated when the system dynamics are perturbed.

Our work on chaos has interesting connections to thermalization in closed quantum systems. We have the evidence that the rate of thermalization in quantum systems is closely related to non-integrability, chaos and symmetries respected by the system. Could we find a ``thermalization witness"  based on the measurement record and properties of the covariance matrix? Many-body chaos is another avenue that can be exlpored using the signatures of chaos we have found.

Our work on quantum discord raises many interesting questions and opens new directions. For example, can a state with zero discord be treated as ``classical"? Are states with vanishing quantum discord useful for certain tasks which are not possible classically? For example, the question whether \textit{concordant} computations can be simulated classically has been investigated by Eastin \cite{adv}. A concordant computation is one in which after each stage of computation, the resulting quantum state is diagonal in a product basis, and hence has zero discord. The entire simulation is finding the right product basis, yet Eastin's findings suggest that it might be difficult to simulate such computations efficiently. Therefore, quantum states with vanishing quantum discord might have a non-classical character to them. This leads to a fundamental question about the border between quantum and classical. Is there another way of marking this border? 

Another related direction concerns with the quantification of the amount of non-classicality of quantum states. Any measure of non-classicality associates a number to a quantum state that quantifies the amount of non-classicality present in the state. In this thesis, we have been able to find the operational meaning of that number, when the non-classicality was quantified by quantum discord. Is there another way of quantifying the quantum character of a system?
If the answer is yes, then what is the operational significance of such a metric?

 








\end{document}